\documentclass[12pt]{article}
\pdfoutput=1

\usepackage{putex}
\usepackage{amsmath,amssymb,amsfonts}
\usepackage{psfrag}
\usepackage{footmisc}
\usepackage{url}
\usepackage{mathtools}
\usepackage{color}

\usepackage{tikz}
    \usepackage{amssymb,amsfonts,amsmath}
    \usepackage{tkz-euclide}
        \usetikzlibrary{arrows,calc,patterns}
\usepackage{pgfplots}

\definecolor{darkblue}{rgb}{0.1,0.1,.7}
\definecolor{purple}{rgb}{0.6,0,0.6}
\definecolor{orange}{rgb}{0.9,0.6,0}
\usepackage[colorlinks, linkcolor=darkblue, citecolor=darkblue, urlcolor=darkblue, linktocpage]{hyperref} 
\usepackage[square, comma, compress,numbers]{natbib}
\usepackage[]{amsmath}
\usepackage[]{graphicx}
\usepackage[]{latexsym}
\usepackage[utf8]{inputenc}
\usepackage{geometry}
\usepackage{amscd}
\usepackage[all,cmtip]{xy}
\usepackage{mathrsfs}

\usepackage[margin=10pt,font=small,labelfont=bf]{caption}
\usepackage{dsdshorthand}
\usepackage{changepage}
\usepackage{setspace}
\usepackage{tikz}
\usepackage{subcaption}

\usepackage[titles]{tocloft}
\usetikzlibrary{arrows,positioning}

\def\SL2{\widetilde{SL}(2,\mathbb R)}

\newcommand\mR{\mathbb{R}}
\newcommand\mZ{\mathbb{Z}}

\newcommand\MSU{M_{SU(1,1|1)}}

\numberwithin{equation}{section}

\newcommand{\grp}[1]{\mathrm{#1}}

\newcommand {\bes} {\begin {equation*}}
\newcommand {\ees} {\end {equation*}}
\newcommand {\beq} {\begin {equation}}
\newcommand {\eeq} {\end {equation}}
\newcommand {\bea} {\begin {eqnarray}}
\newcommand {\ea} {\end {eqnarray}}
\newcommand {\eea} {\end {eqnarray}}

\newcommand{\Sch}{\text{Sch}}

\usepackage{capt-of}

\newcommand{\NSch}{\mathcal{N}=2 \text{ Schw}}

\numberwithin{equation}{section}

\def\<{\langle}
\def\>{\rangle}

\tikzset{
    >=stealth',
    punkt/.style={
           rectangle,
           rounded corners,
           draw=black, very thick,
           text width=15em,
           minimum height=2em,
           text centered},
    pil/.style={
           ->,
           thick,
           shorten <=2pt,
           shorten >=2pt,}
}

 \def\ie{\begin{equation}\begin{aligned}}
\def\fe{\end{aligned}\end{equation}}

\begin{document}

\institution{Warsaw}{${}^1$ Department of Physics, University of Warsaw, ul. Pasteura 5, 02-093 Warsaw, Poland}
\institution{IAS}{${}^2$  Institute for Advanced Study, Princeton, NJ 08540, USA}
\institution{PU}{${}^3$ Joseph Henry Laboratories, Princeton University, Princeton, NJ 08544, USA}
    \institution{SU}{${}^4$ Stanford Institute for Theoretical Physics, Stanford University, Stanford, CA 94305, USA}

\title{
BPS and near-BPS black holes in AdS$_5$ \\ and their spectrum in $\mathcal{N}=4$ SYM 
}

\authors{J. Boruch${}^1$, M. Heydeman${}^{2,3}$, L. V. Iliesiu${}^4$, G. J. Turiaci${}^2$ }

\abstract{
We study quantum corrections in the gravitational path integral around nearly $1/16$-BPS black holes in asymptotically $AdS_5 \times S^5$ space, dual to heavy states in 4D $\cN=4$ super Yang-Mills. The analysis provides a gravitational explanation of why $1/16$-BPS black holes exhibit an exact degeneracy at large $N$ and why all such states have the same charges, confirming the belief that the superconformal index precisely counts the entropy of extremal black holes. We show the presence of a gap of order $N^{-2}$ between the $1/16$-BPS black holes and the lightest near-BPS black holes within the same charge sector. This is the first example of such a gap for black holes states within the context of AdS$_5$ holography. We also derive the spectrum of near-BPS states that lie above this gap. Our computation relies on finding the correct version of the $\cN=2$ super-Schwarzian theory which captures the breaking of the $SU(1, 1|1)$ symmetry when the black hole has finite temperature and non-zero chemical potential. Finally, we comment on possible stringy and non-perturbative corrections that can affect the black hole spectrum.}

\date{\begin{center}
    {\it Dedicated to the people of Ukraine}
\end{center}}
\maketitle
\tableofcontents

\newpage

\section{Introduction}

$1/16$-BPS states have recently played a crucial role in analyzing the duality between Type-IIB string theory in AdS$_5 \times S_5$ and 4D $\cN=4$ super-Yang Mills \cite{Maldacena:1997re,Gubser:1998bc,Witten:1998qj, Gutowski:2004ez, Gutowski:2004yv, Chong:2005hr, Kunduri:2006ek,   Cabo-Bizet:2018ehj, Choi:2018hmj, Benini:2018ywd, Honda:2019cio, ArabiArdehali:2019tdm, Cabo-Bizet:2019eaf, ArabiArdehali:2019orz, Murthy:2020rbd, Agarwal:2020zwm, Benini:2020gjh, GonzalezLezcano:2020yeb, Copetti:2020dil,   Cabo-Bizet:2020ewf, ArabiArdehali:2021nsx, Cassani:2021fyv, Aharony:2021zkr}. Such states can be accurately counted by computing the superconformal index, a grand canonical partition function with multiple chemical potentials for the black hole angular momenta and $R$-charges turned on. On the boundary side the superconformal index can be obtained exactly since its independent of the coupling \cite{Romelsberger:2005eg, Kinney:2005ej}. Expanding the exact answer in the large $N$ limit, most contributions to the superconformal index were matched with a corresponding Euclidean gravity saddles, with the dominant contributions given by well-known supersymmetric black hole solutions \cite{Aharony:2021zkr}. The computation of the superconformal index, on both the boundary and the bulk side, has provided a detailed check of holography and a detailed counting of black hole micro-states. 
 
  A lot less is known about black hole states that are not protected by supersymmetry. In particular, an important issue in classical black hole thermodynamics is the breakdown of the statistical description of black holes \cite{Preskill:1991tb,Michelson:1999kn}, occurring at low temperatures whose scale is power-law suppressed in the number of degrees of freedom describing the black hole. When quantum effects are included, two resolutions to this thermodynamic breakdown were found. First, for non-supersymmetric black holes in flatspace or AdS, a recent computation \cite{Iliesiu:2020qvm} which accounted for quantum effects occurring in the near-horizon region of near-extremal black holes showed an effective continuum of states with a strongly modified spectrum at the scale identified in \cite{Preskill:1991tb}. In the second resolution, due to the addition of fermionic degrees of freedom in the computation of quantum effects, the spectrum of nearly-supersymmetric black holes in supergravity in 4D flatspace or $(4, 4)$ supergravity in AdS$_3$  was shown to be drastically different \cite{Heydeman:2020hhw}: there is an exact degeneracy of supersymmetric black holes at extremality followed by a gap precisely at the energy scale at which a failure of black hole thermodynamics is predicted. Thus, quantum effects in the near-horizon region proved to be important in both cases. However, exact degeneracies at extremality and gaps between extremal and near-extremal states which were predicted in stringy constructions \cite{Strominger:1996sh, Maldacena:1996ds, Maldacena:1997ih} were only found in supersymmetric theories.
  
  In the case of supersymmetric black holes in AdS$_5$, we want to ask whether there is a gap (with an appropriate power-law suppression in $N$) in the spectrum of masses between the $1/16$-BPS black holes and the lightest unprotected black hole state (which we call near-BPS) in a sector with the same angular momenta and R-charge quantum numbers. These states preserve less supersymmetry than their flatspace counterparts, and a rigorous understanding of whether a gap is truly present has, up to the point of this paper, not been achieved.\footnote{In the supersymmetric flatspace and $(4,4)$ AdS$_3$ cases the presence of a gap in stringy examples was explained in \cite{Maldacena:1996ds, Maldacena:1997ih}. We could not find a string theory argument about the existence of the gap in the literature.} Even when it comes to the $1/16$-BPS states themselves, not all of their properties are known. Firstly, the superconformal index cannot distinguish whether such states are purely bosonic or a combination of bosonic and fermionic states, which would yield cancellations in the index. In other words, it is unclear whether the entropy associated to the index precisely matches the actual entropy of BPS black holes. Secondly, from the gravitational perspective, the degeneracy of the extremal  $1/16$-BPS states has not been rigorously understood due to the presence of an infinite number of zero modes observed when computing the one-loop determinant for black holes at extremality.\footnote{As we will explain shortly, these zero-modes get lifted when studying the partition function at small but non-zero temperature. This regularizes the one-loop determinant in the finite temperature partition function and this regularization will be responsible for the observed degeneracy among supersymmetric black holes.  }
  
  In this paper, we address all these questions by computing the low-temperature expansion of the partition function of near-BPS black holes. At zero temperature, the leading contribution to the partition function comes from $1/16$ BPS black holes that exhibit an $SU(1, 1|1)$ isometry in their near-horizon region \cite{Sinha:2006sh}. As the temperature is turned on, the $SU(1, 1|1)$ isometry is broken and we explain how the gravitational modes associated to this breaking are effectively captured by the $\cN=2$ super-Schwarzian theory \cite{Fu:2016vas, Stanford:2017thb, Mertens:2017mtv, Stanford:2019vob,Larsen:2019oll, Heydeman:2020hhw}.\footnote{The relation between these black holes and the $\cN=2$ Schwarzian theory was explored at the classical level in \cite{Larsen:2019oll,Choi:2021nnq}.} At small temperatures, the super-Schwarzian theory and the associated modes in supergravity become strongly coupled. Luckily the partition function of the super-Schwarzian can be found exactly and this consequently allows us to reliably compute the low-temperature corrections to the free energy of such black holes, appearing at linear order in $T$ (from classically evaluating the super-Schwarzian action and the black hole action) and logarithmic order in $T$ (from evaluating quantum corrections in the super-Schwarzian or, equivalently, specific set of modes of the graviton, gravitino and gauge fields in the black hole background). This calculation is valid up to temperatures that are much smaller than the one identified in \cite{Preskill:1991tb}. 
  
  Studying these quantum corrections to the black hole spectrum leads to the main results of our paper. We find that there is indeed a gap between the $1/16$-BPS state and the lightest near-BPS black hole states precisely at the energy scale identified in \cite{Preskill:1991tb} (once again, in contrast to the non-supersymmetric case of black holes in AdS$_5$). Above this gap there is a continuum of states whose density we can predict (a precise understanding of the discreteness of the spectrum in this sector would require a better non-perturbative understanding of type IIB string theory). Additionally, we find that the $1/16$ BPS black hole states all have the same charges, to leading order in $N$, confirming that there is no cancellation in the superconformal index. This improves a previous argument for black holes with an emergent $SU(1,1|1)$ symmetry described in \cite{Benini:2015eyy}, that did not take into account the physics from the Schwarzian mode. This is important since there are also theories with an emergent $SU(1,1|1)$ symmetry and vanishing index, and we show this does not happen for the $1/16$-BPS states of $\mathcal{N}=4$ Yang Mills.
  
  Before diving into a more quantitative description of our results, it is useful to first describe some of the properties of the black hole solutions whose spectrum we determine in this paper. When viewed from the 10D perspective, such black hole solutions have five angular momenta: two angular momenta parametrizing rotations in AdS$_5$ as well as three angular momenta parametrizing rotations on the $S^5$.  On the boundary side, the former are the angular momenta of the dual state within the conformal group, which we will denote by $J_{1,2}$ while the latter are three Cartans of the $SO(6)$ R-symmetry group, which for simplicity we will set to all be equal and denote by $R$. The mass of the black hole can be fixed in terms of the temperature in addition to these five charges. For extremal BPS black holes, their mass can be determined in terms of four (of the five) angular momenta since there is an additional non-linear relation between these charges necessary in order for Killing spinor solutions to exist in the geometry while at zero-temperature. For instance, the BPS value of $R$ as well as the mass of the black hole, are uniquely determined at the BPS value by  $J_{1,2}$.

With the quantum numbers of such black holes in mind, our computation of the partition function predicts the density of states of the near-BPS black holes as a function of the scaling dimension $\Delta$ of the corresponding states in $\mathcal{N}=4$ Yang Mills. The answer is given by 
 \begin{align}
  \rho_{R, J_1, J_2}(\Delta) &\sim e^{S^*}\bigg(\underbrace{\delta_{R, R^*}\delta\left(\Delta - \Delta_{BPS}\right)}_{\substack{\text{Degenerate}\\ 1/16-\text{BPS black holes}}}\nn \\ &+
  \,\underbrace{\frac{\sinh\left( \frac{\pi}2 \sqrt{\frac{1}{\Delta_\text{gap}} \left(\Delta -   \Delta_{\text{extremal}}\right)}\right)}{2\pi (\Delta - \Delta_\text{BPS}) }}_{\substack{\text{Continuum density of states}\\ \text{above the gap}}}\, \,\,\underbrace{\Theta \left(\Delta- \Delta_{\text{extremal}}\right)}_{\substack{\text{Gap resulting}\\ \text{from quantum corrections}}} + \underbrace{(R \rightarrow R+1)}_{\substack{\text{Degeneracy within} \\ \text{supermultiplet} }}\bigg)\, .
  \label{eq:density-of-states-N=4-SYM}
\end{align}   
The first line represents the degenerate contribution of extremal states when the R-charge is fixed to its BPS value $R^*$ by the angular momenta $J_{1,2}$.  The degeneracy is precisely found to be $ e^{S^*}$, where $S^*$ is the Bekenstein-Hawking entropy of the extremal supersymmetric black hole.\footnote{This degeneracy is up to corrections logarithmic in $S^*$ or, equivalently, up to logarithmic area corrections in the free energy. } The second line represents the contribution of the continuum of states (with non-perturbative gaps expected to be exponentially small in $N$) starting at the scaling dimension $\Delta_{\text{extremal}}$ which then represents the extremal black hole state. This gives a gap above the BPS scaling dimension $\Delta_\text{BPS}$ (defined for $R=R^\star$), that scales as 
  \be
    \Delta_{\text{extremal}}(R) - \Delta_{\rm BPS} &=\frac{1}{2}(R-R^\star)+ \Delta_\text{gap} \left(2R-2R_*- {1}\right)^2\,,\nn \\
\Delta_\text{gap} &= \frac{\tilde \Delta(\mathcal{J}_1,\, \mathcal{J}_2)}{N^2}   + O\left(\frac{1}{N^3}\right)\,,
    \ee
  where $N$ is related to the 5D Newton constant by $N^2 = \frac{\pi \ell_{AdS_5}^3}{2G_5}$, and where we rescale the two angular momenta $J_{1,2} = N^2 \mathcal{J}_{1,2}$,\footnote{Since $\mathcal{J}_{1,2}$ do not need to scale with any parameter in the theory for the black holes discussed in this paper.} and where $\tilde \Delta(\mathcal J_1, \mathcal J_2)$ is a function that we determine exactly. Since the states that are part of the continuum are unprotected, they come in super-multiplets with charges $R$ and $R+1$. Now we can see the meaning of the gap scale $\Delta_{\rm gap}$. For any charge $R\neq R^\star$, the spectrum starts at $ \Delta_{\text{extremal}}(R)$ with no gap. But for $R=R^\star$, we have the BPS states at scaling dimension $\Delta_\text{BPS}$ and the first excited state above them in this charge sector start at $\Delta_\text{BPS}+\Delta_\text{gap}$ (the continuum sector at $R=R^\star$ comes from supermultiplets with highest R-charge $R^\star$ and $R^\star+1$). This is a large-$N$ analysis and therefore we cannot rule out the possibility of an order $O(1)$ number of states between $\Delta_\text{BPS}$ and $\Delta_\text{BPS}+\Delta_\text{gap}$. We conjecture that $\Delta_{\rm gap}$ is the true gap of the theory at large $N$.

  When the chemical potential of the $R$-symmetry is fixed appropriately, the grand canonical partition function can also be used to compute the superconformal index of such black holes and the only contribution comes from the first line of \eqref{eq:density-of-states-N=4-SYM}. This result can be obtained by directly computing the corresponding supersymmetric index in the $\cN=2$ super-Schwarzian theory. This matches the result from the leading gravity saddle found in \cite{Cabo-Bizet:2018ehj, Aharony:2021zkr} and, additionally, shows that the one-loop determinant from the gravitational theory matches the computation of the index in the boundary theory. An interesting feature that comes out of our analysis is that the IR R-charge of the $\mathcal{N}=2$ super-Schwarzian theory is shifted by a known value $R^*$ compared to the UV R-charge of $\mathcal{N}=4$ Yang Mills. 
  
  We would like to emphasize that since we work in a mostly canonical ensemble where almost all charges are fixed, we avoid the subtleties encountered in the grand canonical ensemble regarding which sheet the superconformal index is computed \cite{Cassani:2021fyv}. In our formalism, these issues would appear from attempting to sum over charges to construct the grand canonical answer.  
  
  Additionally, we compute the leading (in $\alpha'$) non-zero stringy correction to the black hole spectrum. While as expected the scaling dimension and charges of BPS states remains unaffected, $\Delta_\text{gap}$ and the scaling dimension of the extremal state $ \Delta_\text{extremal}(R)$ are affected, and are pushed towards lower energies. Nevertheless, because the $\cN=2$ super-Schwarzian theory remains a good effective description even in the presence of stringy corrections, the overall dependence of the density of states on  $\Delta_\text{gap}$ and $ \Delta_\text{extremal}(R)$ remains as in \eqref{eq:density-of-states-N=4-SYM}.

    A technical difficulty in our computation comes from identifying the correct parameters of the $\cN=2$ super-Schwarzian, that capture the correct breaking of the $SU(1, 1|1)$ isometry as the temperature of the near-$1/16$ BPS black holes is increased. In particular, there are multiple version of the $\cN=2$ super-Schwarzian since the R-symmetry group associted to the breaking of the near-horizon isometry is $U(1)$. One can change the radius of the associated $U(1)$ mode in the super-Schwarzian by choosing a different value for the fundamental charge in the theory, and, additionally, one can add a topological $\theta$-angle term associated to this $U(1)$ mode in the super-Schwarzian action. Depending on this fundamental charge and on the value of $\theta$, the $\cN=2$ super-Schwarzian density of states can have widely different behaviors:\footnote{This is closely related to how the spectrum of a particle moving on a circle  is also controlled by the radius of the circle and by the $\theta$-angle term that one can also add to the action \cite{Gaiotto:2017yup}.  } for instance, in some cases there is gap and the BPS states are purely bosonic, while in others the theory has no such gap and has degenerate bosonic and fermionic BPS states (leading to possible cancellations in the superconformal index). For black hole in AdS$_5$, we explain how each parameter is fixed from the perspective of the $\cN=4$ SYM boundary theory and of the bulk supergravity theory. In particular, we find that the fundamental $R$-charge in the super-Schwarzian is fixed to be $1$ from the quantization of the $SO(6)$ R-symmetry charges and two angular momenta along $S^3$ and find that  $\theta = 0$ by carefully evaluating the 10D supergravity action. Fixing these two parameters determines the value of the gap and density of states that we described above. Moreover, we also determine the Schwarzian energy and R-charge in the IR, in terms of the $\mathcal{N}=4$ SYM scaling dimensions $\Delta$ and R-charge $R$.

 One might however wonder whether there are  near-BPS black holes in AdS/CFT whose density of states is also controlled by the $\cN=2$ super-Schwarzian but with different values for the fundamental charge and $\theta$-angle. For instance, one might ponder whether there are situations when there is no thermodynamic mass gap above the BPS state. We construct such near-BPS black holes in AdS$_3$ for $\cN=(2,2)$ and $\cN=(2, 0)$ supergravity. We explain how, in such a case, the density of states predicted by the super-Schwarzian can also be obtained from modular invariance in the boundary $2D$ $\cN=(2,2)$ superconformal field theory by making mild assumptions about the boundary theory (i.e.~that it has large central charge and a non-vanishing twist gap).\footnote{As noticed in \cite{Mertens:2017mtv,Lam:2018pvp} in theories with other amounts of supersymmetry, this connection originates from the equality between the partition function of the (super)Schwarzian and the semiclassical limit (large central charge) of the vacuum (super)Virasoro character. }

    The rest of this paper, is organized as follows. In section \ref{sec:N=2-super-Schwarzian} we first describe the connection between the Schwarzian theory and the spectrum of near-extremal black holes more broadly. We then describe the various features of the $\cN=2$ super-Schwarzian theories which will be needed to extract the spectrum of near-BPS black holes. In section \ref{sec:BPS-and-near-BPS-BHs}, we give a detailed description of the BPS and near-BPS black hole solutions in AdS$_5$. We work in a mixed ensemble with some charges fixed, and argue that in this ensemble the index is always in a deconfined phase. We follow this by an analysis focused on the steps necessary to determine the coupling, fundamental charge and $\theta$-angle in the corresponding $\cN=2$ super-Schwarzian theory by analyzing the low-temperature expansion of the action and the super-algebra of the near-horizon isometry present for BPS black holes. Putting these results together, we give a detailed discussion of the spectrum of near-BPS black holes. We conclude with a discussion about the leading stringy corrections to the spectrum of such black holes.  We discuss further appearance of the $\cN=2$ super-Schwarzian within holography in section \ref{sec:(2,2)-sugra-examples} for black holes in $(2,2)$ supergravity in AdS$_3$, dual to heavy states in 2D $(2,2)$ SCFTs. Finally, in section \ref{sec:discussion} we discuss possible non-perturbative corrections to the spectrum of $\cN=4$ SYM and 2D $(2,2)$ SCFTs and conjecture that the gap persists in the spectrum even when including states that exhibit string excitations in the black hole background. 

\vspace{0cm}

\section{The $\cN=2$ super-Schwarzian}

\label{sec:N=2-super-Schwarzian}

\subsection{From the Schwarzian to near-extremal black holes}
\label{sec:schwarzian-philosophy} 

The connection between the dynamics of near-extremal black holes and the one-dimensional Schwarzian theory  has been extensively studied \cite{Sachdev:2015efa, Almheiri:2016fws, Anninos:2017cnw, Turiaci:2017zwd, Nayak:2018qej, Moitra:2018jqs, Hadar:2018izi, Castro:2018ffi, Larsen:2018cts,  Moitra:2019bub, Sachdev:2019bjn, Hong:2019tsx, Castro:2019crn, Charles:2019tiu,Larsen:2020lhg, Iliesiu:2020qvm,Heydeman:2020hhw}. Through this connection, a better understanding of the spectrum of near-extremal black holes has been achieved and the previously  open question about the existence of a thermodynamic mass gap has been resolved for a variety of near-extremal black holes \cite{Iliesiu:2020qvm,Heydeman:2020hhw}. Nevertheless, there are still near-extremal black holes whose connection to the Schwarzian has not yet been completely understood and consequently, their spectrum near-extremality remains unknown. One such example, are the near-BPS black holes in AdS$_5$ whose connection to the Schwarzian will be extensively discussed in section \ref{sec:BPS-and-near-BPS-BHs}. It is useful to first review the general mechanism through which the spectrum of near-extremal black holes is related to that of the Schwarzian with various amounts of supersymmetry. 

 The clear-cut way to understand the relation between the two spectra is by performing a dimensional reduction of the full gravitational theory in the near-horizon region to AdS$_2$ and isolate the low-energy modes whose one-loop determinant solely is temperature $T$ dependent at leading order.   In the limit in which the inverse-temperature $\beta$ is much larger than the horizon size of the black hole at extremality $r_0$ ($\beta \gg r_0$) and under the assumption that the extremal entropy $S_0$ is the largest dimensionless parameter in the problem, the theory that results from the dimensional reduction is a two-dimensional theory of gravity coupled to a dilaton field (which parametrizes the area of the sphere on which the reduction is performed). In addition, the two-dimensional theory includes gauge fields that result either from the s-wave reduction of gauge fields (whose gauge group we will denote by $G$) in the original theory or as Kaluza-Klein modes that capture the isometry group of the space on which the dimensional reduction was performed (whose gauge group we will denote by $ G_\text{iso}$). In theories of supergravity which include fermions in higher dimensions, an analogous procedure leads to a coupling between a dilatino field and the gravitino  in the 2D theory. Expanding this action in $S_0$, one finds that the theory can be approximated by JT gravity plus a BF theory whose gauge group is $G_{BF} = G \times G_\text{iso}$. For higher dimensional theories of supergravity, the dilaton and metric as well as the gauge field entering in the BF theory all couple to the gravitino and dilatino.  Fluctuations in the region outside of the near-horizon region can be captured by a boundary term for the JT gravity action and for the BF action computed along the curve that separates the near-horizon region from the asymptotic region. 

For concreteness, we will briefly review the example of large near-extremal Reissner-Nordstrom black holes in (bosonic) Einstein-Maxwell theory in AdS$_5$ (i.e.~whose horizon size at extremality $r_0$ is much larger than the AdS$_5$ size $r_0\gg \ell_{\text{AdS}_5}$ and whose inverse-temperature $\beta \gg r_0$). The spectrum of such black holes was extensively studied in \cite{Iliesiu:2020qvm}. Following the steps outlined above, the canonical partition function of such black holes can be expressed as
\be 
Z(\beta&,\, Q, \, J_{1,2}=0) = \Tr_Q ~e^{-\beta H}  \nn \\ &= S_0(Q)^{\#} ~e^{S_0(Q) - \beta E_0(Q)}  \int D\phi  Dg_{\mu \nu} e^{-\int_{\text{AdS}_2}\phi(R + \Lambda_{\text{AdS}_2}) - \frac{\phi_b(Q)}{\epsilon} \int_{\partial \text{AdS}_2} (K-1)} \nn   \bigg[1+ O\left({1}/{S_0}\right)\bigg] \nn \\ &= S_0(Q)^{\#} ~e^{S_0(Q) - \beta E_0(Q)}  \int \frac{Df(\tau)}{SL(2, \mR)} \,e^{\phi_b(Q)\int_0^\beta d\tau  \text{Sch}(f,\tau)} \bigg[1+ O\left({1}/{S_0}\right)\bigg] \nn \\ &= \underbrace{S_0(Q)^{\#}}_{\substack{\text{Sen's logarithmic} \\ \text{corrections}}} \underbrace{\left(\frac{1}{M_{SL(2)}\beta}\right)^{3/2}}_{\substack{\text{Schwarzian} \\ \text{all-loop correction}}} \,\,\,\exp\underbrace{\left({S_0(Q) - \beta M_0(Q) + \frac{2\pi^2 }{M_{SL(2)}\beta}} \right)}_{\substack{\text{Extremal entropy,}\\ \text{extremal energy,}\\ \text{ and the Schwarzian }\sim 1/\beta \text{ correction}\\ \text{to the action} }} \bigg[1+ O\left(\frac{1}{S_0}\right)\bigg]\,,
\label{eq:near-extremal-RN-part-function}
\ee
where the extremal mass $M_0(Q)$, the ``extremal entropy'' $S_0(Q)$ and the Schwarzian coupling denoted by $\phi_b(Q)$ (or by the more physically meaningful notation $M_{SL(2)}^{-1}$ which we shall explain shortly) are given by\footnote{Here, $Q$ can be viewed as the charge associated to the supergravity gauge field $A$, whose conventions we set in \eqref{eq:5DLag} and thereafter. For reader convenience, in this section we leave the radius of AdS$_5$ to be arbitrary and set to $\ell_{\text{AdS}_5}$.  }
\be 
\label{eq:M0-S0-phiB-largeRN}
M_0(Q) \sim \frac{G_5^{1/3} (|Q|)^{4/3}}{\ell_{\text{AdS}_5}^2}\,,  \qquad  S_0(Q) \sim  |Q| \,, \qquad  \phi_b(Q) = M_{SL(2)}^{-1} \sim \frac{\ell_{\text{AdS}_5}^2 |Q|^{\frac{2}3}}{G_5^{\frac{1}3}}\, ,
\ee
where $\ell_{\text{AdS}_5}$ is the radius of $AdS_5$ and $G_5$ is the Newton constant.
To obtain the first line of \eqref{eq:near-extremal-RN-part-function}, we peform a dimensional reduction on $S^3$ in the $AdS_2\times S^3$ near-horizon region and then expand the resulting action at large $S_0$. For large black holes, the radius of $AdS_2$ is the same as the radius of $AdS_5$, which fixes $\Lambda_{{\rm AdS}_2} = 6/(\ell_{\text{AdS}_5}^2)$. The leading order result yields the action of JT gravity, whose degrees of freedom are the 2D metric of the near-horizon region and the dilaton $\phi$ which parametrizes the size of $S^3$. Above, the BF terms associated to the gauge group $G_{BF}=U(1) \times SO(4)$ do not give a non-trivial contribution since we are fixing both the $U(1)$ and $SO(4)$ fluxes when fixing the charge of the black hole to $Q$ and its angular momenta $J_1=J_2 = 0$ (this will be contrasted with the case of the grand-canonical partition function below) \cite{Iliesiu:2019lfc}. There is additionally a Gibbons-Hawking-York (GHY) boundary term at the edge of the near-horizon region where the dilaton (fixed to $\phi|_{\partial AdS_2} = \phi_b/\epsilon$) as well as the induced boundary metric are fixed (to have a proper boundary length $\beta/\epsilon$). 

In addition to these these modes there are also KK modes obtained from the dimensional reduction on $S^3$. The one-loop determinant of most KK modes corresponding to massless fields in the original theory yield logarithmic corrections to the extremal entropy, given by $S_0^\#$ which can in principle be computed using the methods in \cite{Banerjee:2010qc, Banerjee:2011jp, Sen:2011ba, Sen:2012cj, Iliesiu:2020qvm}.\footnote{For fields that are not massless in the original gravitational theory the one-loop determinant yields an answer that, to leading order, is independent of the extremal entropy $S_0$.} Here, the exact value of $\#$ depends on the massless field content in the original gravitational theory and is unimportant in the analysis of this paper since it does not affect the energy dependence of the density of states but only its overall scaling.  The remaining modes which are not taken into account are precisely the JT gravity modes which we have separated in the first line of \eqref{eq:near-extremal-RN-part-function}. At zero temperature, these are the zero-modes that are ubiquitous when computing one-loop determinants in black hole backgrounds and are given by the set of large diffeomorphisms which do not vanish close to the boundary of the near-horizon region.\footnote{See for instance \cite{Sen:2011ba} for a treatment of these modes.} However, when working at finite temperature, these zero-modes are lifted and become the boundary modes of the near-horizon region weighed by the Schwarzian action. This can be seen when going from the first to the second line of  \eqref{eq:near-extremal-RN-part-function} by integrating out the dilaton and rewriting the GHY term in terms of a field $f(\tau)$ which parametrizes the set of possible large diffeomorphisms. Equivalently,  $f(\tau)$ parametrizes the shape of the boundary of the near-horizon region. The Schwarzian theory can be seen to be weekly-coupled when $T\gg M_{SL(2)}$ and becomes strongly coupled when $T\sim M_{SL(2)}$. Finally, to go from the second to the third line and compute the partition function for any coupling, one can use the fact that the path integral of the Schwarzian theory is exactly solvable and is in fact one-loop exact. This one-loop determinant can be obtained by accounting for the three remaining bosonic zero-modes that survive even at finite temperature and are due to the near-horizon $SL(2, \mR)$ isometry.

The low-temperature expansion of the action is sufficient to obtain the full partition function  in \eqref{eq:near-extremal-RN-part-function}. This is because the Schwarzian theory can be viewed as the effective theory for the breaking of the near-horizon $SL(2, \mR)$ isometry, as the temperature is turned-on. This motivates denoting the Schwarzian coupling by $M_{SL(2)}$.

The result in \eqref{eq:near-extremal-RN-part-function} implies that the density of states obtained by Laplace transforming \eqref{eq:near-extremal-RN-part-function},  
\be 
\rho(E, Q) = S_0(Q)^{\#} ~e^{S_0(Q)} \sinh\left(2\pi \sqrt{ \frac{E}{M_{SL(2)}} }\right) \Theta(E-M_0) \,.
\label{eq:density-of-states}
\ee
Before we proceed, it is worth pondering the meaning of the continuum of states in \eqref{eq:density-of-states}. The partition function \eqref{eq:near-extremal-RN-part-function} is obtained in a $1/S_0$ expansion and we have only kept the leading order term. Consequently, the density of states \eqref{eq:density-of-states} also receives perturbative and non-perturbative corrections in $S_0$. Such corrections, which we do not have control over, could  in principle lead to a discretum of states as one would expect from the perspective of the boundary dual CFT. Nevertheless, \eqref{eq:density-of-states} is still useful: when integrating the density of states within some energy interval we should obtain the number of black hole microstates within that energy interval, once again, up to $1/S_0$ corrections.

For the purposes of this paper, it is also useful to study the partition function in the grand-canonical ensemble, imposing that the holonomy of the $U(1)$ gauge field is fixed to $e^{\oint A} = e^{-\beta \mu}$  at the boundary of the spacetime, along the thermal circle. The partition function can be instructively rewritten as, 
\begin{align}
\label{eq:example-grand-canonical-part-function}
Z(\beta, \mu) &= \Tr \,e^{-\beta H-\beta \mu Q } \nn\\  & = \frac{S_0^{\#}  }{\left(M_{SL(2)}\beta \right)^{3/2}} e^{S_0-\beta \mu Q_* -\beta M_0 + \frac{2\pi^2 }{M_{SL(2)}\beta}}\sum_{q \in \mZ } e^{2\pi \xi q -\beta M_{U(1)} \frac{q^2}{4}}   \nn \\ &= {S_0^{\#} } e^{S_0-\beta \mu Q_* -\beta M_0 + \frac{2\pi^2 }{M_{SL(2)}\beta}} \sum_{n\in \mZ}  \frac{1}{M_{SL(2)}^{3/2} M_{U(1)}^{1/2}\beta^2} e^{\frac{4\pi^2(\xi + i n)^2}{M_{U(1)}\beta} }	\,,
\end{align}
The grand-canonical partition function can be obtained from   \eqref{eq:near-extremal-RN-part-function} by summing over fixed charges. This sum is dominated by the charge $Q_*$, which in terms of the chemical potential $\mu$ is given by the solution of $\mu = \partial_Q M_0(Q)|_{Q=Q_*}$.\footnote{$S_0$, $M_0$, $M_{SL(2)}$ and $M_{U(1)}$ in \eqref{eq:example-grand-canonical-part-function} depend on the charge $Q=Q_*$ as in \eqref{eq:M0-S0-phiB-largeRN} and \eqref{eq:U(1)-mode-parameters}.} Above, $q$ can be viewed as the quadratic fluctuation around the charge $Q_*$. Similarly, $\xi/\beta$ can be viewed as the effective $U(1)$ chemical potential in the near-horizon region, while $M_{U(1)}$ can be viewed as the the coupling of the $U(1)$ mode associated to the s-wave gauge field in the near-horizon region. For concreteness,  these are given by the thermodynamic relations
\be 
\label{eq:U(1)-mode-parameters}
2\pi\xi=  \left(\frac{\partial S_0}{\partial Q}\right)_{T=0}\,, \qquad M_{U(1)} =  \frac{2}{\left({\partial Q}/{\partial \mu}\right)_{T=0}} \,.
\ee

 After Poisson resummation in $q$, one can obtain the sum over $n$ in \eqref{eq:example-grand-canonical-part-function}. The sum over $n$ represents a sum over classical saddles for the solutions of the gauge field. The partition function need only depend on the holonomy associated to the chemical potential $\xi$ imposed in the near-horizon region. Thus, the partition function should be periodic under shift of $\xi \to \xi + i n$ with $n \in \mZ$, which is precisely what the sum over classical saddles implements. Similar to the Schwarzian, excitations of the gauge field in the black hole background are captured by large $U(1)$ gauge transformations that do not vanish at the boundary of the near-horizon region. Such transformations can be parametrized by a $U(1)$ mode that captures the breaking of the $U(1)$ symmetry as the chemical potential is turned on. Like the Schwarzian, this theory is one-loop exact, and its on-shell action and one-loop determinant are captured in the sum over $n$ in \eqref{eq:example-grand-canonical-part-function}. Once again, the full partition function of this effective theory can be read off from the low-temperature expansion of the on-shell action. 

We can additionally fix the angular velocities of the black hole instead of its angular momenta. This can be done by setting the holonomy of the $SO(4)$ gauge field that is obtained from the dimensional reduction to $e^{i \oint B} = e^{-\beta \omega_1 \hat J_1 -\beta \omega_2 \hat J_2}$ where $\hat J_1$ and $\hat J_2$ are two Cartans of $SO(4)$ whose eigenvalues are $J_1$ and $J_2$. Similar to the $U(1)$ mode, excitations of the $SO(4)$ gauge field can be captured by an $SO(4)$ valued field which parametrizes non-vanishing gauge transformations at the boundary of the near-horizon region. As above, there are multiple saddles for this  $SO(4)$-mode which is once again due to the fact that the partition function does not explicitly depend on $\omega_1$ or $\omega_2$, but rather on the holonomy $e^{-\beta \omega_1 \hat J_1 -\beta \omega_2 \hat J_2}$. 

The black holes described above should also appear in the bosonic truncation of a supergravity theory. Nevertheless, black holes which at extremality preserve some amount of supersymmetry are instead described by different versions of the super-Schwarzian theory. This is because the effective theory which computes the low-temperature expansion of the partition function does not only capture the breaking of the near-horizon bosonic isometry  group $SL(2, \mR) \times G_{BF}$; instead, it captures the breaking of a super-group. 
As mentioned above, in such a case there are additional modes that remain massless in the near-horizon region, which contribute in the low-temperature expansion of the partition function - these are the dilatino and the gravitino. These modes do not decouple from the bosonic modes and consequently the sum over $n$ in \eqref{eq:example-grand-canonical-part-function} has a more complicated dependence with $\beta$ and $n$. Nevertheless, as we will explain in section \ref{sec:SchwarzianSolution} the couplings determining the low-temperature expansion of the partition function can still be read-off from the on-shell action, as was the case in \eqref{eq:near-extremal-RN-part-function} or \eqref{eq:example-grand-canonical-part-function}. 

Thus, instead of performing the full-dimensional reduction, in this paper we will take the simpler approach of reading off the effective action which captures the low-temperature expansion of the partition function. As discussed above, we will identify this effective theory by (a) the near-horizon isometry that gets broken when the temperature, chemical potential and angular velocities are turned on and (b) the low-temperature expansion of the on-shell action. This will determine the correct version of the super-Schwarzian theory needed to describe the near-BPS black holes discussed in this paper. Before diving into that identification, it is useful to first discuss the properties of the effective theory that will end up being important in our analysis, the $\cN=2$ super-Schwarzian theory.

\subsection{The model}
\label{sec:Scwharzianmodel}

The $\cN=2$ super-Schwarzian theory was described in detail in \cite{Fu:2016vas}. It is a theory that is described by $\cN=2$ super-reparametrizations.  Just like in the bosonic case, the bulk origin of these  super-reparametrizations is given by the action of super-diffeomorphisms on the metric, gravitino and gauge field at  the boundary of the near-horizon region. After imposing the appropriate chirality constrains, these super-reparametrizations can be described in super-space coordinates $(\tau, \theta, \bar \theta) \to (\tau', \theta', \bar \theta')$ in terms of two time-dependent bosonic fields $f(\tau)$ and $e^{i r \sigma(\tau)}\in U(1)$, where $r$ is a normalization constant we will discuss below, as well as two fermionic fields $\eta(\tau)$ and $\bar{\eta}(\tau)$:
\bea
\label{eq:N=2-super-reparam}
\tau' &=& f(\tau) + \ldots,\\
\theta' &=& e^{i r \sigma(\tau)} \sqrt{f'(\tau)} \theta + \eta(\tau) + \ldots \\
\bar{\theta}' &=& e^{-i r \sigma(\tau)} \sqrt{f'(\tau)} \bar{\theta} + \bar{\eta}(\tau)+ \ldots,
\eea
where the dots can be obtained by explicitly solving the super-reparametrizations constraints. 
The fields entering in these super-reparametrizations become the degrees of freedom of the $\cN=2$ super-Schwarzian theory.  The Schwarzian derivative is then given by
\beq
S(f,\sigma, \eta,\bar{\eta}) = \frac{\partial_\tau \bar{D} \bar{\theta}'}{\bar{D}\bar{\theta}'} - \frac{\partial_\tau D \theta'}{D \theta'} - 2 \frac{\partial_\tau \theta' \partial_\tau \bar{\theta}'}{(\bar{D} \bar{\theta}' )(D\theta')}= \ldots + \theta \bar{\theta} S_b(f,\sigma,\eta,\bar{\eta})\,,
\eeq
from which one can write the $\cN=2$ super-Schwarzian action in terms of super-coordinates
 \bea 
 \label{eq:N=2-action-super-coords}
 I_{\NSch}&=&-\frac{1}{\MSU }\int d\tau d\theta d\bar{\theta} S = - \frac{1}{\MSU } \int d\tau S_b\,,\\
 &=&\frac{1}{\MSU }\int \Sch(f,\tau) + 2 (\partial_\tau \sigma)^2 + ({\rm fermions})\,.\label{eq:N=2-action}
 \ea 
 Just like in \eqref{eq:near-extremal-RN-part-function}, the action of the $\cN=2$ super-Schwarzian can be obtained from an $\cN=2$ bulk JT theory which in addition to the metric and dilton present in the bosonic theory described in section \ref{sec:schwarzian-philosophy}, also contains fermionic super-partners, the gravitino and dilatino as well as a $U(1)$ R-symmetry gauge field coupled to a zero-form Lagrange multiplier. After integrating-out the dilaton, dilatino and $U(1)$ Lagrange multiplier what remains are only the large super-diffeomorphisms describing the transformations of the metric, gravitino and  $U(1)$ gauge field at the boundary of the near-horizon region. These can be again parametrized by the fields $f(\tau)$, $\sigma(\tau)$, $\eta(\tau)$ and $\bar\eta(\tau)$ entering in \eqref{eq:N=2-super-reparam}.  Once again, just like for the bosonic counter-part \eqref{eq:near-extremal-RN-part-function}, these super-diffeomorphisms enter in the boundary terms necessary for the $\cN=2$ JT theory to satisfy the variational principle when imposing Dirichlet boundary conditions for all the aforementioned fields. Plugging in the large super-diffeomorphisms parametrized by $f(\tau)$, $\sigma(\tau)$, $\eta(\tau)$ and $\bar \eta(\tau)$, these boundary terms are then precisely equal to the super-Schwarzian action \eqref{eq:N=2-action-super-coords}. When performing the path integral over the set of large super-diffeomorphisms which are weighed by the boundary term, one should be careful to not over-count identical super-geometries. For this reason, we should quotient the path integral over these modes by whatever super-isometry is present in the bulk. In $\cN=2$ JT gravity this is $SU(1,1|1)$. These super-isometries can be explicitly identified in \eqref{eq:N=2-action} as a global $SU(1,1|1)$ acting on the fields $f,\sigma, \eta,\bar{\eta}$ \cite{Fu:2016vas}. Consequently, we should quotient the space of configurations of $f,\,\sigma,\, \eta,\,\bar{\eta}$ by such transformations.

In addition, one can add a topological term in the bulk associated to the $U(1)$ R-symmetry gauge field. On the boundary, one should add to the action an associated topological term to the $U(1)$ mode, $\sigma$:
\beq 
\label{eq:topological-term}
I_\text{topological} = i \theta r \int d\tau\,( \partial_\tau \sigma)\,, 
\eeq
Due to its topological nature this term is also invariant under the $SU(1,1|1)$ transformations. Because $e^{i r \sigma(\tau)} \in U(1)$ and thus $\sigma \sim \sigma + 2 \pi/r$, we can identify theories with $\theta \sim \theta + 2\pi$ and will therefore restrict to $\theta \in [0, 2\pi)$. From the bulk perspective, since the boundary holonomy of the $U(1)$ R-symmetry gauge field is given by $e^{i (\sigma(\beta+\tau) - \sigma(\tau))} $, $r$ is determined  by the smallest R-charge among any field that could possibly be coupled to the $U(1)$ R-symmetry gauge field in $\cN=2$ JT.\footnote{In the context of SYK, $r = 1/\hat q$ in \cite{Fu:2016vas} and represents the number of fermionic fields that are needed to construct the $\cN=2$ SYK supercharges. However, in section \ref{sec:BPS-and-near-BPS-BHs} we will find that the relevant effective theory has $r=1$ which is not realized by any known SYK model.  }

As in \eqref{eq:near-extremal-RN-part-function}, we will next review the features of the resulting partition function 
\beq
\label{eq:path-integral-N=2-Schwarzian}
Z(\beta,\alpha) = \int \frac{\mathcal{D}f \mathcal{D}\sigma \mathcal{D}\eta \mathcal{D}\bar{\eta}}{SU(1,1|1)} e^{-I_{\NSch} - I_{\text{topological}}} \, ,
\eeq
where the partition function depends on the inverse temperature $\beta$ which sets the length of the thermal circle and on a $U(1)$ chemical potential $\alpha$ associated to the mode $\sigma$, under which fermionic fields are also charged. Thus, the periodicity conditions for the resulting path integral become $f(\tau+\beta)=f(\tau)$, $e^{i\sigma(\tau+\beta)}=e^{2 \pi i \alpha} e^{i \sigma(\tau)}$, $\eta(\tau+\beta)=-e^{2\pi i r \alpha}\eta(\tau)$ and similarly for $\bar{\eta}$.

\subsection{Exact partition function and its spectrum}
\label{sec:SchwarzianSolution}

The partition function of the $\cN=2$ super-Schwarzian theory can be computed exactly \cite{Stanford:2017thb,Mertens:2017mtv}. We work in conventions where the $U(1)_R$ charge of the complex supercharge is one and the minimal R-charge of fundamental fields is fractional and given by $r$. Consistency of the spectrum with $\mathcal{N}=2$ supersymmetry requires that $1/r$ is an integer, since states related by applying a supercharge should both be in the spectrum. The partition function is one-loop exact and given by \cite{Stanford:2017thb, Mertens:2017mtv}
\beq\label{N2sumsaddles}
Z_{\NSch}(\beta,\alpha) =\sum_{n\in \frac{1}{r} \cdot \mathbb{Z}} e^{i \,r \theta n } ~\frac{2\cos\left( \pi  (\alpha+ n)\right)}{\pi\left(1-4 (\alpha+ n)^2\right)} ~e^{S_0+\frac{2\pi^2}{\beta \MSU}(1-4 (\alpha+n)^2)}
.
\eeq
The sum over $n$ is a sum over saddles analogous to the previous result \eqref{eq:example-grand-canonical-part-function}, where the $U(1)_R$ mode $\sigma(\tau)$ has different windings around the thermal circle, $\sigma(\tau+\beta) = \sigma(\tau) + 2\pi n/r$. The exponential terms all come from the evaluation of the classical action on the saddle point configurations. Finally the prefactor of the exponentials is the one loop determinant of the Schwarzian mode, $U(1)_R$ mode, and the fermions. The non-trivial dependence on the chemical potential comes form the fermions, while the possible $\beta$ dependence in the one loop determinant cancels since $SU(1,1|1)$ has four bosonic and four fermionic generators. When the theta angle in \eqref{eq:topological-term} vanishes the partition function is invariant under $\alpha \sim \alpha + 1/r$. Thus, for simplicity we can restrict to $\alpha \in [0, 1/r)$.   
The theory is also charge conjugation invariant only when $\theta=0$ or $\theta=\pi$, since only then is the partition function real.

\textbf{Supersymmetry in the on-shell action.} The contribution of the $SL(2, \mR)$ Schwarzian mode to the action is independent of the winding mode $n$ and gives $\frac{2\pi^2 }{\beta \MSU}$, while the contribution of the $U(1)_R$ mode is given by  $-\frac{8\pi^2}{\beta \MSU} (\alpha+n)^2$. Thus, the equality of coupling seen in \eqref{eq:N=2-action}, translates to a relation between the $\alpha$-dependent terms in the on-shell and the $\alpha$ independent terms, in the convention in which $\alpha \sim \alpha + 1/r$.

\textbf{Density of states within each supermultiplet. }We can Fourier transform the above result to obtain the decomposition of $Z_{\NSch}(\beta,\alpha)$ as a sum over $U(1)_R$ charges $Z$, with the smallest charge equal to $r$.  Because of supersymmetry the spectrum should organize itself in supermultiplets $(Z)\oplus (Z-1)$ for states with $E\neq 0$ and solely charge $Z$ for states with $E=0$. Based on symmetry principles the partition function should thus be decomposed as
\beq
\label{eq:N=2-part-function-grand-can}
Z_{\NSch}(\beta,\alpha) =\sum_{Z} e^{2\pi i \alpha Z} \rho_{\rm ext}(Z)+ \sum_{Z}\int dE e^{-\beta E} \left( e^{i 2\pi \alpha Z} + e^{i 2\pi \alpha (Z-1)} \right) \rho_{\rm cont}(Z,E).
\eeq
In contrast to Schwarzian theories with smaller amounts of supersymmetry ($\cN=0$ and $\cN=1$), the density of states for $\mathcal{N}=2$ splits into an extremal piece and a continuous piece. Since the first term in \eqref{eq:N=2-part-function-grand-can} is temperature independent this implies that those states are extremal and yield an exact Dirac delta-function in the density of states at $E=0$. Rewriting \eqref{N2sumsaddles} as in \eqref{eq:N=2-part-function-grand-can}, gives \cite{Mertens:2017mtv}
\begin{align}
\label{eq:density-of-states-charge-decomp}
&Z_{\NSch}(\alpha,\beta)= \sum_{\substack{Z\in r \cdot \mathbb{Z}-\frac{r \theta}{2\pi},\\ |Z|<\frac{1}{2}} } e^{2 \pi i \alpha Z} ~e^{S_0}~r  \cos \left(\pi Z\right) \\ 
&+ \! \! \!\sum_{Z\in r \cdot \mathbb{Z}-\frac{r \theta}{2\pi}}r \left( e^{2\pi i \alpha Z } + e^{2\pi i \alpha (Z-1)}\right) \int_{E_\text{gap}(Z)}^\infty dE e^{-\beta E}\frac{e^{S_0} \sinh{\left(2\pi \sqrt{\frac{2}{\MSU}(E-E_\text{gap}(Z))}\right)}}{2\pi E},\nonumber
\end{align}
where $E_\text{gap}(Z) \equiv \frac{\MSU}{8}(Z-\frac{1}{2})^2$ denote the energy gap for a supermultiplet labeled by the charge $Z$. The first line in \eqref{eq:density-of-states-charge-decomp} captures the contribution of BPS states that have $E=0$ and whose R-charges $Z$ can thus take the values $Z\in r \cdot \mathbb{Z}-\frac{r \theta}{2\pi}$, with the constrain $|Z|<\frac{1}{2}$ which ensures the density of states is positive. The contribution of the second line captures the non-BPS states which have a continuous density.  

\textbf{The BPS to non-BPS energy gap.} When does such a theory have a gap between the BPS states and lightest non-BPS state? If $Z=1/2$ is included in the sum over charges in \eqref{eq:density-of-states-charge-decomp}, then $E_\text{gap}(Z = 1/2) = 0$ which implies that there is a continuum of states starting at $E=0$ and that there is no gap in the spectrum. For $\theta = 0$, there is no gap if $1/r$ is even and there is a gap if $1/r$ is odd. For $\theta = \pi$, the opposite is true: there is no gap if $1/r$ is even and the gap is present if $1/r$ is odd. If $\theta \neq \{0 \text{ or } \pi \}$, then the gap is always present.

\textbf{The supersymmetric index.} For particular values of the $U(1)_R$ chemical potential $\alpha$, the grand canonical partition function computes a supersymmetric index. 
In particular, for any $\alpha$ such that $e^{i 2 \pi \alpha} = -1$ the fermionic degrees of freedom $\eta$ and $\bar \eta$ become periodic and consequently, turning on this fugacity is equivalent to the insertion of $(-1)^F$ in the super-Schwarzian path integral. This works for any $\alpha_k = k-1/2$ with $k =1,\ldots, r^{-1}$. For $r=1$ there is only one index (with $\alpha=1/2$) and if we fix $-\pi \leq \theta \leq \pi$ then it is given by 
\beq
\label{eq:SchwarizanIndex}
\mathcal{I} = Z_{\NSch}(\beta,\alpha=1/2) = e^{S_0} e^{-i \frac{\theta}2} \cos \left(\frac{\theta}{2}\right).
\eeq
In such a case,  note that $Z(\beta \to \infty, \alpha \to 0) = Z(\beta, \alpha \to \frac{1}2 )$ which implies that in such a case the ``ground state'' of the $\cN=2$ super-Schwarzian are purely bosonic. 

\textbf{The 't Hooft anomaly.} Finally, we will comment on the symmetries of the particular theories with $\theta=0$ and $\theta=\pi$, for which charge conjugation invariance appears, at least classically. For a related toy model see Appendix D of \cite{Gaiotto:2017yup}. When $\theta=0$ and $1/r\in\mathbb{Z}$ the theory presents both charge conjugation and $U(1)_R$ symmetry. This is evident from the fact the partition function \eqref{N2sumsaddles} is real and that the charges are integer multiples of $r$. Instead, the theory with $\theta=\pi$ has a 't Hooft anomaly between charge conjugation and $U(1)_R$ symmetries. For example, the localization calculation leading to \eqref{N2sumsaddles} is manifestly charge conjugation invariant, since it is real. Nevertheless the spectrum in \eqref{eq:density-of-states-charge-decomp} is shifted by a half-integer unit of charge, inconsistent with $U(1)_R$ symmetry. This issue can be fixed by shifting the charge by a constant $Z\to Z + r/2$, making the charge operator still commutes with the Hamiltonian; however, the price to pay is to break charge conjugation invariance. For this reason the theory with $\theta=\pi$ has a 't Hooft anomaly, while $\theta=0$ preserves the symmetries at the quantum level.

\section{BPS and near-$1/16$ BPS black holes in AdS$_5$}
\label{sec:BPS-and-near-BPS-BHs}

In the previous section, we have reviewed the role Jackiw-Teitelboim gravity and the Schwarzian theory play in determining the form of the spectrum of near extremal black holes in general. It can be thought of as the soft mode coming from the broken symmetries that emerge in the near extremal limit. We also consider in particular the case of a broken $SU(1,1|1)$ symmetry which is described by $\mathcal{N}=2$ Schwarzian theory. 

 In this section we are going to apply this to $1/16$-BPS black holes in AdS$_5 \times S^5$, which AdS/CFT predicts are dual to $1/16$ BPS states in $\mathcal{N}=4$ Super Yang-Mills. We will begin by reviewing general properties of these black holes in sections \ref{sec:AdS5black hole} and \ref{sec:BPSvsSUSYvsExt}, taken from \cite{Chong:2005hr}, but see also \cite{Hawking:1998kw,Gutowski:2004ez,Gutowski:2004yv, Chong:2005da, Kunduri:2006ek}. Black holes in AdS$_5$ have a-priori several related but distinct limits:
\begin{itemize}
\item The near-extremal limit in which $T\rightarrow 0$.
    \item The extremal black holes in which $T= 0$.
    \item The supersymmetric limit in which the geometry has Killing spinors.
    \item The BPS limit which has $T=0$ and an enhanced set of Killing spinors.
\end{itemize}
By adjusting the parameters of the solution, one may move independently off of extremal and supersymmetric surfaces in parameter space \cite{Hosseini:2017mds,Cabo-Bizet:2018ehj}. We adopt the convention of these references to refer to the intersection of these surfaces as the \emph{BPS limit}. 

 Recently the Bekenstein-Hawking entropy of exactly $1/16$-BPS black holes was reproduced from the superconformal index of $\mathcal{N}=4$ Yang Mills \cite{Romelsberger:2005eg,Kinney:2005ej,Cabo-Bizet:2018ehj,Choi:2018hmj,Benini:2018ywd}. The goal of this section is to reproduce this result from the point of view of the gravitational path integral, and in the process predict the spectrum of excited black holes states above the BPS ones; these states are \emph{not} protected by exact supersymmetry. In section \ref{subsec-ensembles} we specify a mixed ensemble where some charges are fixed that will allow us to use the results of section \ref{sec:AdS5black hole} without requiring the most general black hole solution. In section \ref{sec:extremal10D} we analyze the BPS limit and show the black holes have an $AdS_2 \times S^3 \times S^5$ throat with an emergent $SU(1,1|1)$ symmetry. We identify then the $\mathcal{N}=2$ Schwarzian theory controlling the near BPS spectrum from the explicit breaking of this superconformal group. In section \ref{sec:low-T-expansion} we verify this at the level of the classical action of the black hole saddle and in section \ref{sec:SpectrumN4} we put everything together and give a picture of the spectrum of nearly $1/16$-BPS states in $\mathcal{N}=4$ Yang Mills.

\subsection{General black hole solutions in AdS$_5$ }\label{sec:AdS5black hole}

The goal of this paper is to extract information about the spectrum of nearly $1/16$-BPS states in $\mathcal{N}=4$ Yang Mills. According to AdS/CFT, these states are dual to black holes and we will study their spectrum using the gravitational path integral. Since we want to know the energy and charges of these states, we will compute and derive the spectrum from the dual field theory partition function on $S^1_\beta \times S^3$ with angular velocities and chemical potentials conjugate to the $SO(6)$ R-charges turned on. From the bulk perspective, the gravitational path integral instructs us to sum over all geometries satisfying boundary conditions appropriate for our choice of ensemble in asymptotically $AdS_5$. Therefore we need to know black hole solutions carrying these charges, which we briefly review next, mostly to establish our conventions. 

The theory of gravity in asymptotically $AdS_5$ space dual to four dimensional $\mathcal{N}=4$ Yang Mills arises from a dimensional reduction of ten dimensional type IIB supergravity on $AdS_5 \times S^5$. From this perspective the $SO(6)$ gauge symmetry in the bulk is identified with the isometries of $S^5$, with the $SO(6)$ charges associated to rotations along the $S^5$. Since $SO(6)$ has rank three it leads to three gauge fields $A_{1,2,3}$ and charges $R_{1,2,3}$. The resulting theory in $AdS_5$ is quite complicated, but simplifies drastically when the rotation happens symmetrically along the three Cartan directions on $S^5$, meaning $A\equiv A_1=A_2=A_3$ and $R\equiv R_1=R_2=R_3$. We will restrict only to that case in this paper, the reduction leads to a simple theory of minimal gauged supergravity on $AdS_5$ with a single gauge field\footnote{Note that, as in almost all string constructions of AdS/CFT, the radius of the sphere is of the same order as the AdS radius. The five-dimensional theory studied here is a consistent truncation of the 10D theory, and all black hole solutions we study may be uplifted to 10D type IIB, as we show in \ref{sec:extremal10D}.}. The  bosonic part of the action is given by (e.g. \cite{Chamblin:1999tk,Gauntlett:2003fk}):
\be 
\label{eq:5DLag}
  I \supset \frac{1}{16 \pi G_5}\int \left[ (R+12) \ast 1 - \frac{2}{3} F \wedge \ast F + \frac{8}{27} F \wedge F \wedge A \,\right] .
\ee
We work in units in which $\ell_{\textrm{AdS}_5} = 1$ and the gauge coupling $g$ has been set to $1$. When this effective action is obtained from string theory, the dimensionful parameters will appear in the full action through the 5D Newton constant $G_5$. The Chern-Simons term affects both the definition of asymptotic electric charges and the value of the on-shell action. 

Before writing the solutions we should determine the boundary conditions. We will parametrize the bulk by coordinates $(\tau,r,\theta,\phi,\psi)$ where $(\phi,\psi)$ are $2\pi$ periodic and $\theta\in [0,\frac{\pi}{2}]$. The coordinates $(\tau,\theta,\phi,\psi)$ also parametrize the boundary $S^1\times S^3$ at large radius $r$. This means that the metric should behave asymptotically, after a rescaling of $r$, as $ds^2 = \frac{dr^2}{r^2} + r^2 \left(d\tau^2 +d\Omega_3^2 \right) +\mathcal{O}(r^0)$, with the unit $S^3$ element $d\Omega_3^2= d\theta^2 + \sin^2\theta d\phi^2 + \cos^2\theta d\psi^2$. In order to fix the R-charge chemical potential to be $\Phi$, we fix the gauge field holonomy at large $r$ through $A = - \Phi d\tau +\mathcal{O}(r^{-1})$. The above electric potential is related to the $R$-charge chemical potentials as $\Phi_1 =\Phi_2=\Phi_3 = \frac{2}{3}\Phi $ \cite{Aharony:2021zkr}. Following \cite{Gibbons:1976ue} we fix the angular velocities along the $S^3$ and the temperature through the identification
\begin{equation}
    (\tau,\theta,\phi,\psi) \sim (\tau + \beta, \theta, \phi - i \Omega_1 \beta, \psi - i \Omega_2 \beta).
\end{equation}
A general solution with this asymptotic behavior is a charged rotating black hole with unequal angular momenta. The solution, which we write in Lorentzian signature for simplicity defining $t=-i\tau$, was found in \cite{Chong:2005hr}, but we follow the conventions of \cite{Aharony:2021zkr}. In terms of asymptotically static coordinates the metric is given by 
\begin{align}
ds^2 = 
\label{eq-5d-metric}
&-\frac{\Delta_{\theta } dt \left(2 \nu  q+\rho ^2 \left(
r^2+1\right) dt\right)}{\rho ^2 \Xi _a \Xi _b}
+\frac{2 \nu  q w}{\rho ^2}+\frac{\rho ^2 dr^2}{\Delta _r}
+\frac{f_t \left(\frac{\Delta _{\theta } dt}{\Xi _a \Xi _b}
-w\right){}^2}{\rho ^4} 
\\
&+\frac{\rho ^2 d\theta ^2}{\Delta _{\theta }}
+\frac{\left(a^2+r^2\right) \sin ^2\theta  d\phi ^2}{\Xi _a}
+\frac{\left(b^2+r^2\right) \cos ^2\theta  d\psi ^2}{\Xi_b} ,\nn \\
A =&
\frac{3 q }{2 \rho ^2} \left(\frac{\Delta _{\theta } dt}{\Xi _a \Xi_b}-w\right) - \Phi dt \label{eq-5d-gauge}
,
\end{align}
where
\begin{align}
\nu &=a \cos ^2\theta  d\psi +b \sin ^2\theta  d\phi ,
\qquad 
\rho =\sqrt{a^2 \cos ^2\theta +b^2 \sin ^2\theta +r^2},
\\
\Delta _r &=\frac{\left(a^2+r^2\right) \left(b^2+r^2\right) \left(
r^2+1\right)+2 a b q+q^2}{r^2}-2m,\\
f_t&=2 a b  \rho ^2 q+2 m \rho ^2-q^2, \qquad
\Delta _{\theta }=1 -a^2 
\cos ^2\theta -b^2 
\sin ^2\theta ,\\
\Xi _a&=1-a^2 
, \qquad
\Xi _b=1-b^2 
, \qquad
w=\frac{a \sin ^2\theta  d\phi }{\Xi _a}+\frac{b\cos ^2\theta  d\psi }{\Xi _b} .
\end{align}

The above metric gives a four parameter family of solutions for charged and rotating black holes in five dimensions, parametrized by $(m , q ,a ,b)$, where we will restrict to $0<a,b<1$. These parameters are related to $(m,q,a,b) \to (\beta, \Phi, \Omega_1,\Omega_2)$ by imposing the solution is smooth at the Euclidean horizon, located at the radius $r_+$ defined as the largest positive root of $\Delta_r (r_+) = 0$. We trade from now on the variable $m$ by $r_+$ through 
\be
m=\frac{\left(a^2+r_+^2\right) \left(b^2+r_+^2\right) \left(
r_+^2+1\right)+2 a b q+q^2}{2 r_+^2} .
\ee
One can show the solution is smooth with appropriate choice of parameters following the methods in \cite{Gibbons:1976ue}. The cycle becoming contractible at the horizon is generated by the vector field $V = \frac{\partial}{\partial t} + \Omega_1 \frac{\partial}{\partial \phi} + \Omega_2 \frac{\partial}{\partial \psi}$, where  
\be
\Omega_1 = 
\frac{a \left(b^2+r_+^2\right) \left(
r_+^2+1\right)+b q}{\left(a^2+r_+^2\right) \left(b^2+r_+^2\right)+a b q}
,\qquad
\Omega_2 = 
\frac{b \left(a^2+r_+^2\right) \left(
r_+^2+1\right)+a q}{\left(a^2+r_+^2\right) \left(b^2+r_+^2\right)+a b q}
,
\ee
denote angular velocities on the horizon. Smoothness also determines the temperature to be
\begin{equation}
   \beta^{-1}= T = 
\frac{r_+^4 \left(
\left(a^2+b^2+2 r_+^2\right)+1\right)-(a b+q)^2}{2 \pi  r_+ \left(\left(a^2+r_+^2\right) \left(b^2+r_+^2\right)+a b q\right)}.
\end{equation}

The chemical potential at the boundary is fixed by the asymptotic value of $A\to -\Phi dt$, when $r\to\infty$. The relation between $\Phi$ and $q$ is determined by demanding the solution is regular at the Euclidean horizon. Finally smoothness of the gauge potential $A$ at the horizon $A_\mu V^\mu|_{r_+}=0$ gives the final relation 
\begin{equation}
    \Phi=\frac{3 
q r_+^2}{2 [ \left(a^2+r_+^2\right) \left(b^2+r_+^2\right)+a b q ]}.
\end{equation}
Altogether, smoothness gave us the four relations needed to solve for $(r_+,q,a,b)$ in terms of $(\beta, \Phi, \Omega_1,\Omega_2)$. Interestingly, in addition to the solution we have just discussed, there are other distinct solutions with the same boundary conditions obtained by integer shifts of the chemical potentials an angular velocities, first analyzed in the context of $AdS_5$ black holes in \cite{Aharony:2021zkr}. We will comment on these solutions in section \ref{subsec-ensembles}, where they play an important role in our analysis.

Alternatively, these black holes can be identified through specifying the values of their four conserved charges $(E,R,J_1,J_2)$ canonically conjugate to $(\beta, \Phi, \Omega_1,\Omega_2)$. The charges can be defined by the ADM procedure which gives 
\begin{align}
E &= \frac{\pi  \left(2 a b 
q \left(\Xi _a+\Xi _b\right)+m \left(- \Xi _a \Xi _b +2 \Xi _a+2 \Xi _b\right)\right)}{4 G_5 \Xi _a^2 \Xi _b^2}+\frac{3\pi}{32 G_5} \, , \, \, \, \, R = \frac{\pi q}{2 
G_5 \Xi _a \Xi _b} ,
\\
J_1 &= \frac{\pi  \left(b q \left(a^2 
+1\right)+2 a m\right)}{4 G_5 \Xi _a^2 \Xi _b} \, , \, \, \, \, 
J_2 = \frac{\pi  \left(a q \left(b^2 
+1\right)+2 b m\right)}{4 G_5 \Xi _a \Xi _b^2}
.
\label{eq:masses-charge-am}
\end{align}
The energy of vacuum $AdS_5$ is given by $E_0 = \frac{3\pi}{32 G_5}$. We will denote the black hole mass by $M\equiv E-E_0$. To reiterate, $R$ is proportional to the angular momenta along $S^5$ distributed symmetrically along the Cartan directions.\\

Having found the solution of the equations of motion filling the boundary conditions, we can approximate the partition function in this grand-canonical ensemble by the exponential of the classical action
\be 
Z(\beta, \Omega_{1},\Omega_2, \Phi_{1}, \Phi_2 , \Phi_3) \sim e^{-I_{GCE}(\beta, \Omega_{1},\Omega_2, \Phi_{1}, \Phi_2 , \Phi_3)}.
\ee
As mentioned above, there are other saddles which are relevant in the near-extremal limit which we discuss in section \ref{subsec-ensembles} but ignore for now. The action $I_{GCE}(\beta, \Omega_{1,2}, \Phi_{1,2,3})$, including now the GHY boundary term and the holographic counterterm \cite{Balasubramanian:1999re} appropriate to this ensemble where we fix the metric and gauge potential at infinity, is given by
\begin{equation}
    I = \frac{1}{16 \pi G_5}\int d^5x \sqrt{-g} \mathcal{L} + \frac{1}{8 \pi G_5}\int_{\partial}d^4x \sqrt{-h} K - \frac{3}{8 \pi G_5}\int_{\partial} d^4x \sqrt{-h} \left(1+ \frac{R_{h}}{12} \right) ,
\label{eq-full-action-with-ct}
\end{equation}
where $h= g|_{\rm bdy}$ and $R_h$ is the Ricci curvature scalar of $h$. Evaluating all contributions to the on-shell action \eqref{eq-full-action-with-ct}, together with the Chern-Simons term, one obtains \cite{Chen:2005zj,Cassani:2019mms}
\be 
I_{GCE} = \frac{\pi \beta}{4G_5 \Xi_a \Xi_b} \left(
m - 
(r_+^2 + a^2)(r_+^2 + b^2) - 
\frac{q^2 r_+^2}{(r_+^2 +a^2)(r_+^2 +b^2) + a b q}
\right) 
 + \frac{3\pi \beta}{32 G_5} .
\ee
Defining the Bekenstein-Hawking entropy as the area of the horizon
\begin{equation}
S=\frac{A}{4G_5}=\frac{\pi ^2 \left(\left(a^2+r_+^2\right) \left(b^2+r_+^2\right)+a b q\right)}{2 G_5 r_+ \Xi _a \Xi _b} \, ,
\end{equation} the on-shell action satisfies the so-called quantum statistical relation, 
\be 
I_{GCE} = \beta (E- TS - \Omega_1 J_1 - \Omega_2 J_2 - \Phi R) ,
\ee
which can explicitly be checked using the expressions for the charges and potentials. Finally, if we want to compute the partition function in a fixed charge sector we need to not only write $(r_+,q,a,b)$ in terms of the charges but also add the apropriate boundary terms in the action to make the variational principle well-defined \cite{Hawking:1995ap}. For example, if we wanted to fix the charge $R$ this amounts to adding an extra $  I \to I + \beta \Phi R$ term in the action, understanding that $\Phi$ should be written as a function of charges, and similarly for angular momentum.

\subsection{The $1/16$-BPS black hole solution}
\label{sec:BPSvsSUSYvsExt}

As discussed in the introduction of section \ref{sec:BPS-and-near-BPS-BHs}, the general AdS$_5$ black hole solution has different limits corresponding to extremality ($T\rightarrow 0$) and supersymmetry (existence of a Killing spinor). Written in terms of charges, the supersymmetric (but not yet extremal) condition is~\cite{Cvetic:2005zi}
\be 
\label{eq:susychargecondition}
M - \left(\frac{3}{2}R + J_1 +  J_2  \right) \bigg\vert_{\text{SUSY}} = 0 ,
\ee
which after inserting the explicit expressions for $M$, $R$, and $J_{1,2}$ valid for both BPS and non-BPS black holes, takes the form
\be \label{eqsusyparam}
q = \frac{m}{1+ a +b} .
\ee
We next want to impose extremality, meaning that the solution has zero temperature. This can be achieved by imposing the further condition 
\be \label{eqctcparam}
m = (a+b)(1+a)(1+b)(1+a+b) .
\ee
The size of the horizon for these BPS black holes is given by the simple expression 
\be 
r_+=r^* \equiv \sqrt{(a+b+a b )} . 
\ee
In terms of $r^*$ the constraint \eqref{eqsusyparam} becomes $q=q^*=(a+b)(1+a)(1+b)$. It is easy to check that \eqref{eqsusyparam} together with \eqref{eqctcparam} implies that $T=0$. This means that, while in general one could have supersymmetric non-extremal ($T\neq 0$) black holes, imposing extremality leads to the \emph{BPS condition}, which is the intersection of the supersymmetric and extremal surfaces. Note that this also means that BPS black holes are now labeled by only two parameters $(a,b)$, through the relations $(r^*(a,b) ,q^* (a,b))$.
In the above, following \cite{Cabo-Bizet:2018ehj}, we introduced the $(\phantom{r})^*$--notation, which from now on will denote quantities evaluated after imposing supersymmetry and extremality. 

The black hole solutions that are both extremal and supersymmetric are the ones dual the $1/16$-BPS states in $ U(N)$  $\mathcal{N}=4$ Yang Mills at large $N$. In terms of the parameters $(a,b)$ the BPS value of the Bekenstein-Hawking entropy is 
\be 
S^* = \frac{\pi ^2 (a+b) \sqrt{a b +a+b}}{2  G_5 (1- a ) (1- b )} .
\ee
Importantly, we cannot yet determine from gravity whether this is the true entropy of BPS states without including quantum effects from the gravity path integral, as explained in section \ref{sec:N=2-super-Schwarzian}. The field theory calculation of the index supports this interpretation of $S^*$ which we will verify from gravity as well below. The BPS values for the energy and charges are
\begin{eqnarray}
M^* &=& -\frac{\pi  \left(2 a^2 b^2 +a^3  (b +1)+a \left(b^3 -3\right)+b \left(b^2 -3\right)\right)}{4  G_5 (a -1)^2 (b -1)^2}
,~~R^* = \frac{\pi  (a+b)}{2  G_5
   (1-a ) (1-b )}
,\\ 
J_1^* &=& \frac{\pi  (a+b) (a (b +2)+b)}{4  G_5 (a -1)^2 (1-b)}
,~~~~~J_2^* = \frac{\pi  (a+b) (a b +a+2 b)}{4  G_5 (1-a) (b -1)^2}
.
\end{eqnarray}
It is easy to verify the relation $M^* - \frac{3}{2} R^* -  J_1^* -  J_2^* = 0 $. Thus, due to the BPS constraints, given the angular momenta $J_1^*$ and $J_2^*$ we uniquely determine the equal $U(1)^3 \in SO(6)$ charge $R^*$ for  the BPS black hole. A field theory explanation of this constraint from the point of view of $\mathcal{N}=4$ Yang Mills was suggested in \cite{Berkooz:2006wc, Berkooz:2008gc}. 
The chemical potentials potentials associated to the BPS black holes are given by $\Phi^* = \frac{3}{2}$, $\Omega_1^* =\Omega_2^*= 1$.

This restriction on parameters by the BPS condition is not unfamiliar. In the case of black holes in four dimensional ungauged supergravity, supersymmetry only implies that the mass is equal to the charge, while BPS states satisfy the further condition that the angular momentum vanishes. A motivation for this is obvious in this example, black holes with real angular momentum that are supersymmetric have a naked singularity in Lorentzian signature\footnote{From a Euclidean perspective, having supersymmetric solutions with imaginary angular momentum and non-zero temperature is perfectly fine, giving a real smooth metric, and just computes the index from gravity \cite{Iliesiu:2021are}.}. The issue with supersymmetric non-extremal $AdS_5$ black holes is instead the presence of closed timelike curves (CTC) in Lorentzian signature outside the horizon \cite{Chong:2005hr}, and imposing extremality together with supersymmetry removes this pathology.

\subsection{What we want to compute}
\label{subsec-ensembles}

In this section we clarify two issues that arise from the discussion so far. The first is that in the grand-canonical ensemble the solution reviewed in section \ref{sec:AdS5black hole} is not the only one. Take for example the gauge potential $A$. While we demanded that $A_\mu V^\mu |_{\textrm{horizon}} = 0$, the gauge invariant statement that the holonomy is trivial along a contractible cycle allows for more general values of the potential at the horizon, which can be removed by a large gauge transformation but lead to physically distinct solutions. The other saddles are mostly subleading but become important at low temperatures, which is the regime of interest of this paper. They were considered in the near extremal limit in the context of a different black hole solution in \cite{Iliesiu:2020qvm,Heydeman:2020hhw}, and we will discuss them in the context of AdS$_3$ in section \ref{sec:(2,2)-sugra-examples}.

The second issue is related to the fact that the supergravity theory  considered above and the corresponding electrically charged black holes arise from the dimensional reduction of Type IIB supergravity solutions on $S^5$ in the sector in which the solutions have equal Kaluza-Klein momenta $R\equiv R_1=R_2=R_3$ on the $S^5$ factor. In a grand canonical ensemble, even if we fix the chemical potentials conjugate to the R-charges $R_i$ to be equal, these new solutions discussed in the previous paragraph will inevitably involve configurations that do not respect this symmetry. This would require dealing with the much more complicated solution of type IIB with scalar fields turned on \cite{Wu:2011gq}, which we will not attempt in this paper.

We will address these issues in the following way. First, we will use information about the UV completion of the theory given by $\mathcal{N}=4$ Yang Mills to determine the full space of saddle point configurations. Second, we will choose an ensemble for which only configurations with $R_1=R_2=R_3$ contribute. Our conventions in this subsection closely follow those of \cite{Aharony:2021zkr}.

We begin by identifying the new solutions of the equations of motion. The most obvious way to generate new solutions is the following. Looking at the boundary conditions specified in section \ref{sec:AdS5black hole} it is clear that any geometry obtained by an integer shift $\Omega_i \to \Omega_i + \frac{2\pi i \mathbb{Z}}{\beta}$ solves the same equations with the same boundary condition, while being physically distinct. Therefore the gravitational path integral instructs us to sum over all of them. Something similar is true for the gauge potential. Two configurations related by $A \to A + \frac{2\pi i \mathbb{Z}}{\beta} dt$ satisfy the same gauge invariant boundary condition at infinity while continuing to be smooth since the extra integer flux can be undone by a large gauge transformation. This is not surprising since the R-charge chemical potential in $AdS_5$ corresponds to angular velocity along $S^5$ direction in ten dimensions. The partition function with all chemical potentials fixed is given in the large $N$ limit by a sum over saddles
\begin{align}
Z (\beta , \Omega_1 , \Omega_2 , \Phi_i  ) & = 
\Tr[ e^{-\beta H + \beta \Omega_1 J_1 + \beta \Omega_2 J_2 + \frac{1}{2} \beta \Phi_1 R_1 
+ \frac{1}{2} \beta \Phi_2 R_2 + \frac{1}{2} \beta \Phi_3 R_3
} ] 
\nonumber\\
&\hspace{-1cm}\to \sum_{
n_1 , n_2 , m_1 , m_2 , m_3
} 
Z_{\text{one-loop}} \, e^{-I_{GCE}(\beta, \Omega_1 + \frac{2\pi i n_1}{\beta} , \Omega_2
+ \frac{2\pi i n_2}{\beta}
, \Phi_1 +  \frac{2\pi i m_1}{\beta}, \Phi_2 +  \frac{2\pi i m_2}{\beta}, \Phi_3
+  \frac{2\pi i m_3}{\beta} )},\label{eqnssaddlesumads5}
\end{align}
where $I_{GCE}(\beta,\Omega_1,\Omega_2,\Phi_1,\Phi_2,\Phi_3)$ is the action of the black hole considered in \cite{Wu:2011gq} with Dirichlet boundary conditions. When $\Phi_1=\Phi_2=\Phi_3=\frac{2}{3} \Phi$ (which along with the factors of $\frac12$ in \eqref{eqnssaddlesumads5} defines our normalization for $\Phi_i$) this action above becomes equivalent to the action computed in section \ref{sec:AdS5black hole}. We will discuss the one-loop contribution later. At this point we need to make a choice regarding the range of integers that are allowed in the sum \eqref{eqnssaddlesumads5}. One option is to consider more carefully the global nature of the gauge groups in five dimensions from a ten dimensional perspective given by Type IIB. A simpler option is to use AdS/CFT and analyze the properties of this partition function given by $\mathcal{N}=4$ Yang Mills, which we turn to next.

The $\mathcal{N}=4$ Yang Mills theory has the superconformal symmetry $PSU(2,2|4)$, and parallel to the gravity analysis we decompose the Lorentz and R-symmetry factors to the Cartan subalgebra $U(1)_{J_1} \times U(1)_{J_2} \times U(1)^3_{R_i}$, with $i=1,2,3$. In $\mathcal{N}=1$ language, this theory is a particular instance of a vector multiplet coupled to a triplet of adjoint chiral multiplets $X_i$, where we normalize the R-charges such that $R_i$ assigns charge 2 to the corresponding chiral multiplet. This means that scalars have even integer R-charge and fermions have odd integer R-charge. States of the theory are also labeled by the two angular momenta $J_1$, $J_2$ which as usual are integer or half-integer and satisfy spin statistics. With these conventions, the fermion number operator, is given by $F= 2 J_{1,2} = R_i$ mod 2. This constraint implies a particular periodicity of the grand canonical partition function under shifts of the chemical potential \cite{Aharony:2021zkr}, which we can use to deduce the allowed saddles we need to include. Comparing with the gravitational answer \eqref{eqnssaddlesumads5} the constraint implies the following restriction on the solutions:
\be \label{eq:restrictionintegers}
m_1 + m_2 + m_3 + n_1 + n_2 = 2 \mathbb{Z} .
\ee
This determines the saddles that should in principle be included based on smoothness and global properties of the gauge groups\footnote{Actually there should be more restrictions on these integers since otherwise the sum strictly diverges. A rationale for which saddles should be included or not is currently under investigation \cite{HIOT}. Additional restrictions besides \eqref{eq:restrictionintegers} also were observed in \cite{Aharony:2021zkr} when computing the superconformal index. These issues are not relevant in the limit considered in this paper.}. The need to sum over saddles was pointed out in this context in \cite{Aharony:2021zkr} for supersymmetric configurations only, but is true more broadly. 

This brings us to our second issue. We see clearly now that even if we restrict to $\Phi_1=\Phi_2=\Phi_3=\frac{2}{3} \Phi$, the sum over integers will necessarily involve configurations outside the scope of section \ref{sec:AdS5black hole}. In order to resolve this we will go to a mixed ensemble where some charges will be fixed. First of all we will introduce some notation, writing the grand-canonical partition function as 
\begin{equation}
    Z (\beta , \Omega_1 , \Omega_2 , \Phi_i  )=\Tr[ e^{-\beta \{ \mathcal{Q},\mathcal{Q}^\dagger \} + \beta (\Omega_1 - 1) J_1 + 
\beta (\Omega_2 - 1) J_2  + \frac{1}{2} \beta (\Phi_1 - 1) R_1 
+ \frac{1}{2} \beta (\Phi_2 - 1) R_2  + \frac{1}{2} \beta (\Phi_3 - 1) R_3 
} ] ,
\end{equation}
where $\{ \mathcal{Q},\mathcal{Q}^\dagger \} = H - J_1 - J_2 - \frac{1}{2} (R_1 + R_2 +R_3) $ and $\mathcal{Q},\mathcal{Q}^\dagger$ is the supercharge preserved by the $1/16$-BPS states at zero temperature. This will allow us later to make a more direct connection with the superconformal index \cite{Kinney:2005ej}. The definition of $\{\cal{Q}, \cal{Q}^\dagger \}$ is chosen to match the supersymmetry condition, \eqref{eq:susychargecondition}, when all $R_i$ are equal. We redefine the chemical potentials to track their temperature scaled deviations from their BPS values \cite{Cabo-Bizet:2018ehj}, as well as new flavor charges 
\be 
\mathfrak{j}_{1,2} \equiv J_{1,2} + \frac{R_3}{2},~~~\mathfrak{q}_{1,2}\equiv \frac{R_{1,2}-R_3}{2}\,,
\ee
which commute with the $\mathcal{N}=1$ subalgebra, as well as
\be 
\Delta_{1,2} = \frac{\beta}{2\pi i} (\Phi_{1,2}-1) ,
\quad 
\omega_{1,2} = \frac{\beta}{2\pi i} (\Omega_{1,2}-1) ,
\quad
\alpha = \frac{\beta}{4\pi i} \varphi ,
\ee
where \be 
\varphi = \Phi_1 + \Phi_2 +\Phi_3 - \Omega_1 - \Omega_2 - 1\,.
\ee In terms of these, the grand canonical partition function becomes
\begin{align}
Z(\beta,\omega_1 , \omega_2, \Delta_1 ,\Delta_2 , \varphi) = 
\Tr[ e^{-\beta \{ \mathcal{Q},\mathcal{Q}^\dagger \} + 
2\pi i \omega_1 \mathfrak{j}_{1} + 
2\pi i \omega_2 \mathfrak{j}_{2}
+ 2\pi i \Delta_1 \mathfrak{q}_1
+ 2\pi i \Delta_2 \mathfrak{q}_2 
+ 2\pi i \alpha R_3
} ] 
.
\end{align}
These variables have the advantage of removing the constraints between the integer shifts appearing in the different solutions. Indeed, the partition function \eqref{eqnssaddlesumads5} with a constraint \eqref{eq:restrictionintegers} can be equivalently written as 
\be 
Z (\beta , \omega_1 , \omega_2 , 
\Delta_1 , \Delta_2 , \alpha )
= 
\sum_{n_1 , n_2 , m , m_1 , m_2} 
Z_{\text{one-loop}} \, e^{-I_{\rm GCE}(\beta, \omega_1 + n_1, \omega_2 +n_2,\Delta_1 + m_1 ,\Delta_2 + m_2 , \alpha + m)} 
,
\ee
with unconstrained integers $(m,m_1,m_2,n_1,n_2)$. This in part motivates these redefinitions. If we expect to obtain an emergent $\mathcal{N}=2$ Schwarzian mode in the nearly $1/16$-BPS limit, it should come with a sum over saddles involving an unconstrained integer. 

Avoiding black hole solutions with unequal R-charges suggests we use a mixed ensemble in which we fix the inverse temperature $\beta$ and the BPS chemical potential $\alpha$, but Laplace transform with respect to the other potentials to obtain a trace over charges. For simplicity we will also work in an ensemble of fixed $\mathfrak{j}_{1,2}$ charges
\begin{align}
Z (\beta, \mathfrak{j}_1,\mathfrak{j}_2, \mathfrak{q}_1 , \mathfrak{q}_2 , \alpha ) &= 
\Tr_{\mathfrak{j}_1, \mathfrak{j}_2,\mathfrak{q}_1,\mathfrak{q}_2}[ e^{-\beta \{ \mathcal{Q},\mathcal{Q}^\dagger\}  
+ 2\pi i \alpha R_3
} ] 
\\
 = 
\int_0^1 d\omega_1 \int_0^1 d\omega_2 
\int_0^1 d\Delta_1 \int_0^1 d\Delta_2 & \,
e^{- 2\pi i \Delta_1 \mathfrak{q}_1 
- 2\pi i \Delta_2 \mathfrak{q}_2
-2\pi i \omega_1 \mathfrak{j}_1
-2\pi i \omega_2 \mathfrak{j}_2}
Z (\beta, \omega_1 , \omega_2, \Delta_1 , \Delta_2 , \alpha )
.
\end{align}
To ease the notation we will use the same letter to denote the generators $\mathfrak{j}_{1,2}$, $\mathfrak{q}_{1,2}$ and their fixed numerical values in the ensemble. To work with equal charges $R_1 = R_2 = R_3$, we set $\mathfrak{q}_1 = \mathfrak{q}_2 = 0$, which leaves us with one free $U(1)$ charge which we have previously called $R$ in the gravity solution. As mentioned above, solutions with more general charges are increasingly difficult to construct \cite{Cvetic:2004ny,Cvetic:2005zi,Chong:2005da,Kunduri:2006ek, Wu:2011gq} because one must include scalar moduli corresponding to deformations of the $S^5$ metric, and are outside the scope of the present work. The mixed ensemble at equal $U(1)^3$ charges is then:
\begin{align}
Z (\beta, \mathfrak{j}_1 , \mathfrak{j}_2, \alpha ) &= 
\Tr_{\mathfrak{j}_{1,2},\mathfrak{q}_{1,2} = 0}[ e^{-\beta \{ \mathcal{Q},\mathcal{Q}^\dagger \}  
+ 2\pi i \alpha R_3
} ] 
\\
 = 
\int_0^1 &d\omega_1  \int_0^1 d\omega_2 
\int_0^1 d\Delta_1 \int_0^1 d\Delta_2  \,
e^{
-2\pi i \omega_1 \mathfrak{j}_1
-2\pi i \omega_2 \mathfrak{j}_2}
Z (\beta, \omega_1 , \omega_2, \Delta_1 , \Delta_2 , \alpha )
.
\end{align}
Now we can put everything together and write an expression for the semiclassical limit of the partition function in the mixed ensemble we are interested in, with  
\be
\label{eq:fixedchargesmixed}
\text{fixed } \alpha,\,\, \mathfrak{j}_{1,2} \text{ and }  R_1-R_2=R_1-R_3 = 0\,.
\ee 
The answer is a sum over a single integer-valued family of saddles 
\begin{align}\label{eqnfinalmixedads5}
Z (\beta, \mathfrak{j}_1,\mathfrak{j}_2, \alpha ) 
&= \sum_{n\in\mathbb{Z}} Z_{\text{one-loop}}
e^{-I_{\rm GCE}(\mathfrak{q}_1=0,\mathfrak{q}_2=0,  \mathfrak{j}_1 , \mathfrak{j}_2  , \alpha + n)
-2\pi i \omega_1 \mathfrak{j}_1
-2\pi i \omega_2 \mathfrak{j}_2
} 
\\
&\equiv \sum_{n\in\mathbb{Z}} Z_{\text{one-loop}}
e^{-I_{\rm ME}(\beta, \mathfrak{j}_1 , \mathfrak{j}_2, \alpha+n )} 
.
\end{align}
In the first line we write the action in this mixed ensemble as the grand-canonical action $I_{\rm GCE}(\mathfrak{q}_1=0,\mathfrak{q}_2=0,  \mathfrak{j}_1 , \mathfrak{j}_2 , \alpha + n)$ plus the evaluation of the boundary terms needed to make the variational problem well-defined (adapting \cite{Hawking:1995ap} to our problem) resulting in the extra terms $2\pi i \omega_1 \mathfrak{j}_1
+2\pi i \omega_2 \mathfrak{j}_2$. In these last two terms $\omega_1$ and $\omega_2$ should be written in terms of the quantities that are fixed in this ensemble. In the second line we defined the total value of the action as $I_{\rm ME}(\beta, \mathfrak{j}_1,\mathfrak{j}_2,\alpha)$.

We now have only the sum over the integer shifts of the holonomy corresponding to the BPS potential in \eqref{eqnfinalmixedads5}. The sum over saddles has the schematic form of \eqref{N2sumsaddles} in which the integer parameter $r=1$. Different values of $m$ can be interpreted as different winding sectors of the Schwarzian $U(1)$ boson (seen in section \ref{sec:SchwarzianSolution}), or in the bulk, as sectors of an abelian gauge field in JT gravity. Here however we see they are the holonomies of a particular linear combination of 10D Kaluza-Klein gauge fields. 

Either from the Schwarzian theory perspective or from the AdS$_5$/$\mathcal{N}=4$ Yang Mills perspective, a black hole where we set $\alpha=1/2$ is supersymmetric and this is equivalent to inserting $(-1)^F=e^{i \pi R_3}$ in the path integral. At these values of chemical potentials we will see the gravity answer matches with the superconformal index. We would like to emphasize that the index (evaluated in the ensemble we are working with) is always in a deconfined phase in the large $N$ limit. The standard superconformal index is studied in the fully grand-canonical ensemble, defined as  
\beq
\widetilde{\mathcal{I}} (\beta, \omega_1 , \omega_2, \Delta_1 , \Delta_2 ) \equiv\Tr[ e^{i \pi R_3} e^{-\beta \{ \mathcal{Q},\mathcal{Q}^\dagger \} + 
2\pi i \omega_1 \mathfrak{j}_{1} + 
2\pi i \omega_2 \mathfrak{j}_{2}
+ 2\pi i \Delta_1 \mathfrak{q}_1
+ 2\pi i \Delta_2 \mathfrak{q}_2 
} ] ,
\eeq
and computed for $\mathcal{N}=4$ Yang Mills in \cite{Romelsberger:2005eg,Kinney:2005ej}. In our ensemble, the index obtained from evaluating $\mathcal{I} (\beta, \mathfrak{j}_1 , \mathfrak{j}_2)\equiv Z(\beta,\mathfrak{j}_1,\mathfrak{j}_2,\alpha=1/2)$ is related to the superconformal index of \cite{Romelsberger:2005eg,Kinney:2005ej} by the following exact relation\footnote{We can generalize this calculation to an ensemble with fixed $\mathfrak{q}_{1,2}\neq 0$ by a similar formula. This should also display always a deconfined phase, although the bulk analysis is more complicated.}
\begin{align}
\mathcal{I} (\beta, \mathfrak{j}_1 , \mathfrak{j}_2) &= 
\Tr_{\mathfrak{j}_{1,2},\mathfrak{q}_{1,2} = 0}[(-1)^F e^{-\beta \{ \mathcal{Q},\mathcal{Q}^\dagger \}  
} ] 
\\
 = 
\int_0^1 &d\omega_1  \int_0^1 d\omega_2 
\int_0^1 d\Delta_1 \int_0^1 d\Delta_2  \,
e^{
-2\pi i \omega_1 \mathfrak{j}_1
-2\pi i \omega_2 \mathfrak{j}_2}
\widetilde{\mathcal{I}}  (\beta, \omega_1 , \omega_2, \Delta_1 , \Delta_2 )
.
\end{align}
While the superconformal index $\widetilde{\mathcal{I}} (\beta, \omega_1 , \omega_2, \Delta_1 , \Delta_2 )$ can be in a confined phase for some ranges of parameters, hiding the black hole behavior \cite{ArabiArdehali:2021nsx,Cassani:2021fyv,Copetti:2020dil}, the index $\mathcal{I} (\beta, \mathfrak{j}_1 , \mathfrak{j}_2)$ defined above (with some charges fixed) is always dominated by the black hole saddle. We leave for future work verifying this from a boundary perspective.

In section \ref{sec:low-T-expansion}, we will see that not only does \eqref{eqnfinalmixedads5} share the same sum over $U(1)$ saddles of the $\cN=2$ super-Schwarzian partition function, but it also has the same classical action as a function of $\beta$ and $\alpha$ in the limit of low temperatures and fixed $\alpha$. From this we will determine the one-loop determinant around each saddle in section \ref{sec:SpectrumN4}. 

\subsection{Near horizon geometry and supersymmetries}
\label{sec:extremal10D}
Having discussed the general black hole solution, the mixed ensemble partition function, and the saddles that contribute to it, we turn our attention to the near-extremal (and specifically near-BPS) limits.   For the AdS$_5$ black hole \eqref{eq-5d-metric}, the exact supersymmetric and extremal limits have a near-horizon geometry described by AdS$_2$, and upon uplifting to 10D, the solution is a specific fibration of AdS$_2 \times S^3 \times S^5$ which we describe in detail.

In the strict BPS near-horizon limit, we show the solution develops emergent superconformal isometries. On the bosonic side, there is both an emergent $SL(2,\mathbb{R})$ for the AdS$_2$ as well as a $U(1)$ which rotates the solution along a particular combination of angles on $S^3 \times S^5$. For the fermions, we find as expected a doubling of the number of supersymmetries corresponding to the appearance of a conformal Killing spinor. 

Altogether, the BPS gravity theory has a local $SU(1,1|1)$ superconformal algebra of symmetries in the near horizon region, but finite temperature quantum corrections explicitly break this local symmetry to a global $SU(1,1|1)$ which acts as isometries of the near-AdS$_2$ saddles. The $\mathcal{N}=2$ Schwarzian theory captures the breaking of these symmetries. In order to demonstrate that we have identified the correct pattern of symmetry breaking (and thus the correct Schwarzian), we consider the BPS limit of the leading saddle and compute explicitly the global super-isometries of the 10D solution explained above.

In contrast to other parts of section \ref{sec:BPS-and-near-BPS-BHs}, in this subsection we will set the rotation parameters $a=b$ for simplicity. This will not change the general structure of the symmetry algebra, but it will impact the specific form of the Killing vectors and Killing spinors. Much of this section is based on the analysis of \cite{Sinha:2006sh}, but see also \cite{David:2020ems}. Details of the supergravity analysis are given in Appendix \ref{app:details-killing-spinors}.

In the original coordinates $(t,r,\theta,\phi,\psi)$ the metric is asymptotically static and the Killing horizon is generated by 
\be 
V = \frac{\partial}{\partial t} + \Omega_1 \frac{\partial}{\partial \phi} + \Omega_2 \frac{\partial}{\partial \psi}.
\ee
For the purpose of determining the near horizon geometry, it is convenient to switch to corotating coordinates, in which $V = \frac{\partial}{\partial t}$ becomes null at the horizon. This amounts to changing angles $(\phi,\psi)$ to $(\tilde{\phi},\tilde{\psi})$, such that $\phi = \tilde{\phi} + \Omega_1 \, t$ and $\psi = \tilde{\psi} + \Omega_2 \, t $. In these coordinates the horizon becomes static and the asymptotic metric is now rotating. In the BPS limit, going to corotating coordinates amounts to
\be 
\phi = \tilde{\phi} + \Omega_1^* \, t = \tilde{\phi} + t , \qquad
\psi = \tilde{\psi} + \Omega_2^* \, t = \tilde{\psi} + t
.
\ee
The near-horizon limit is now taken by setting 
\be 
r \rightarrow \sqrt{a(a+2)} + r , 
\ee
and expanding for $r \ll r^*$. In what follows, we introduce parameters 
\be
\omega = \sqrt{\frac{2a}{1-a}} \, , \qquad \lambda = \sqrt{1+3 \omega^2} \, 
.
\ee
We also make a further general coordinate transformation in which we scale 
\begin{eqnarray}
r = \frac{\omega}{4 \lambda} (1+a)\sqrt{\frac{a}{2+a}} \, \tilde{r} \, ,
\end{eqnarray} then drop the tilde for notational convenience. This leads to a particular fibration of AdS$_2 \times S^3$:
\be 
ds^2_5 &= \left (\frac{\omega}{2\lambda} \right)^2 \left (-r^2 dt^2 
+ \frac{dr^2}{r^2} \right ) + 
3 \left(\frac{\omega ^2}{4}  \sigma _3^L+\frac{\omega}{4 \lambda } r dt\right)^2
+\omega ^2 d\Omega_3^2 , 
\\
A &= -\frac{3}{2} \left(\frac{\omega ^2}{4}  \sigma _3^L+\frac{\omega}{4 \lambda} r dt\right) 
,
\ee
where $d\Omega_3^2 =  d\theta ^2 +  \cos ^2 \theta d\tilde{\psi} ^2+\sin ^2 \theta  d\tilde{\phi}^2$ is the metric on $S^3$, and $\sigma_3^L = 2 (\cos ^2 \theta  d\tilde{\psi} +\sin ^2 \theta  d\tilde{\phi} )$. 

To determine the Killing spinors and Killing vectors in this geometry, we follow \cite{Sinha:2006sh} and work with a 10D lift of the above metric. While it is not strictly necessary to work in 10D, this presentation serves to set our conventions and gives a geometrical origin for all the symmetries of the solution. Working with the leading order Type IIB supergravity theory, the massless fields are the NS-NS sector fields $(G_{MN}, B^{(2)}_{MN}, \Phi)$, the RR-form fields $(C^{(0)}, C^{(2)},C^{(4)})$, as well as complex Weyl spinors and gravitinos $(\lambda, \Psi_M)$ of the same chirality. The field strength $F^{(5)} = dC^{(4)}$ is self-dual, $F^{(5)} = \star_{(10)}F^{(5)}$. In the empty AdS$_5$ $\times$ S$^5$ as well as the $1/16$-BPS AdS backgrounds of IIB supergravity, we may set the axio-dilaton $(\Phi, C^{(0)})$ to a constant, all fermions $(\lambda, \Psi_M)$ to zero, and only turn on a supersymmetric background for the metric and the 5-form flux $(G_{MN}, F_{MNPQR})$. In Einstein frame, there is no explicit dependence on the scalars, no Chern-Simons terms, and the action takes the simple form:
\begin{equation}
    S_{IIB} = \frac{1}{16 \pi G_{10}}\int d^{10}x \sqrt{-G} \left (\mathcal{R}_{(10)} -\frac{1}{4\cdot 5!}F_{M_1\dots M_5}F^{M_1 \dots M_5}\right ) \, .
\end{equation}
As is standard for theories with self-dual gauge fields, above we wrote the kinetic term for $F^{(5)}$, but one should always impose the $F = \ast F$ equation of motion by hand\footnote{In practice, one can work with the covariant equations of motion\cite{Schwarz:1983qr}, perform a dimensional reduction on $S^5$ and then use the 5D action \eqref{eq:5DLag}. The generation of the Chern-Simons term comes from reducing the $F^{(5)}$ equation of motion\cite{Pernici:1985ju,Gunaydin:1985cu}.}.

The near-horizon metric and gauge field may now be uplifted to a 10D solution. This is now given by \cite{Chamblin:1999tk,Cvetic:1999xp}:
\begin{align}
ds^2_{10} 
&=   ds^2_5 +  \sum_{i=1}^3 \left[ (d\mu_i)^2 + \mu_i^2 \left( d\xi_i^2 - \frac{2}{3} A \right)^2  \right]
, \label{eq:near-horizon-metric-10d-lift}\\
F^{(5)} &=
(1+ \ast_{(10)} )
\left[ 
-4 \, \text{vol}_{(5)} - \frac{1}{3} 
\sum_{i=1}^{3} d(\mu_i^2) \wedge 
d\xi_i \wedge \ast_{(5)} F^{(2)} 
\right]
,
\end{align}
where we introduced $\mu_1 = \sin \tilde{\alpha} ,$ $ \mu_2 = \cos \tilde{\alpha} \sin \tilde{\beta} ,$ $\mu_3 = \cos \tilde{\alpha} \cos \tilde{\beta} $. The angles $\tilde{\alpha} \in [0, \frac{\pi}{2}]$, $\tilde{\beta} \in [0, \frac{\pi}{2}]$, $\xi_i \in [0,2\pi]$, parametrize the sphere $S^5$.

In the 10D background, the supersymmetry transformation of the gravitino is
\begin{equation}
\delta_\epsilon \Psi_M = (\partial_M +\frac14 \omega^{AB}_M \Gamma_{AB}) \epsilon + \frac{i}{16\cdot 5!}F_{N_1 \dots N_5}\Gamma^{N_1 \dots N_5}\Gamma_M \epsilon \, .
\end{equation}
Using this, the independent Killing spinors are found to be 
\begin{align}
\epsilon_1 &= \sqrt{\frac{8 \lambda}{\omega^2 r}} 
e^{-\frac{i}{2}(\xi_1 + \xi_2 + \xi_3)} \left[ 
\epsilon_0^+ - \left(
 \frac{i r t }{2} + \frac{3\omega}{2\lambda}
\right) \Gamma^4 \Gamma^9 \epsilon_0^+ 
\right]
,\\
\epsilon_2 &= \sqrt{\frac{\omega^2 r}{8 \lambda}} e^{-\frac{i}{2}(\xi_1 + \xi_2 + \xi_3)} \epsilon_0^- 
,
\label{eq:killing-spinors}
\end{align}

where $\epsilon_0 ^+$ and $\epsilon_0 ^-$ are constant Majorana-Weyl spinors determined from integrability condition for Killing spinors \eqref{appendix-eq-integrability condition}. They are parametrized by two arbitrary real parameters each, giving four real parameters in total. This means that the near-horizon region preserves four supersymmetries, in contrast to full geometry of the $\frac{1}{16}$ BPS black hole we started with \cite{Gauntlett:2004cm,Chamseddine:1996pi}. \\
Knowing the Killing spinors of the near-horizon metric, we can immediately find its Killing vectors by computing linearly independent Killing spinor bilinears $\bar{\epsilon}_I \Gamma^a \epsilon_J$. After appropriately normalizing the constant spinors $\epsilon_0^\pm$ we find Killing vectors
\begin{align}
D &= 
-t \partial_t + r \partial_r 
\label{eq:killing-vector-J}
,\\
\mathcal{Z} &= 
\frac{i \omega^2}{\lambda^2} (
\partial_{\xi_1}+\partial_{\xi_2}+\partial_{\xi_3} 
) - \frac{i}{2\lambda^2} (\partial_{\tilde{\phi}} + \partial_{\tilde{\psi}})
,
\label{eq:killing-vector-Z}
\\
E_+ &= 
- \frac{i \omega^2}{4\lambda} \partial_t 
,
\label{eq:killing-vector-Ep}
\\ 
E_- &= 
\frac{8 i \lambda }{\omega^2} r t \partial_r 
+ \left( 
- \frac{4 i \lambda }{\omega^2}  t^2
+ \frac{ 4i   (9\omega^2 - 4\lambda^2 )}{\omega^2 \lambda}\frac{1}{r^2}
\right) \partial_t 
\\
&  
+ \frac{8i }{\lambda^2 \omega }\frac{1}{r} (\partial_{\xi_1}
+\partial_{\xi_2} + \partial_{\xi_3}
) + \frac{12 i  }{\lambda^2 \omega }\frac{1}{r} 
(\partial_{\tilde{\phi}} + \partial_{\tilde{\psi}} )
.
\label{eq:killing-vector-Em}
\end{align}
One can verify that with the above normalization they satisfy the following commutation relations 
\begin{align}
[D,E_{\pm}] &= \pm E_{\pm} , & [D,\mathcal{Z}] &= 0
,\\
[\mathcal{Z}, E_{\pm}] &= 0
,&
[E_+ , E_- ]&= 2 D \,.
\end{align}
We can thus identify $D,\,E_{\pm}$ as the generators of $SL(2, \mR)$ and $\mathcal{Z}$ as a  $U(1)$ R-symmetry generator. Using a standard procedure of determining spacetime isometry superalgebras \cite{Beck:2017wpm,Gauntlett:1998kc,Gauntlett:1998fz,Figueroa-OFarrill:1999klq} we can interpret the above commutation relations as giving the bosonic part of the isometry superalgebra through
\be
[\mathcal{Q}_B (k_i) , \mathcal{Q}_B (k_j)] &= \mathcal{Q}_B ([k_i,k_j]) ,
\ee
with $\mathcal{Q}_B (k_i)$ denoting the bosonic generator associated to Killing vector $k_i$. To determine the rest of the superalgebra we follow the prescription 
\begin{align}
[\mathcal{Q}_B (k) , \mathcal{Q}_F (\epsilon) ] &= \mathcal{Q}_F (\mathbb{L}_k \epsilon) ,
\\
\{ \mathcal{Q}_F (\epsilon_I ) , \mathcal{Q}_F (\epsilon_J ) \} &= \mathcal{Q}_B (\epsilon_I \Gamma \epsilon_J) 
,
\end{align}
where $\mathcal{Q}_F(\epsilon_I)$ denotes now the fermionic generators associated to Killing spinor $\epsilon_I$, and the Killing vectors act on the Killing spinors through the spinorial Lie derivative \cite{Figueroa-OFarrill:1999klq}, shown explicitly in \eqref{appendix-eq-spinorial-Lie-derivative}. With that, we obtain
\begin{align}
\{ \bar{\mathcal{Q}}_2,\mathcal{Q}_1 \} &= \mathcal{Z} + D , & \{ \bar{\mathcal{Q}}_1,\mathcal{Q}_2 \} &= \mathcal{Z} - D ,
\\ 
\{ \bar{\mathcal{Q}}_2,\mathcal{Q}_2 \} &= E_+ , & \{ \bar{\mathcal{Q}}_1,\mathcal{Q}_1 \} &= E_- ,
\end{align} 
together with
\begin{align}
[\mathcal{Q}_1,\mathcal{Z}] &= -\frac{1}{2}\mathcal{Q}_1 , & [\bar{\mathcal{Q}}_1,\mathcal{Z}] &= \frac{1}{2} \bar{\mathcal{Q}}_1 , &
[\mathcal{Q}_2,\mathcal{Z}] &= -\frac{1}{2}\mathcal{Q}_2 , & [\bar{\mathcal{Q}}_2,\mathcal{Z}] &= \frac{1}{2} \bar{\mathcal{Q}}_2 , \nn \\
[\mathcal{Q}_1,D] &= \frac{1}{2}\mathcal{Q}_1 , & [\bar{\mathcal{Q}}_1,D] &= \frac{1}{2} \bar{\mathcal{Q}}_1 , &
[\mathcal{Q}_2,D] &= -\frac{1}{2}\mathcal{Q}_2 , & [\bar{\mathcal{Q}}_2,D] &= -\frac{1}{2} \bar{\mathcal{Q}}_2 ,\nn \\
[\mathcal{Q}_1,E_+] &= \mathcal{Q}_2 , & [\bar{\mathcal{Q}}_1,E_+] &= -\bar{\mathcal{Q}}_2 , &
[\mathcal{Q}_2,E_+] &= 0 , & [\bar{\mathcal{Q}}_2,E_+] &=0 ,\nn \\
[\mathcal{Q}_1,E_-] &= 0 , & [\bar{\mathcal{Q}}_1,E_-] &= 0 , &
[\mathcal{Q}_2,E_-] &= \mathcal{Q}_1 , & [\bar{\mathcal{Q}}_2,E_-] &= -\bar{\mathcal{Q}}_1\,,
\end{align}
which can be identified as the $SU(1,1|1)$ superalgebra that the extremal black holes exhibit in their near-horizon region.

It is worth mentioning, as explained in \cite{Sinha:2006sh} (which was specialized to the case $J_1 = J_2$, $R_i = R$), that the solution has additional bosonic isometries $SU(2) \times U(3)$ which do not mix with $SU(1,1|1)$. In principle we could introduce more general charges or chemical potentials corresponding to these symmetries, but as we have already explained in the discussion surrounding \eqref{eq:fixedchargesmixed}, we are working in a mixed ensemble in which some combination of the rotations and R-charges is held fixed, so these extra symmetries play no role in the thermodynamics. 

In the standard example of AdS$_2$ at low but non-zero temperature, the boundary modes of the metric become strongly coupled, and the gravitational path integral reduces to that of JT gravity at leading order in $e^{S_0}$ and $\beta$. JT gravity may itself by written as an $SL(2,\mathbb{R})$ $BF$ theory in the first order formulation of gravity, subject to the correct boundary conditions from the second order formalism \cite{Saad:2019lba}. In the present example, superconformal symmetry of the 10D solution implies that both the boundary modes of the metric and its superpartners become strongly coupled, and the effective two-dimensional theory in the AdS$_2$ throat should instead be based on a $SU(1,1|1)$ $BF$ theory. In the present work, we have not attempted to derive the full two-dimensional dilaton supergravity, however, this point of view is not essential for our argument. A reader interested in dimensional reductions to JT supergravity as well as the relation to super-$BF$ theory may consult \cite{Heydeman:2020hhw,Stanford:2019vob,Forste:2020xwx, Cardenas:2018krd,Iliesiu:2019xuh, Blommaert:2018iqz, Fan:2021wsb}.

\subsection{The low-temperature expansion of the action}
\label{sec:low-T-expansion}

The purpose of this section is to show the emergence of the $\mathcal{N}=2$ Jackiw-Teitelboim mode in the near $1/16$-BPS limit. This was anticipated in the previous section based on the pattern of symmetry breaking of nearly $1/16$-BPS states. We will work in an ensemble of fixed $\mathfrak{j}_1$, $\mathfrak{j}_2$ and $\mathfrak{q}_{1}=\mathfrak{q}_{2}=0$. For simplicity we focus first on saddles with integer $n=0$ since the general case can be obtained by a simple shift of the chemical potential $\alpha \to \alpha + n$, with $n\in \mathbb{Z}$. Instead of implementing a dimensional reduction of type IIB supergravity near the horizon to obtain JT gravity, we will take a shortcut and match the classical action. We leave a full treatement of the reduction for future work.

We now turn to expansion of the on-shell action around the $1/16$-BPS values of parameters $T=0$ and $\varphi = 0$. We turn on non-zero values of $T$ and chemical potential $\varphi$ and assume $T,\varphi\ll 1$, keeping $\alpha = \frac{\beta}{4\pi i} \varphi$ fixed. As explained above, in the fixed $(\beta, \mathfrak{j}_{1,2}, \mathfrak{q}_{1,2}=0, \alpha)$ ensemble the classical action involves new boundary terms compared with the grand canonical ensemble 
\begin{equation}
 I_{\rm ME}(\beta, \mathfrak{j}_{1},\mathfrak{j}_{2}, \alpha) =  I_{\rm GCE}(\mathfrak{q}_1=0,\mathfrak{q}_2=0,  \mathfrak{j}_1 , \mathfrak{j}_2  , \alpha)
+2\pi i \omega_1 \mathfrak{j}_1
+2\pi i \omega_2 \mathfrak{j}_2,
\end{equation}
where the GCE action is evaluated at necessary chemical potentials and angular momenta such that it is dominates by the quantum numbers $\mathfrak{q}_1=0,\,\mathfrak{q}_2=0, \, \mathfrak{j}_1,$ and $ \mathfrak{j}_2  $. 
To set up the calculation, we begin picking $a^\star,b^\star$ to correspond to the zero temperature $1/16$-BPS black hole. The parameters $a^\star,b^\star$ determines the fixed values of $\mathfrak{j}_{1,2}$ in the mixed ensemble through 
\beq
\mathfrak{j}_1 =N^2 \frac{  \left(a^*+b^*\right) \left(a^* +1\right) \left(b^* +1\right)}{2 \left(a^* -1\right)^2 \left(1-b^* \right)},~~~~\mathfrak{j}_2 =N^2 \frac{  \left(a^*+b^*\right) \left(a^* +1\right) \left(b^* +1\right)}{2 \left(1-a^*\right) \left(b^* -1\right)^2},
\eeq
where we restored the $N$ dependence by replacing $G_5 = \frac{\pi}{2N^2}$ to make the comparison with $\mathcal{N}=4$ Yang Mills more transparent. Next we go away from the $1/16$-BPS black hole by expanding around these values $(r^{\star },q^\star,a^\star,b^\star)\to (r^{\star }+\delta r,q^\star+\delta q,a^\star+\delta a,b^\star+\delta b)$ where $(r^{\star },q^\star,a^\star,b^\star)$ are determined in terms of $\mathfrak{j_1}$ and $\mathfrak{j_2}$ due to the nonlinear relation imposed for BPS black holes. Imposing then that we work with fixed $(T,\varphi,\mathfrak{j}_1 ,\mathfrak{j}_2 )$, fixes the expansion parameters $\delta r$, $\delta q$, $\delta a$ , $\delta b$ in terms of $T$ and $\varphi$, using the expressions given in section \ref{sec:AdS5black hole}. We carry this out in detail in Appendix \ref{app:low-temp-expansion}. It is convenient to write all formulas in terms of $a^\star$ and $b^\star$, understanding they should be thought of as functions of $\mathfrak{j}_1$ and $\mathfrak{j}_2$. 

The final answer for the action in the low temperature expansion in the mixed ensemble is given by 
\begin{eqnarray}
\label{eq:action-low-temp-expansion}
 I_{\rm ME}(\beta, \mathfrak{j}_{1},\mathfrak{j}_{2}, \alpha) &=& 
\beta E_0-S^*
-  \frac{1}{2} \beta \varphi R^*
-  
\frac{2\pi^2}{\beta \MSU} \left(  1 + 
\left(\frac{\beta \varphi}{2\pi } \right)^2
\right) 
,\nonumber\\
&=&\beta E_0 -S^* - 2\pi i \alpha R^* 
- \frac{2\pi^2}{\beta \MSU} 
(1- 4 \alpha^2) 
,
\end{eqnarray}
where we defined the BPS Bekenstein-Hawking entropy and R-charge by the expressions
\begin{align}
S^* &\equiv N^2 \frac{\pi (a^* + b^*) \sqrt{a^* b^* +a^* + b^*}}{(1- a^* ) (1- b^* )}
, 
\\ 
R^* &\equiv N^2 \frac{(a^* + b^* )}{
(1-a^* ) (1-b^* )}
, 
\end{align}
and as in \eqref{eq:near-extremal-RN-part-function}, we defined the parameter $\MSU$ explicitly as
\be
\label{eq:Phi-r-formula}
\frac{1}{\MSU}&= 
N^2 \frac{ \left(a^*+b^*\right)^2 \left(3-a^*  b^* +a^*+b^*\right)}
{8 \left(1-a^*\right) \left(1-b^*\right) \left(3 a^*  b^*
   +3 a^*+ a^{*2} + b^{*2} +3 b^* +1\right)} 
.
\ee
Other evidence of the dynamics in the near-extremal limit being controlled by the breaking of $SU(1,1|1)$ was given in \cite{Larsen:2019oll}, which also identified \eqref{eq:Phi-r-formula} as the relevant energy scale. 

Now we can put everything together and write a formula giving an approximation of the mixed ensemble partition function in the nearly $1/16$-BPS limit after including the sum over saddles 
\begin{equation}\label{eq:part-function-partial-can-ens-just-action-at-low-T}
    Z (\beta, \mathfrak{j}_1,\mathfrak{j}_2, \alpha ) 
=e^{2\pi i\alpha R^\star} e^{-\beta E_0} \sum_{n\in\mathbb{Z}} Z_{\text{one-loop}}
~e^{S^\star + \frac{2\pi^2}{\beta \MSU}(1-4(\alpha+n)^2)},
\end{equation}
where for now we ignored the one-loop corrections. This is precisely the classical partition function as computed by the $\mathcal{N}=2$ Schwarzian action, as given in \eqref{N2sumsaddles}. In this identification $S^\star$ corresponds to the topological term $S_0$ in the Schwarzian theory. The only difference is the first term $e^{2\pi i \alpha R^\star}$. Since for the theory under consideration the $R$ charge is an integer, that phase is not affected by the sum over saddles and can be pulled in front of the sum in \eqref{eq:part-function-partial-can-ens-just-action-at-low-T}. This term comes from the fact that there is a mismatch between $R$, (the R-charge in the UV) and the Schwarzian $U(1)_R$ charge $Z$ given by $Z = R-R^\star$. This is analogous to how in section \ref{sec:schwarzian-philosophy}, the grand-canonical ensemble was dominated by some charge $Q_*$ around which excitations were captured by a $U(1)$ whose contribution once again consisted of a sum over saddles \eqref{eq:example-grand-canonical-part-function}. Additionally, we observe that the R-charge of the supercharge is the same as the fundamental charge, so we have $r=1$ compared to \eqref{N2sumsaddles}.  Finally, there are no Schwarzian $\theta$ terms arising from our UV theory which would have appeared through an imaginary contribution to the action, $i \theta n$ for each saddle in \eqref{eq:part-function-partial-can-ens-just-action-at-low-T}. Such a topological term might have been present due to the Chern-Simons term in  the 5d action \eqref{eq:5DLag}, but turns out to be absent here.

In the general case, the equations above depend non-trivially on the angular momenta $\mathfrak{j}_{1,2}$ through the complicated dependence on $a^\star$ and $b^\star$. This dependence becomes simpler in the limit where for example $a^\star \to 1$ and $b^\star \to 1$. In this limit $\mathcal{J}_{1,2} \equiv \mathfrak{j}_{1,2}/N^2 \to \infty$, and we get 
\begin{equation}
    S^\star \sim N^2 \sqrt{3} \Big(\frac{\mathcal{J}_1 \mathcal{J}_2}{2}\Big)^{1/3},~~~R^\star \sim N^2  \Big(\frac{\mathcal{J}_1 \mathcal{J}_2}{2}\Big)^{1/3},~~~\frac{1}{\MSU} \sim  N^2 \frac{2}{3}  \Big(\frac{\mathcal{J}_1 \mathcal{J}_2}{2}\Big)^{1/3} \sim \frac{2 R^\star}{3} .
\end{equation}
Since in this limit $R^\star \sim N^2 \mathcal{O}(  \mathcal{J}^{2/3})$, we see that in defining $\mathfrak{j}_{1,2}$, the contribution from $R^\star$ is subleading compared with $J_{1,2}^\star$. Therefore in this limit we can also identify $\mathcal{J}_{1,2}$ with the angular momentum along $S^3\subset AdS_5$ directly.

\subsection{The black hole spectrum including quantum corrections}\label{sec:SpectrumN4}

The classical analysis above determined that the relevant effective theory which describes the near-horizon region (as a function of the temperature and chemical potential) is the $\mathcal N=2$ super-Schwarzian theory. Following the conventions in section \ref{sec:SchwarzianSolution}, the low-temperature expansion of the action yields the coupling $\MSU^{-1} $ in \eqref{eq:Phi-r-formula}, as well as the $r$ unit-charge and $\theta$-angle:
\be 
\ r=1\,, \qquad \theta=0\,.
\ee
Combining the classical analysis above with the quantum corrections from the Schwarzian mode gives the following prediction for the mixed ensemble partition function from gravity:
\beq\label{N2sumsaddlesAdS55}
  Z (\beta, \mathfrak{j}_1,\mathfrak{j}_2, \alpha ) 
=e^{2\pi i\alpha R^\star}e^{-\beta E_0} \sum_{n\in\mathbb{Z}} \frac{2\cos\left( \pi  (\alpha+ n)\right)}{\pi\left(1-4 (\alpha+n)^2\right)}~e^{S^\star + \frac{2\pi^2}{\beta \MSU}(1-4(\alpha+n)^2)} .
\eeq
As explained in section \ref{sec:schwarzian-philosophy}, if we work in an ensemble where we fix all charges except the one that couples to the Schwarzian mode, the one-loop determinant coming from all additional fields is a constant proportional to $N^\#$. These corrections were recently studied in the context of AdS$_5$ black holes in, for example, \cite{David:2021qaa}. These appear in the exponential as $\log N$ corrections. These will not be the focus of our discussion and therefore in the equation above we absorbed them into $S^\star$ itself. Following \ref{sec:SchwarzianSolution}, the quantum corrected partition function can also be written  
\begin{align}
\label{eq:density-of-states-charge-decomp-AdS5}
&  Z (\beta, \mathfrak{j}_1,\mathfrak{j}_2, \alpha )=  e^{-\beta E_0}e^{2 \pi i \alpha R^\star}\sum_{Z_{\rm Sch}\in \mathbb{Z}}\Big(\delta_{Z_{\rm Sch},0}~e^{S^\star}\\ 
&+  \left( e^{2\pi i \alpha Z_{\rm Sch} } + e^{2\pi i \alpha (Z_{\rm Sch}-1)}\right) \int_{E_\text{gap}(Z_{\rm Sch})}^\infty dE_{\rm Sch} e^{-\beta E_{\rm Sch}}\frac{e^{S^\star} \sinh{\big(2\pi \sqrt{\frac{2}{\MSU}(E_{\rm Sch}-E_\text{gap}(Z_{\rm Sch}))}\big)}}{2\pi E_{\rm Sch}}\Big),\nonumber
\end{align}
where we defined $E_{\rm gap}(Z_{\rm Sch}) = \frac{\MSU}{8} (Z_{\rm Sch} -\frac{1}{2})^2$. We introduce in the equation above the variables $E_{\rm Sch}$ and $Z_{\rm Sch}$ when performing the Laplace transform. These are to be interpreted as the energy and charge respectively of the effective $\mathcal{N}=2$ Schwarzian mode arising in the IR. 

We can relate the Schwarzian charge and energy defined above to $\mathcal{N}=4$ Yang Mills data in a simple way. First by taking into account in \eqref{eq:density-of-states-charge-decomp-AdS5} the $e^{2 \pi i \alpha R^\star}$ prefactor, and remembering from the discussion in section \ref{subsec-ensembles} that $\alpha$ couples to $R$, we see the $\mathcal{N}=4$ Yang Mills R-charge is shifted by $R^\star$ with respect to the Schwarzian charge. Second, we observe that temperature in \eqref{eq:density-of-states-charge-decomp-AdS5} couples to $E_0 + E_{\rm Sch}$.  We can use $E=E_0 +\Delta$ to relate the Schwarzian energy to the $\mathcal{N}=4$ Yang Mills scaling dimension. It is useful to first define 
\beq
\Delta_{\rm BPS} \equiv \mathfrak{j}_1 + \mathfrak{j}_2 + \frac{1}{2}R^\star,
\eeq
which is the scaling dimension of the $1/16$-BPS extremal states. It is also useful to define the scaling dimension obtained from imposing the supersymmetric constraints as 
\beq
\Delta_{\rm SUSY}(R) \equiv \mathfrak{j}_1 + \mathfrak{j}_2 + \frac{1}{2}R,
\eeq
with $\Delta_{\rm SUSY}(R^\star) = \Delta_{\rm BPS}$. Combining these definitions with the fact that temperature in $Z(\beta,\mathfrak{j}_1,\mathfrak{j}_2,\alpha)$ couples to $E- J_1-J_2 - \frac{1}{2}(R_1+R_2+R_3)$ allows us to find $\Delta$ in terms of $E_{\rm Sch}$. To summarize, the $\mathcal{N}=4$ Yang Mills data is given by 
\bea
\Delta &=&\Delta_{BPS}+E_{\rm Sch}  + \frac{1}{2} Z_{\rm Sch},\\
R&=& R^\star + Z_{\rm Sch}.
\ea
Using this identification, we can interpret both lines of equation \eqref{eq:density-of-states-charge-decomp-AdS5} in terms of the spectrum of nearly $1/16$-BPS states in $\mathcal{N}=4$ Yang Mills. The first line represent the $1/16$-BPS states. They have $Z_{\rm Sch}=0$ and therefore $R=R^\star$ and $\Delta = \Delta_{\rm BPS}$, and their degeneracy is equal to $\exp{\left(S^\star\right)}$ in the large $N$ limit. The superconformal index corresponds to $\alpha=1/2$, and this only gets contributions from the BPS states since otherwise the two states in the supermultiplet with charge $Z_{\rm Sch}$ and $Z_{\rm Sch}-1$ cancel in the second line of \eqref{eq:density-of-states-charge-decomp-AdS5}. The answer then for the index is, as usually assumed, given by
\bea
\mathcal{I}(\beta,\mathfrak{j}_1,\mathfrak{j}_2) &\equiv& e^{\beta E_0}Z(\beta,\mathfrak{j}_1,\mathfrak{j}_2,\alpha=1/2),\\
&=&(-1)^{R^\star} e^{S^\star(\mathfrak{j}_1,\mathfrak{j}_2)}.
\ea
The first term is the overall sign depending on whether $R^\star$ is even or odd. Consequently, if $R^\star$ is even we find that the BPS states are bosonic, while when  $R^\star$ is odd we find that the BPS states are fermionic. In either case, there is no cancellation (at least to this order in $N$) in the superconformal index and thus the matching between the index and the entropy of BPS black holes is accurate. The change in sign of the index as $R^\star$ goes from even to odd was also observed explicitly in the boundary theory, in the calculations of \cite{Murthy:2020rbd}.
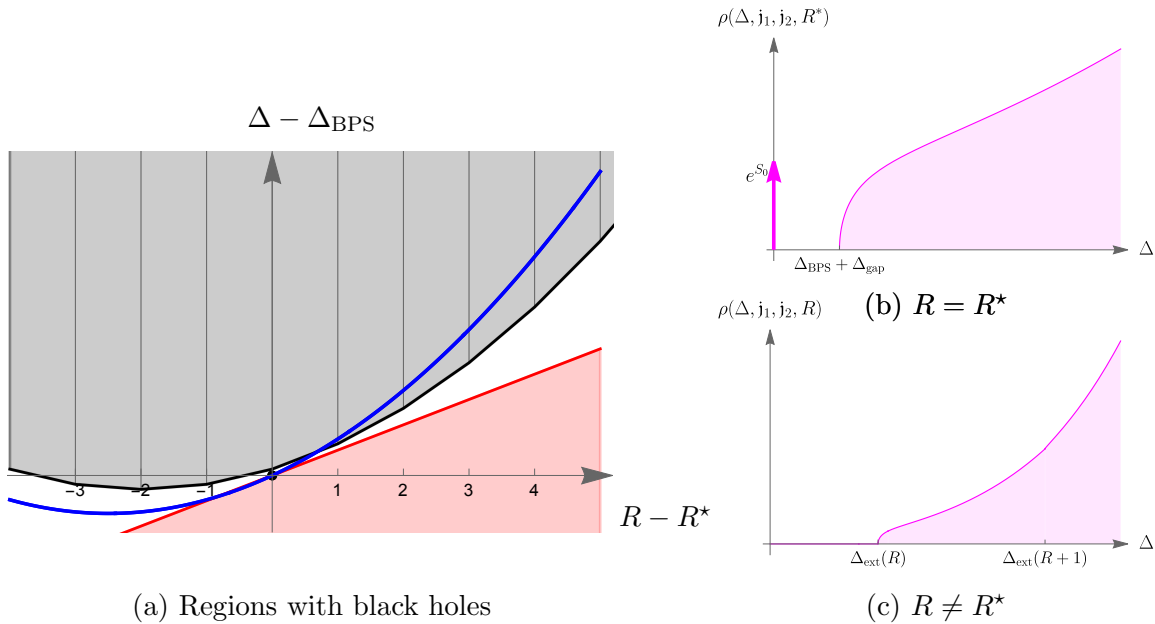
\begin{figure}[t!]
    \centering
\begin{tikzpicture}
     \node[inner sep=0pt] (regions) at (0,0)
    {\includegraphics[width=.47\textwidth]{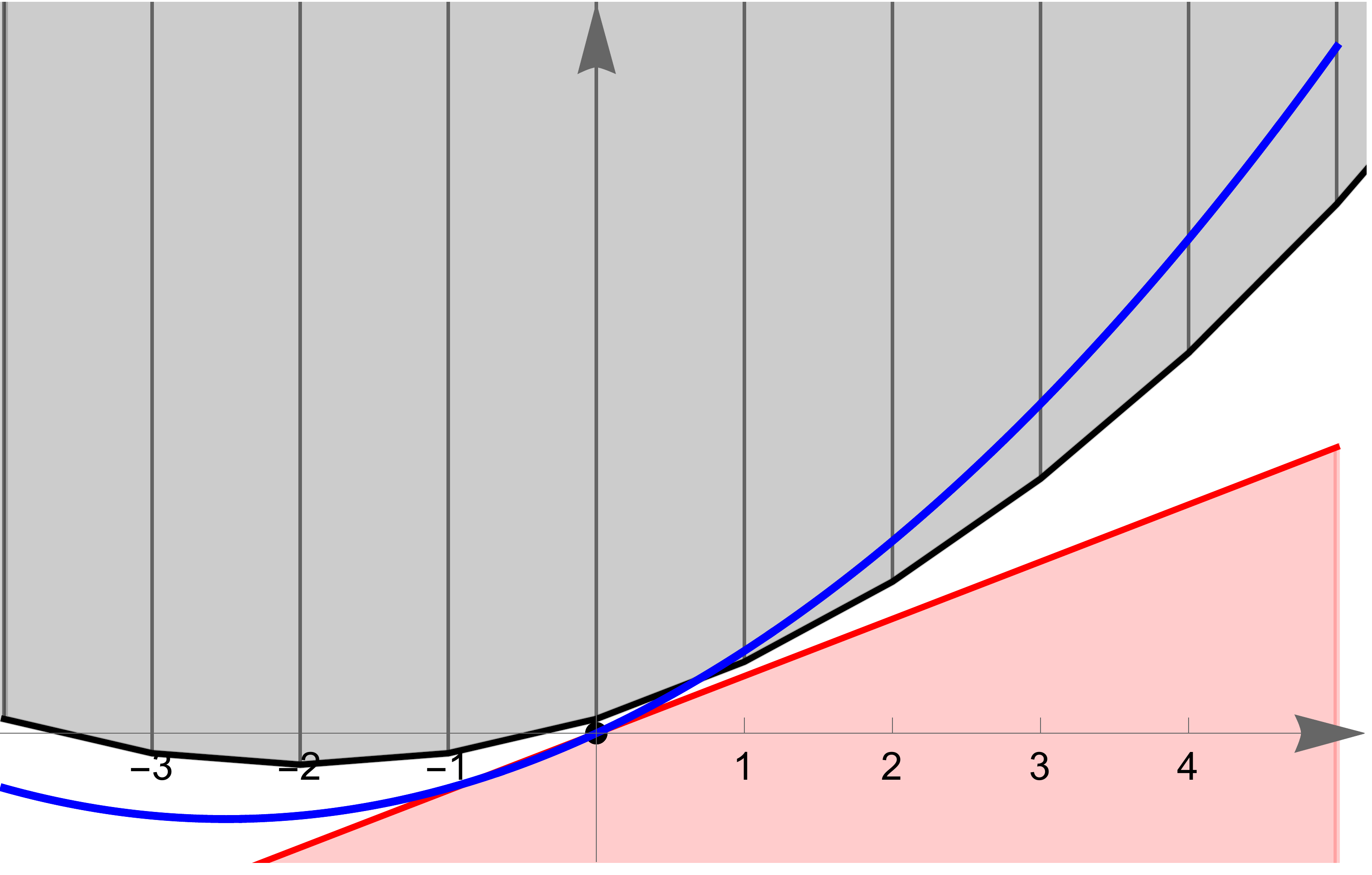}};
     \draw (4.7,-2.3) node  {\small $R-R^\star$};
     \draw (0,3) node  {\small $\Delta-\Delta_{\rm BPS}$};
      \draw (8.3,0.5) node  {\small (b) $R=R^\star$};
    \draw (-0,-3.5) node  {\small (a) Regions with black holes};
    \node[inner sep=0pt] (BPS) at (8.3,2.7)
    {\includegraphics[width=.34\textwidth]{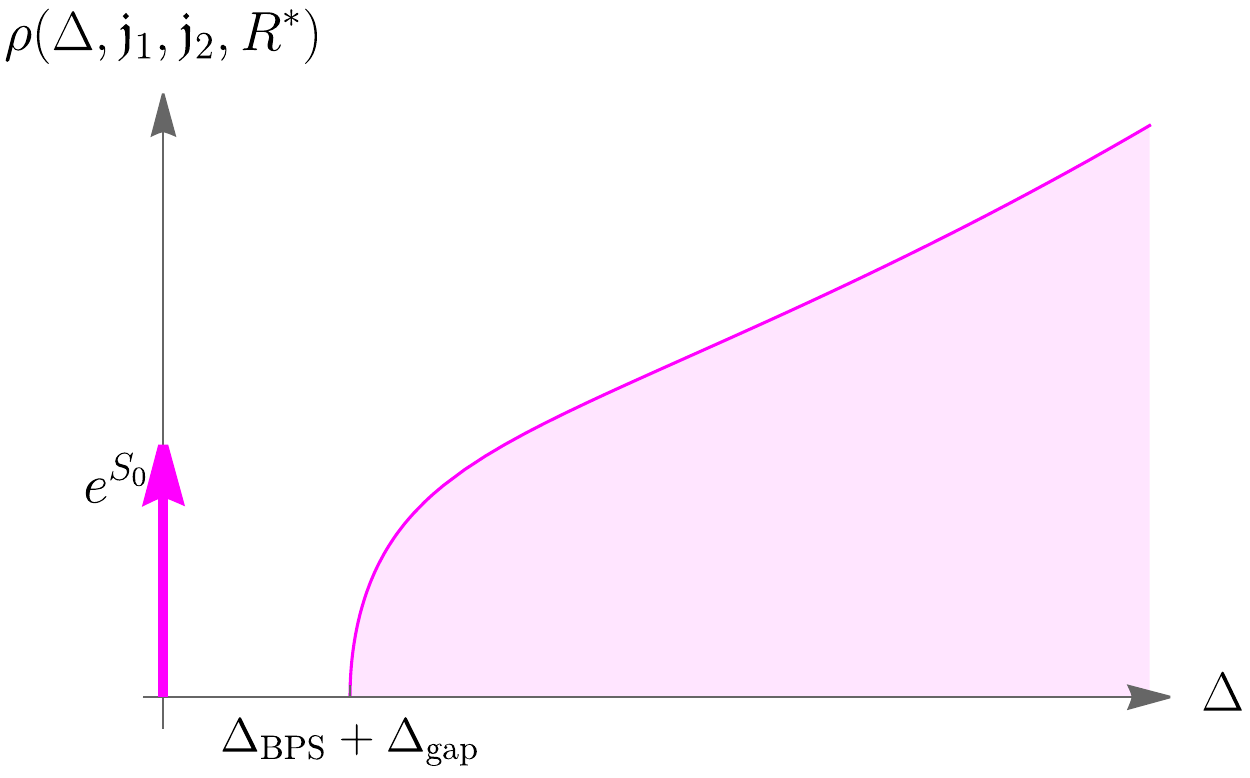}};
     \draw (8.3,0.5) node  {\small (b) $R=R^\star$};
    \node[inner sep=0pt] (nonBPS) at (8.3,-1.2)
    {\includegraphics[width=.34\textwidth]{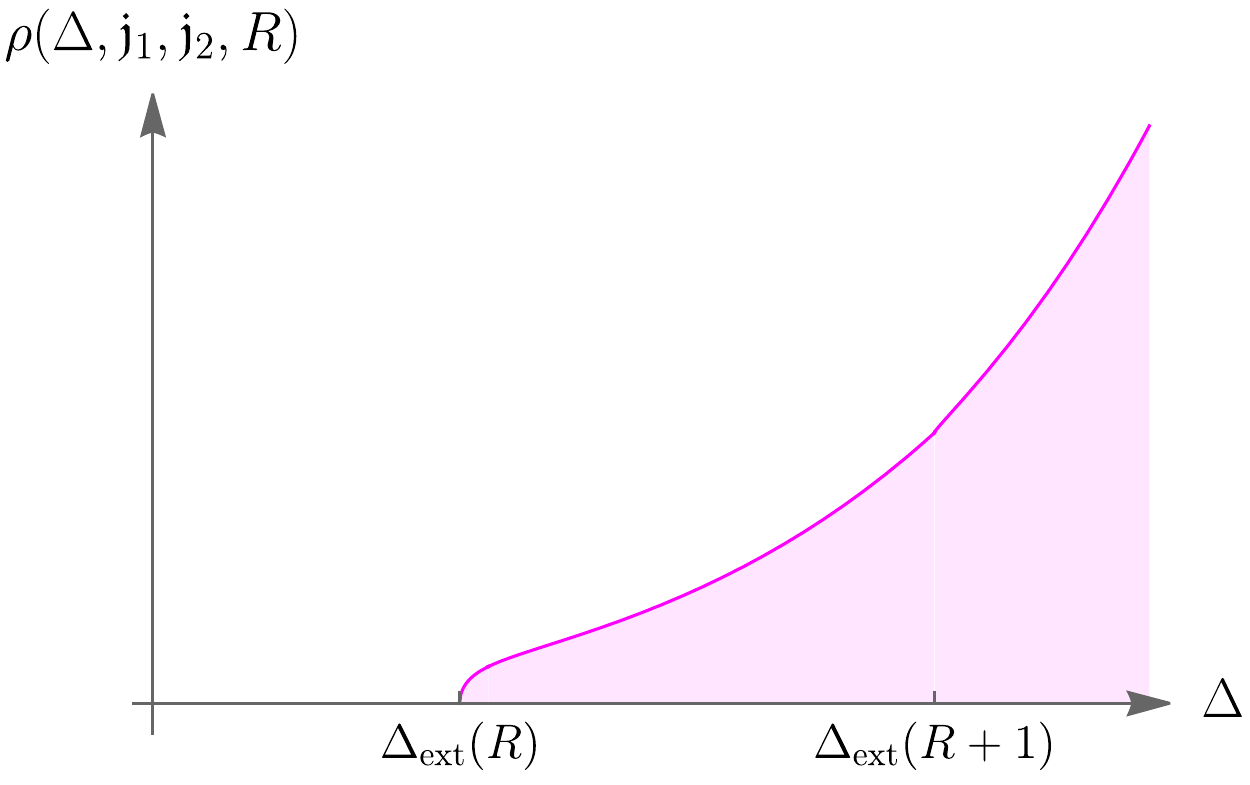}};
     \draw (8.3,-3.5) node  {\small (c) $R\neq R^\star$};
\end{tikzpicture}
      \caption{(a) The figure shows a sketch of the $(R,\Delta)$ plane for fixed $\mathfrak{j}_1$ and $\mathfrak{j}_2$. The red region indicates the forbidden region with $\Delta<\Delta_{\rm SUSY}(R)$. The vertical gray lines denote the continuum sector of the black hole spectrum starting at the quantum corrected extremal value $\Delta_{\rm extremal}(R)$ for $R\neq R^\star$. The blue line denotes the classical extremality bound. When $R=R^\star$ the lightest black hole has dimension $\Delta_{\rm BPS}$ (black dot) and then a continuum above a gap $\Delta_{\rm gap}$. (b) Density of states with charge $R=R^\star$ corresponding to the vertical line at the origin of figure a. (c) Density of state for other charges $R\neq R^\star$.}
    \label{fig:regions}
\end{figure}

Going beyond the BPS states, the second line of  \eqref{eq:density-of-states-charge-decomp-AdS5} gives the spectrum of non supersymmetric black holes. These can be extremal or not. For example when $R\neq R^\star$ the spectrum presents extremal black holes when 
\bea
\Delta_{\rm extremal}(R\neq R^\star) = {\rm min }_{\pm}\left[ \Delta_{\rm BPS} + \frac{1}{2}\Big(R-R^\star +\frac{1}{2}\pm \frac{1}{2} \Big)+ \frac{\MSU}{8 }\Big(R-R^\star \pm \frac{1}{2}\Big)^2 \right] ,
\label{eq:extremal-energy-w-qc}
\ea
that have zero-temperature but are not supersymmetric and only when $R=R^\star$ the two notions match $\Delta_{\rm extremal}(R=R^\star)=\Delta_{\rm BPS}$. The minimum is taken between the two supermultiplets with charge $R$ states. The degeneracy for these black holes predicted by \eqref{eq:density-of-states-charge-decomp-AdS5} goes to zero instead of being exponential in $N^2$.\footnote{This is similar to the behavior for non-supersymmetric black holes studied in \cite{Iliesiu:2020qvm}.} The result for the energy of the extremal states should be compared to the naive classical answer, which in an $(R-R^\star)/R^\star$ expansion is given by, 
\be 
\Delta_\text{extremal}^\text{classical} = \Delta_{\rm BPS} + \frac{1}2 (R-R^\star)+\frac{\MSU}8 (R-R^\star)^2 .
\label{eq:extremal-energy-classical}
\ee
Interestingly, for instance when $R\geq R^\star$, the quantum corrected extremal energy in \eqref{eq:extremal-energy-w-qc} is lower than the naive classical value in \eqref{eq:extremal-energy-classical}.\footnote{This also occurs for sufficiently negative values of $R-R^\star$. } Having a quantum corrected energy ground state below the extremality bound has a possible interpretation in the context of the weak gravity conjecture \cite{Arkani-Hamed:2006emk}. While usually such corrections are seen to come from higher-derivative terms in the action\footnote{See \cite{Harlow:2022gzl} and references therein for a review.} (which we will in fact discuss in section \ref{sec:stringcorrections}), in this case, the correction comes from the temperature dependence of the one-loop determinant in the gravitational path integral. We summarize all these results  about the spectrum of extremal black holes states in figure \ref{fig:regions}(a).

With the expression above we can derive the result quoted in the introduction regarding the spectrum of black hole states in $\mathcal{N}=4$ Yang Mills for states with charge $R=R^\star$. These come from the $1/16$-BPS states counted by the first line of \eqref{eq:density-of-states-charge-decomp-AdS5}, but also from supermultiplets in the second line with maximal R-charge $R^\star+1$. These states appear at a scaling dimension in the range
\beq
\Delta > \Delta_{\rm BPS} + \Delta_\text{gap}\,, \qquad \text{ with } \qquad \Delta_{\rm gap} \equiv  \frac{\MSU}{32} \,. 
\eeq

\begin{figure}[t!]
    \centering
    \begin{tikzpicture}
     \node[inner sep=0pt] (energy BPS) at (0,3.23)
    {\includegraphics[width=.49\textwidth]{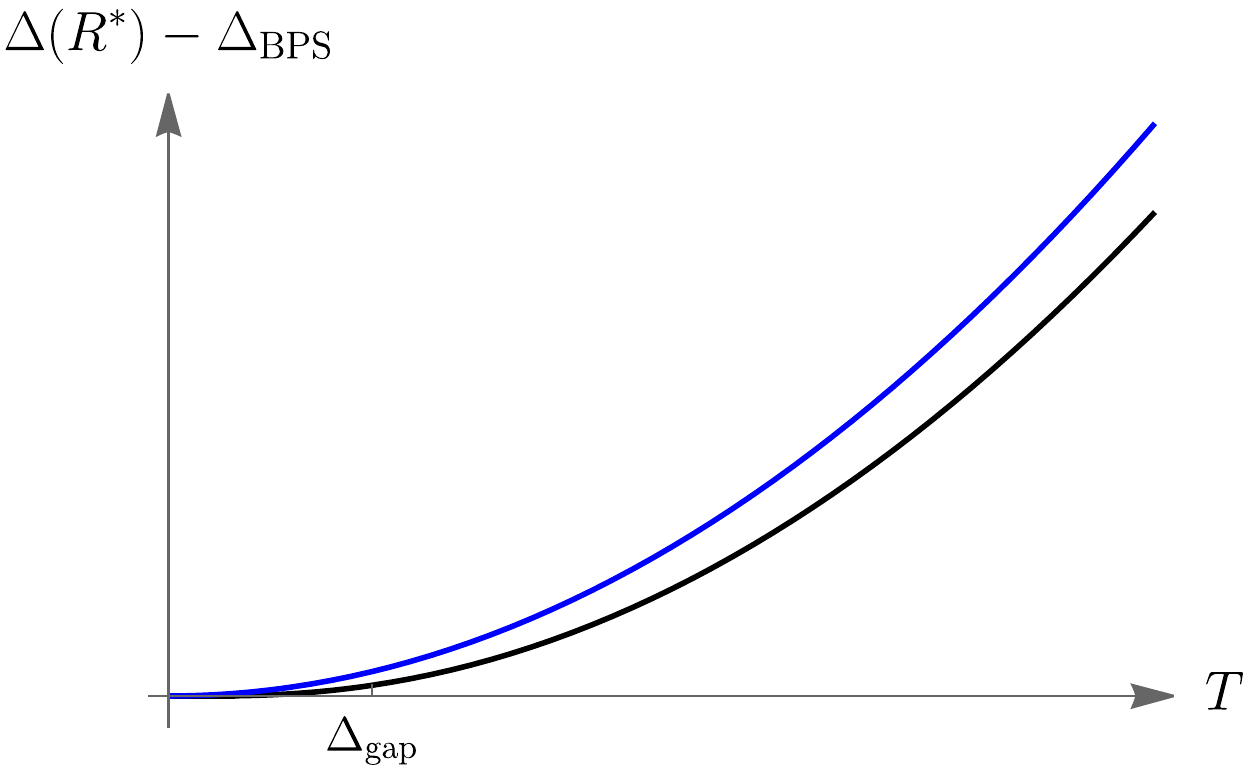}};
    \node[inner sep=0pt] (energy) at (-.28,-2.30)
    {\includegraphics[width=.49\textwidth]{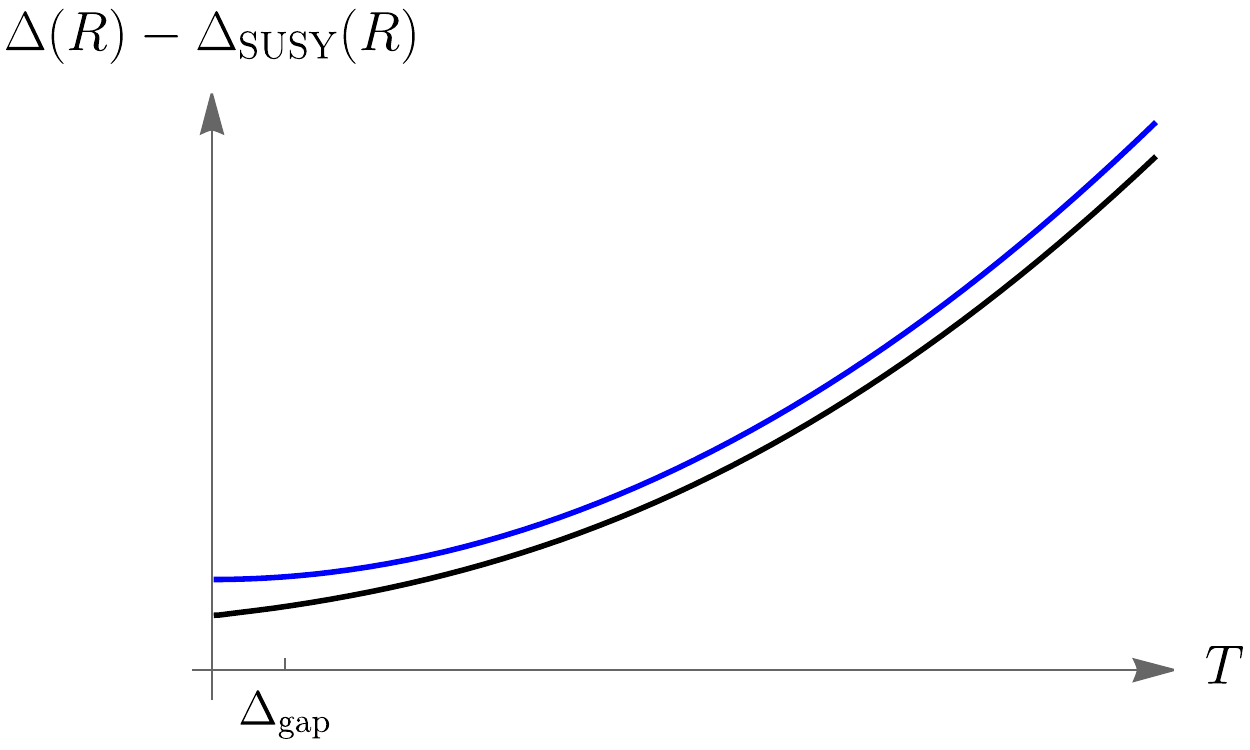}};
    \node[inner sep=0pt] (entropy BPS) at (8.3,3.27)
    {\includegraphics[width=.45\textwidth]{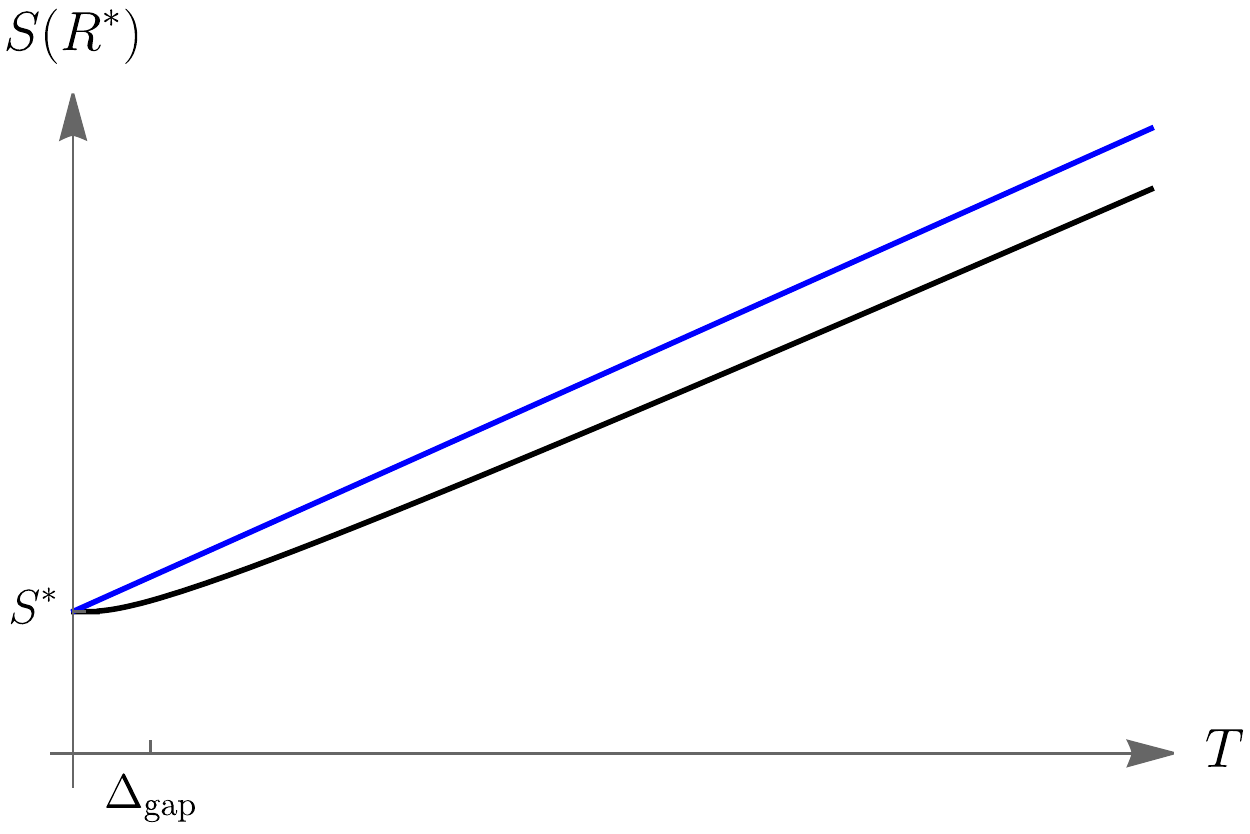}};
    \node[inner sep=0pt] (entropy) at (8.4,-2.17)
    {\includegraphics[width=.45\textwidth]{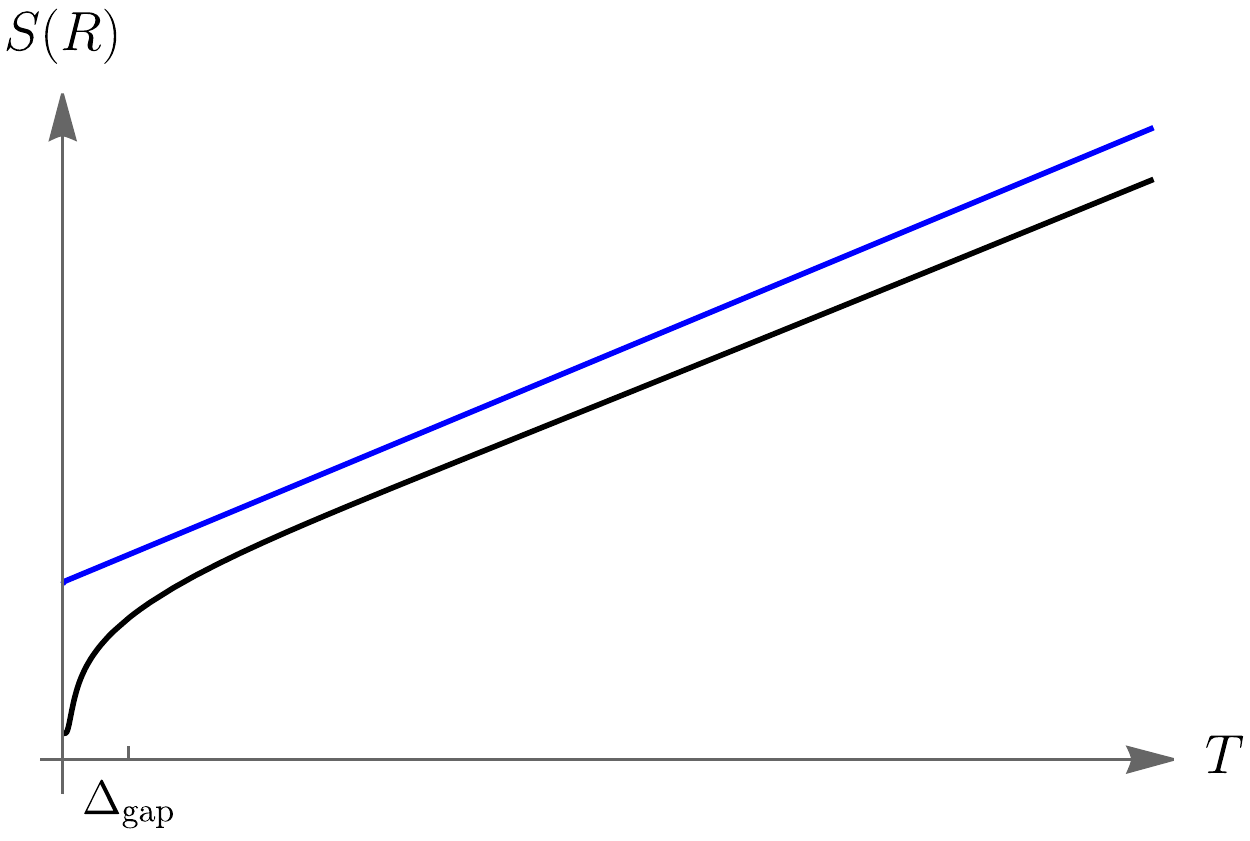}};
\end{tikzpicture}
    \caption{The energy above the supersymmetric bound (left column) and the entropy (right column) as a function of temperature in the canonical ensemble with $R=R^\star$ (top line) and an example of $R \neq R^\star$ (bottom line). The black curve represents the result which includes quantum corrections, while the blue curve is obtained by computing the naive semi-classical answer.}
    \label{fig:energy-and-entropy-in-canonical}
\end{figure}

From this expression we can determine the gap between the $1/16$-BPS black hole and the lowest black hole state with charge $R^\star$, which we denote by $\Delta_{\rm gap}$.  
Using \eqref{eq:Phi-r-formula}, the quantum corrections at low energies in the gravitational path integral thus gives 
\beq
\Delta_{\rm gap}  = \frac{\tilde \Delta(\mathcal{J}_1,\, \mathcal{J}_2)}{N^2},~~~~\tilde \Delta(\mathcal{J}_1,\, \mathcal{J}_2)= \frac{ \left(1-a^*\right) \left(1-b^*\right) \left(3 a^*  b^*
   +3 a^*+ a^{*2} + b^{*2} +3 b^* +1\right)}{4\left(a^*+b^*\right)^2 \left(3-a^*  b^* +a^*+b^*\right)}\,,
\eeq
where $a^\star,b^\star$ are functions of $\mathcal{J}_{1,2} \equiv \mathfrak{j}_{1,2}/N^2$ given above. Moreover, the density of states at charge $R^\star$ above this gap is given by
\beq
\rho(\Delta,\mathfrak{j}_1,\mathfrak{j}_2, R^\star) = e^{S^\star} \delta(\Delta-\Delta_{\rm BPS})+\frac{e^{S^\star}\sinh{\left(\pi \sqrt{\frac{\Delta -\Delta_{\rm BPS}-\Delta_{\rm gap}}{4\Delta_{\rm gap}}}\right)}}{\pi (\Delta -\Delta_{\rm BPS})}\Theta(\Delta-\Delta_{\rm BPS}-\Delta_{\rm gap}) ,
\label{eq:rho1}
\eeq
where $S^\star$, $\Delta_{\rm gap}$ and $\Delta_{\rm BPS}$ are the functions of $\mathfrak{j}_1$ and $\mathfrak{j}_2$ given above. A similar formula can be rewritten for states with $R\neq R^\star$, although the delta function is not present for these states
\bea
\rho(\Delta,\mathfrak{j}_1,\mathfrak{j}_2,R) &=& \frac{e^{S^\star}\sinh{\left(\pi \sqrt{\frac{\Delta-\Delta_{\rm extremal}(R)}{4\Delta_{\rm gap}}}\right)}}{2\pi (\Delta -\Delta_{\rm SUSY}(R))}\Theta(\Delta-\Delta_{\rm extremal}(R))\nonumber\\
&&~~+\frac{e^{S^\star}\sinh{\left(\pi \sqrt{\frac{\Delta-\Delta_{\rm extremal}(R+1)}{4\Delta_{\rm gap}}}\right)}}{2\pi  (\Delta - \Delta_{\rm SUSY}(R+1))}\Theta(\Delta-\Delta_{\rm extremal}(R+1)) ,
\label{eq:rho2}
\ea
where we have suppressed the dependence of all parameters on $\mathfrak{j}_1$ and $\mathfrak{j}_2$. For both cases, we plot the results in figures \ref{fig:regions}(b) and (c). As in section \ref{sec:schwarzian-philosophy}, the continuum part of the density of states \eqref{eq:rho1} and \eqref{eq:rho2} receives perturbative and non-perturbative corrections in $N$. We expect that in the continuum region these corrections could lead to non-perturbative gaps in the spectrum, which we expect to be exponentially small in $N$.

We can also compare the quantum corrected entropies and energies at fixed $R$ with corresponding semiclassical results. We show this in figure \ref{fig:energy-and-entropy-in-canonical}.

As a final comment, it is reasonable to ask whether there are other BPS states with general charge $R$ not necessarily equal to $R^\star$. There is evidence that there are BPS black hole solutions with generic charges \cite{Markeviciute:2018yal,Markeviciute:2018cqs}. These are found by incorporating scalar hair, and can be interpreted as a black hole with a condensate of BPS particles outside the horizon. The entropy of these solutions is expected to be subleading since they do not dominate the superconformal index, but this deserves further study. 

\subsection{Type IIB string corrections to the Schwarzian Result}
\label{sec:stringcorrections}

A basic essential feature of the original AdS/CFT correspondence between Type IIB string theory on AdS$_5 \times S^5$ and $\mathcal{N} = 4$ Super Yang-Mills is that there are two expansion parameters on both sides of the duality. In the bulk, this is because there are two different length scales corresponding to the string scale and the Planck scale:
\begin{eqnarray}
\frac{L_{AdS_5}^4}{\ell_{\textrm{planck}}^4} \sim N \, , \qquad \quad \frac{L_{AdS_5}^4}{\ell_{\textrm{string}}^4} \sim g_{YM}^2 N \equiv \lambda\, .
\end{eqnarray}
With $L_{AdS_5} = 1$, $\ell_{\textrm{string}} = \sqrt{\alpha'}$, we have
\begin{equation}
G_5 = \frac{\pi}{2 N^2}\qquad  \Rightarrow \qquad  G_{10} = 8 \pi^6 g_s^2 (\alpha')^4 \, , \qquad   g_s = \frac{g_{YM}^2}{2 \pi} \, , \qquad  \alpha' = \frac{1}{\sqrt{2 g_{YM}^2 N}} =\frac{1}{\sqrt{2\lambda}}\, .
\end{equation}
In the preceding sections, we have analyzed the euclidean gravity partition function of black holes only in the limit of $N\rightarrow \infty$, $\lambda \rightarrow \infty$. In the bulk, this is the limit of classical supergravity with all higher derivative terms due to string corrections and loop corrections suppressed. To our knowledge, this is the only limit in which the rotating black hole solutions such as those in section \ref{sec:AdS5black hole} are known. The spectrum of these black holes in this limit is dual to the spectrum of $\mathcal{N}=4$ SYM at large $N$ and strong coupling. However, to first order in the `t Hooft expansion, we can actually evaluate the first correction to the Schwarzian result using only the leading gravity solution and knowledge of the corrected 10-dimensional action.

In the 10D uplift of the black hole solution \eqref{eq:near-horizon-metric-10d-lift}, the only IIB fields turned on are the metric and the five form; therefore we may use the $\alpha'$ corrected action: 
\begin{equation}
    S_{IIB} = \frac{1}{16 \pi G_{10}}\int d^{10}x \sqrt{G} \left (\mathcal{R}_{(10)} -\frac{1}{4\cdot 5!}(F_5)^2 + \gamma  \mathcal{W}\right ) \, ,
\end{equation}
where in a particular scheme the 8-derivative term can be written in terms of the Weyl tensor $C$ and:
\begin{eqnarray}
\gamma = \frac{\pi^3}{128} \zeta(3) (\alpha')^3 \, , \qquad \, \, \mathcal{W} \sim C^4 + \textrm{(supersymmetric completion)} \, .
\end{eqnarray}
In \cite{Gubser:1998nz}, the action of this form was used to compute corrections to the free energy of the near-extremal D3-brane, but the full supersymmetric completion of $\mathcal{W}$ was not known. Following earlier work of \cite{Green:2003an,Green:1998by}, \cite{Paulos:2008tn} found $\mathcal{W}$ to be a sum of 20 monomials built from the Weyl tensor $C_{MNPQ}$ and a certain polynomial of derivatives of $F_5$ denoted $\mathcal{T}_{MNPQRS}$. In the application to AdS$_5$ black holes, \cite{Melo:2020amq} carefully computed the effective action and its valuation on the AdS$_5$ black hole. In perturbation theory, the shift in the on-shell action due to the new terms comes entirely from evaluating the leading solution on the perturbation \cite{Reall:2019sah}, and \cite{Melo:2020amq} found the final result in the case of equal rotation to be relatively simple\footnote{The final expression given in equation 16 of \cite{Melo:2020amq} v2 is off by a factor of $N^{-4}$. We thank J. Melo and J. Santos for confirming it.}:
\begin{equation}
\frac{1}{16 \pi G_{10}} \gamma \int   \mathcal{W}\Big|_{BH} =- \frac{\alpha'{}^3}{G_5} \frac{3\pi^4 \zeta(3) \beta(m+q)^2 (m-q-2aq)(m-q+2aq)}{2(1-a^2)^{\frac{1}{2}} (r_+^2 + a^2)^{15/2}} .
\end{equation}

Evaluating this correction in the near-BPS limit in terms of $\mathcal{N}=4$ Yang Mills data we find
\bea
\frac{1}{16 \pi G_{10}} \gamma \int   \mathcal{W}\Big|_{BH} &=& \frac{2\pi^2 \delta  \MSU}{\beta \MSU^2} (1-4\alpha^2) + \ldots ,\\
\frac{\delta \MSU}{\MSU}&=&-\frac{1}{\lambda^{3/2}} \frac{3\pi^3\zeta(3)(1+a^*)^3(1-a^*)^{3/2}}{2a^*{}^{7/2}(3-a^*)} ,
\ea
where $\MSU \sim N^{-2}$ is the answer found above in \eqref{eq:Phi-r-formula} evaluated at $b^*=a^*$.  From this expression it is easy to find the correction to the gap $\delta \Delta_{\rm gap} = \Delta_{\rm gap} \delta \MSU/ \MSU \sim N^{-2} \lambda^{-3/2}$. To write a more explicit expression, we can take the limit $1 - a^\star $ small, such that $\mathcal{J}\equiv \mathfrak{j}_1/N^2 = \mathfrak{j}_2/N^2 \gg 1$. In this limit 
\beq
\Delta_{\rm gap}(\lambda) = \Delta_{\rm gap}(\lambda=\infty)\left(1-\frac{1}{\lambda^{3/2}} \frac{12 \pi^3 \zeta(3)}{\sqrt{\mathcal{J}}}+ \ldots \right) .
\eeq
In particular we find that the gap above the supersymmetric bound in figure \ref{fig:regions}(a) becomes smaller due to $\alpha'$ corrections and the extremal states deviate even further from their naive classical value. As mentioned in the previous section, stringy corrections to the extremality bound, in particular their sign (which in this case proved to be negative) are typically important in the context of the weak gravity conjecture program (once again, see \cite{Harlow:2022gzl} for a review).

To get a sense about the importance of stringy corrections it is useful to compare their magnitude to that of the quantum corrections discussed in section \ref{sec:SpectrumN4}, when it comes to the energy of the extremal black hole states.  Writing 
\be 
\Delta_\text{extremal} = \Delta_{\rm BPS} + \frac{1}2 (R-R^\star)+4\bigg(\Delta_\text{gap} + \underbrace{\delta \Delta_\text{gap}}_{\substack{\text{Stringy}\\ \text{corrections}}}\bigg) \bigg(R-R^\star  \underbrace{ -\,\, \frac{1}2\,}_{\substack{\text{Schwarzian}\\ \text{quantum} \\ \text{corrections}}}\bigg)^2 ,
\ee
for $R>R^\star$.\footnote{There is a similar expression for $R<R^\star$ which will lead to similar conclusions.  }
In such a case, both the quantum corrections and the stringy corrections push the extremal black hole states below the classical extremality bound. Nevertheless, when it comes to the energy at extremality, the stringy corrections can become more important then the Schwarzian quantum corrections, when 
\be 
\frac{|\delta \Delta_\text{gap}|}{\Delta_\text{gap}} > \frac{1}{R-R^\star}\,.
\ee
This can occur when we can scale $R^\star \to \infty$ while making $R-R^\star \gg 1$ with $(R-R^\star)/R^\star \to 0$. Thus, the contribution of the Schwarzian quantum corrections and the stringy corrections to the extremal energy  could exchange dominance when $R-R^\star \sim \lambda^{3/2}$.

\section{Further examples: black holes in $\cN=(2,2)$ supergravity in AdS$_3$}

\label{sec:(2,2)-sugra-examples}

In this section we will consider a simpler theory with similar properties as those of the one studied above. We consider $(2,2)$ supergravity in an asymptotically AdS$_3$ spacetime. Regardless of the matter content, it has rotating and charged black holes solutions which at low temperatures, in a certain charge sector, also present an emergent $SU(1,1|1)$ symmetry in the IR. We will show this is true using the Virasoro symmetry following the approach of \cite{Ghosh:2019rcj,Heydeman:2020hhw}. 

We will focus on the contribution from supergravity only. The reason is that the pure gravity answer is universal and an approximation of the full spectrum for holographic theories, either for large central charge \cite{Hartman:2014oaa} or for large angular momentum sector \cite{Benjamin:2019stq}. We will focus mostly on the role that integer spectral flow has on the spectrum, using the technical results derived in \cite{Eguchi:2003ik}.

\subsection{States and representations of 2D $\mathcal{N}=(2,2)$ CFT}

We consider a theory with a gravity sector in asymptotically AdS$_3$ given by Chern-Simons theory with group $SU(1,1|1)_L \times SU(1,1|1)_R$, and an asymptotic symmetry algebra given by $\mathcal{N}=(2,2)$ Virasoro symmetry. The left- and right-mover generators contain a bosonic Virasoro algebra generated by stress tensor $T_L$ and $T_R$, the complex supercurrent $G^{\pm}_L$ and $G^\pm_R$ and a $U(1)$ current $J_L$ and $J_R$. We denote the charges associated to the $U(1)$ currents by $Q_L$ and $Q_R$ respectively. The Virasoro central charge $c_L=c_R=c$ is given by $\hat{c}\equiv c/3 = 1+2 b^{-2}$, which defines the parameters $\hat{c}$ and $b$. 

We will consider this theory when the boundary is a complexified torus with moduli $\tau$ and $\bar{\tau}$ and define $q=e^{2\pi i \tau}$ and $\bar{q} = e^{-2\pi i \bar{\tau}}$. We also introduce a chemical potential $z$ and $\bar{z}$ conjugated to the $U(1)$ charges $Q_L$ and $Q_R$. A state with charge $(Q_L,Q_R)$ contributes then with a weight $y^{Q_L} \bar{y}^{Q_R}$, where we define $y=e^{2\pi i z}$. 

Before analyzing the black hole spectrum in this theory, we will first quickly review the different types of representations of the $(2,2)$ Virasoro algebra that can appear. We will describe the left-moving case in the NS sector for concreteness. A general representation is labeled by a scaling dimension $\Delta$ and charge $Q$. Unitarity implies $\Delta>|Q|/2$ and $0\leq |Q|<\hat{c}-1$. The character of this representation, after summing over descendants, is given by
\beq
{\rm ch}_{NS}(\Delta,Q;\tau,z) = q^{\Delta-\frac{\hat{c}-1}{8}} y^Q \frac{\theta_3(\tau,z)}{\eta(\tau)^3}.
\eeq
When the bound on dimension is saturated $\Delta = |Q|/2$ and $0<|Q|<\hat{c}$, the representations are BPS and preserve one supercharge. The character of these representations is given by 
\beq
{\rm ch}_{NS}(Q;\tau,z) = q^{\frac{|Q|}{2}-\frac{\hat{c}-1}{8}} y^Q \frac{1}{1+y^{{\rm sgn}(Q)} q^{1/2} }\frac{\theta_3(\tau,z)}{\eta(\tau)^3}
.
\eeq
The denominator comes from the preserved supercharge. Finally, we have the vacuum representation with $\Delta=Q=0$, preserving two supercharges. The character is given by
\beq
{\rm ch}_{NS}(\tau,z) \equiv q^{-\frac{\hat{c}-1}{8}} \frac{1-q}{(1+y q^{1/2})(1+y^{-1}q^{1/2})} \frac{\theta_3 ( q, y)}{\eta(q)^3}.
\eeq
The denominator comes form the preserved supersymetry, while the numerator comes from the preserved ${\rm SL}(2,\mathbb{R})$ symmetry.

This can be extended to other sectors by simple transformations. First of all, we can insert a $(-1)^F$ by simply shifting $y\to -y$. We can also go to the Ramond sector by shifting $z\to z+\tau/2$. In the next section, we will be interested in computing the partition function with Ramond boundary conditions such that fermions are periodic along the spatial circle. Since we will study black holes, this circle is not contractible.

\subsection{Naive Spectrum}

We now move to computing the contribution to the black hole partition function from the supergravity sector. We will study it in the RR sector, as explained above. This can be related to the partition function in vacuum AdS$_3$ by a modular transformation $\tau \to -1/\tau$ and $z\to z/\tau$. After this transformation, the spatial direction becomes time-like and therefore it involves a $(-1)^F$ insertion. On the other hand, time becomes space and the boundary conditions are anti-periodic. Therefore the black hole partition function is given by the vacuum characters in the NS sector, with a $(-1)^F$ insertion $Z_{\rm BTZ, RR} = |e^{-i\pi \frac{\hat{c}z^2}{\tau}} {\rm ch}_{NS}(-1/\tau,z/\tau+1/2)|^2$. The origin of the prefactor is explained in \cite{Kraus:2006wn}. After replacing the explicit form of this character we get 
\beq
Z_{\rm BTZ, RR}= e^{\frac{\pi i}{\tau} \frac{\hat{c}}{4}(1-4 z^2)-\frac{\pi i}{\bar{\tau}} \frac{\hat{c}}{4}(1-4 \bar{z}^2)} ~\Big| \frac{1-q'}{\eta(q')} \frac{1}{\eta(q')} \frac{\theta_3(q',-y')}{\prod_\pm (1-y'{}^{\pm1} q'{}^{1/2})} \Big|^2,
\eeq
where $q'=e^{2\pi i (-1/\tau)}$ and $y'=e^{2\pi i(z/\tau)}$. The first term is the exponential of the classical action on the black hole background. The second term is the one-loop determinant around this solution, where we wrote first the graviton contribution, then the $U(1)_R$ Chern-Simons one, and finally that from the gravitini. This is exact for pure gravity, but also reproduces a universal feature of the spectrum in the large central charge for holographic theories \cite{Hartman:2014oaa} or large angular momentum for theories with a twist gap \cite{Benjamin:2019stq}.

From this expression for the partition function, we can extract the black hole spectrum expanding it in a sum over the characters introduced above. Explicitly, we define the density of states from
\bea\label{eq:2DCFTspectrum}
Z_{{\rm BTZ,RR}} &=& \sum_{E_L,E_R} \sum_{Q_L,Q_R}~ \rho_{Q_L,Q_R}(E_L,E_R)~{\rm ch}_{R}(E_L,Q_L;\tau,z){\rm ch}_{R}(E_R,Q_R;\bar{\tau},\bar{z}),\nonumber\\
&+&\sum_{E_R} \sum_{Q_L,Q_R}~ \rho_{Q_L,Q_R}(E_R)~{\rm ch}_{R}(Q_L;\tau,z){\rm ch}_{R}(E_R,Q_R;\bar{\tau},\bar{z}) + ({\rm L}\leftrightarrow{\rm R}),\nonumber\\
&+&\sum_{Q_L,Q_R}~ \rho_{Q_L,Q_R}~{\rm ch}_{R}(Q_L;\tau,z){\rm ch}_{R}(Q_R;\bar{\tau},\bar{z}).
\ea
We defined the energy of the states in terms of dimensions as $E\equiv \Delta-\hat{c}/8$. There are three types of terms from each line, the first involves left- and right-moving generic representations, the second either left or right BPS representations, and the final one involves both left and right BPS representations. To ease notation, the density of states in each sector is differentiated only by the arguments in $\rho$.

The density of states as a function of charge and energy can be extracted from the following modular transformation for the vacuum character
\bea\label{eq:modtransfcontcharge}
e^{-i\pi \frac{\hat{c}z^2}{\tau}} {\rm ch}_{NS}(-1/\tau,z/\tau+1/2)&=& \int_0^\infty dP \int_{-\infty}^\infty dQ S(P,Q) {\rm ch}_R\Big( \frac{\hat{c}}{8}+\frac{b^2 Q^2}{4}+P^2,Q;\tau,z\Big) \nonumber\\
&&+ 2  \int_{0}^{1} dQ \sin \left[ \pi Q\right] \sum_{n\in \mathbb{Z} } q^{\frac{k}{2}n^2} y^{kn} {\rm ch}_R(Q;\tau,z+n \tau) \,,
\ea
where we defined following \cite{Eguchi:1987sm} the modular S-matrix
\beq
S(P,Q) \equiv \frac{\sinh(2\pi b P) \sinh \left( \frac{2\pi P}{b}\right)}{b^{-1} \sinh\left( \pi b P + \frac{i \pi b^2 \omega}{2} \right) \sinh \left( \pi b P - i \frac{\pi b^2 Q}{2}\right)} .
\eeq

Taking the square of this expression we can compute $\rho_{Q_L,Q_R}(E_L,E_R)$ comparing with \eqref{eq:2DCFTspectrum}. The answer for the fully degenerate states is given by a product of a left- and right-moving modular S-matrices, together with the Jacobian appearing when replacing an integral over $P$ to an integral over energies. The full answer is  
\beq\label{eq:2DCFTcontspec}
\rho_{Q_L,Q_R}(E_L,E_R) = \frac{\sinh\Big(2\pi  \sqrt{\frac{2}{\hat{c}-1}(E_L-\frac{Q_L^2}{2(\hat{c}-1)})}\Big) \sinh \Big( 2\pi \sqrt{\frac{\hat{c}-1}{2}(E_L-\frac{Q_L^2}{2(\hat{c}-1)}})\Big)}{2 \sqrt{\frac{\hat{c}-1}{2}(E_L-\frac{Q_L^2}{2(\hat{c}-1)}})\Big| \sinh\Big( \pi \sqrt{\frac{2}{\hat{c}-1}(E_L-\frac{Q_L^2}{2(\hat{c}-1)})} + \frac{i \pi Q_L}{\hat{c}-1} \Big) \Big|^2}\times({\rm L}\rightarrow{\rm R}),
\eeq
whenever $E_{L/R} > Q_{L/R}^2/(2(\hat{c}-1))$ and zero otherwise. The energy spectrum is continuous but which is fine since we believe that non-perturbative corrections could fix that. 

However, the first issue is that the spectrum involves an integral over $Q_{L/R}\in(-\infty,+\infty)$. This corresponds to an R-symmetry group being $\mathbb{R}$ instead of $U(1)$. This will be resolved in the next section by including additional saddles we so far ignored. This is the problem that \cite{Eguchi:2003ik} address. A second issue is the contribution from the BPS states. We see that it involves a sum over integer spectral flowed characters and its not clear how to interpret that sector of the spectrum. The sum over saddles in the next section will correct both problems. 

Before explaining the resolution of these issues we will take the near extremal limit of the continuum part and show it matches the $\mathcal{N}=2$ Schwarzian answer. For theories with large $\hat{c}$ and at large $E_R$ and fixed $E_L \sim \mathcal{O}(1/\hat{c})$ the density of states is 
\beq
\rho_{Q_L,Q_R}(E_L,E_R) \approx e^{S_0}~ \frac{ \sinh \Big( 2\pi \sqrt{2\Phi_r(E_L-\frac{Q_L^2}{8\Phi_r}})\Big)}{ 2\pi E_L} ,
\eeq
where defined $S_0 = 2\pi \sqrt{ \frac{\hat{c}-1}{2}(E_R-\frac{Q_R^2}{2(\hat{c}-1)})}$ and $\MSU =4/ (\hat{c}-1)$ to make a direct comparison with the $\mathcal{N}=2$ JT gravity answer. Notice that given the standard 2D CFT conventions, the $\mathcal{N}=2$ supermultiplet has charge assignment $(Q+1/2)\oplus(Q-1/2)$, and therefore there is a shift of the charge $Z_{\rm Sch} =Q_{CFT}+1/2$ needed in order to match with the Schwarzian answer.

\subsection{Corrected Spectrum}

As emphasized in \cite{Eguchi:2003ik} this cannot be the whole story since the spectrum derived from this partition function has a continuum value of charges. From the bulk perspective, we need to sum over saddles that recover the discreteness of charge. From the boundary perspective, we need to include the integer spectral flow generator as part of the algebra. 

We will begin by describing the boundary perspective. The spectral flow generators $U_\eta$ are defined more generally, in terms of a real parameter $\eta$, by the following transformation of the currents
\bea
U_\eta^{-1} L_n U_\eta &=& L_n + \eta J_n + \frac{\hat{c}}{2} \eta^2 \delta_{0,n},\\
U_\eta^{-1} J_n U_\eta &=& J_n + \hat{c} \eta \delta_{0,n},\\
U_\eta^{-1} G_s^\pm U_\eta &=& G_{s\pm \eta}^{\pm}. 
\eea
The extended algebra we will consider includes, besides the stress tensor, $U(1)$ current and supergenerators, the integer spectral flow generator $U_{\pm 1/r}$ for integer $1/r \in \mathbb{Z}$. The reason to parametrize this integer by $r$ is to make a connection later with the spectrum described in section \ref{sec:SchwarzianSolution}. It was shown in \cite{Eguchi:2003ik} that the modular properties of representations of this extended algebra are only consistent when the central charge has the form 
\beq
\hat{c} = 1 + 2 r k,~~~k\in\mathbb{Z},
\eeq
where $k$ is a positive integer that can be chosen independently of the integer $1/r$. We will see below that when this generator is included in the algebra the spectrum of charges is fractional $Q\in r \cdot \mathbb{Z}$. Before showing that we will present the new set of extended characters, written in the NS sector for concreteness, given by
\bea
\chi_{NS}(\Delta, Q, s; \tau,z) &=& \sum_{n\in s+r^{-1}\cdot \mathbb{Z}} q^{\frac{\hat{c}n^2}{2}} y^{\hat{c}n}{\rm ch}_{NS}(\Delta,Q;\tau,z+n \tau) , \\
\chi_{NS}(Q,s;\tau,z) &=&\sum_{n\in s+r^{-1}\cdot \mathbb{Z}} q^{\frac{\hat{c}n^2}{2}} y^{\hat{c}n}{\rm ch}_{NS}(Q;\tau,z+n \tau)  ,\\
\chi_{NS}(s;\tau,z) &=& \sum_{n\in s+r^{-1}\cdot \mathbb{Z}} q^{\frac{\hat{c}n^2}{2}} y^{\hat{c}n}{\rm ch}_{NS}(\tau,z+n \tau) ,
\ea
for generic BPS and vacuum representations respectively. The states are further parametrized by an integer $s\in \mathbb{Z}_{1/r}$. Similar formulas can be written for the R sector, and including a $(-1)^F$, after performing half-integer spectral flow and shifting $y\to - y$ respectevely.

So far we discussed the boundary theory. In the bulk theory, the choice of $r$ corresponds to the choice of the size of $U(1)_R$. Picking the correct gauge group is not trivial since it determines the allowed gauge transformations and therefore which set of saddles are to be considered gauge equivalent or not. When we fix $r$, new set of saddles appear under the shift of the chemical potential $z\to z+n$, for $n\in r^{-1} \cdot \mathbb{Z}$. These saddles have the same boundary conditions as $n=0$ up to a global gauge transformation. The sum over these saddles is equivalent to implementing the sum over integer spectral flow. For BTZ in the RR sector we have $Z_{\rm BTZ, RR} = |e^{-i\pi \frac{\hat{c}z^2}{\tau}} \chi_{NS}(0;-1/\tau,z/\tau+1/2)|^2$. This can again be written explicitly in a way that makes it bulk interpretations clear. The partition function is give by
\beq
Z_{\rm BTZ, RR} =\sum_{n,\bar{n}\in r^{-1} \cdot \mathbb{Z}} e^{\frac{\pi i}{\tau} \frac{\hat{c}}{4}(1-4 (z+n)^2)-\frac{\pi i}{\bar{\tau}} \frac{\hat{c}}{4}(1-4 (\bar{z}+\bar{n})^2)} ~\Big| \frac{1-q'}{\eta(q')} \frac{1}{\eta(q')} \frac{\theta_3(q',-y'q'{}^{n})}{\prod_\pm (1-y'{}^{\pm1} q'{}^{\pm n} q'{}^{1/2})} \Big|^2.
\eeq
Again the first term is the exponential of the classical action, while the second term is the one-loop determinant in the new saddles.

As mentioned above, it was shown in \cite{Eguchi:2003ik} that the extended characters have consistent modular transformations when the central charge is fractional. For the vacuum representation case that we need their result is
\bea\label{eq:modtransfextchar}
e^{- i \pi \frac{\hat{c} z^2}{\tau}} \chi_{NS}(0;-1/\tau,z/\tau+1/2) &=& \int_0^\infty dP \sum_{Q\in r\cdot \mathbb{Z}_{\hat{c}-1}}  r~S(P,Q)~\chi^R\Big( \frac{\hat{c}}{8}+\frac{b^2 Q^2}{4}+P^2,Q,0;\tau,z\Big) \nonumber\\
&&+\sum_{Q\in r\cdot \mathbb{Z}_{\hat{c}-1}, 0<Q<1} 2r\sin \left[ \pi Q \right] \sum_{r\in \mathbb{Z}_N }\chi^R(Q,r;\tau,z),
\ea
where $S(P,Q)$ is the same function defined above. Now we see that both problems are resolved. The charge spectrum is discrete with a fractional unit charge $r$ and the modular transformation involves the same representations that are allowed by the extended algebra. The overall factor of $r$ in the right-hand side is required to reproduce \eqref{eq:modtransfcontcharge} in the $r\to0$ limit.

Having the modular properties of the vacuum character of the extended algebra, we can extract the improved spectrum for a finite fractional charge $r$. First of all it is easy to see that other than the fact that charge is discrete the continuum density of states $\rho_{Q_L,Q_R}(E_L,E_R)$ is exactly the same as in the previous case \eqref{eq:2DCFTcontspec}, up to an extra factor of $r$ from \eqref{eq:modtransfextchar} and the final answer is
\beq\label{eq:2DCFTcontspec222}
\rho_{Q_L,Q_R}(E_L,E_R) =r \frac{\sinh\Big(2\pi  \sqrt{\frac{2}{\hat{c}-1}(E_L-\frac{Q_L^2}{2(\hat{c}-1)})}\Big) \sinh \Big( 2\pi \sqrt{\frac{\hat{c}-1}{2}(E_L-\frac{Q_L^2}{2(\hat{c}-1)}})\Big)}{2 \sqrt{\frac{\hat{c}-1}{2}(E_L-\frac{Q_L^2}{2(\hat{c}-1)}})\Big| \sinh\Big( \pi \sqrt{\frac{2}{\hat{c}-1}(E_L-\frac{Q_L^2}{2(\hat{c}-1)})} + \frac{i \pi Q_L}{\hat{c}-1} \Big) \Big|^2}\times({\rm L}\rightarrow{\rm R}),
\eeq
whenever $E_{L/R} > Q_{L/R}^2/(2(\hat{c}-1))$ and $Q_{L/R} \in r \cdot \mathbb{Z}_{\hat{c}-1}$ and zero otherwise. In the near extremal limit, at large $\hat{c}$,  this is exactly the same as the $\mathcal{N}=2$ Schwarzian answer with coupling $\MSU =2/(rk)$, for large $k$. Moreover, the answer is still universal either for holographic theories at large $\hat{c}$ \cite{Hartman:2014oaa} or any theory with a twist gap at large angular momentum \cite{Benjamin:2019stq}. We can also extract the density of states of BPS states, when either $E_R=0$ or $E_L=0$, from the black hole spectrum. Then answer when $E_R=0$ is
\beq
\rho_{Q_L,Q_R}(E_R)=    2 r \sin(\pi Q_L)~2r\frac{\sinh\Big(2\pi  \sqrt{\frac{2}{\hat{c}-1}(E_R-\frac{Q_R^2}{2(\hat{c}-1)})}\Big) \sinh \Big( 2\pi \sqrt{\frac{\hat{c}-1}{2}(E_R-\frac{Q_R^2}{2(\hat{c}-1)}})\Big)}{2 \sqrt{\frac{\hat{c}-1}{2}(E_R-\frac{Q_R^2}{2(\hat{c}-1)}})\Big| \sinh\Big( \pi \sqrt{\frac{2}{\hat{c}-1}(E_R-\frac{Q_R^2}{2(\hat{c}-1)})} + \frac{i \pi Q_R}{\hat{c}-1} \Big) \Big|^2} 
\eeq
The answer when $E_L=0$ is completely analogous. In the near extremal limit with large $E_R$, we reproduce the Schwarzian theory answer $\rho_{Q_L,Q_R}(E_R) \approx e^{S_0}2r \sin(\pi Q_L)$ where $S_0$ is the same function of $E_R$ and $Q_R$ as defined in the previous section. Finally we can look at full BPS states with $E_L=E_R=0$. The answer for pure $\mathcal{N}=(2,2)$ gravity is 
\beq
\rho_{Q_L,Q_R}=   2r \sin(\pi Q_L)~  2r \sin(\pi Q_R) .
\eeq
This example is beyond the near extremal regime and therefore is not universal for $\mathcal{N}=(2,2)$ CFTs either from the regime considered by \cite{Hartman:2014oaa} nor \cite{Benjamin:2019stq}. 

The elliptic genus of these theories vanishes. This can be resolved by defining a refined elliptic genus analogous to \cite{Fu:2016vas}, assuming an exact $\mathbb{Z}_{1/r}$ symmetry. The situation simplifies when $r=1$, since in that case BPS states come only with zero charge and the usual elliptic genus does not vanish anymore. 

Finally, even though this discussion has been done for the case of $\mathcal{N}=(2,2)$ supersymmetry, this can be generalized to cases with $\mathcal{N}=(0,2)$ using the results of \cite{Ghosh:2019rcj} or $\mathcal{N}=(4,2)$ using \cite{Heydeman:2020hhw}.

\section{Discussion and future work}
\label{sec:discussion}

In this paper, we have argued that the spectrum of $1/16$-BPS and near-$1/16$-BPS black holes can be obtained by studying the appropriate version of the $\cN=2$ super-Schwarzian theory. From this, we have shown that the gravitational path integral correctly reproduces the degeneracies and charges of BPS states in $\cN=4$ super Yang-Mills, while at the same time predicting various gaps between these BPS states and a ``dense set'' of near-BPS black hole states. Above this gap we find a continuum of states (to leading order in $S_0$) and predict what the density of states in this region is. There are however numerous open questions, some of which we hope to address in future work.  

\subsection*{Extension to theories with less supersymmetry}

We expect similar conclusions to hold about the spectrum of BPS and near-BPS black holes in bulk theories with less supersymmetry. For example, one can consider type IIB string theory in a spacetime that is asymptotically AdS$_5 \times M_5$, where $M_5$ is a Sasaki–Einstein space. For each such $M_5$, the bulk has a dual description in terms of a  four-dimensional $\mathcal N = 1$ superconformal
field theory. The BPS and near-BPS black hole solutions reviewed in section \ref{sec:AdS5black hole} and \ref{sec:BPSvsSUSYvsExt} are also valid in a space that is asymptotically AdS$_5 \times M_5$ \cite{Buchel:2006gb,Gauntlett:2007ma}; however, in contrast to the case discussed in \ref{sec:AdS5black hole} the bulk does not have an $SO(6)$ R-symmetry gauge field with the three Cartans whose eigenvalues we labeled by $R_{1,2,3}$, but instead generically has a single $U(1)$ R-symmetry gauge field. Since in section \ref{subsec-ensembles} -- \ref{sec:low-T-expansion} we have solely focused on the case when $R_{1,2,3} = R$, the analysis of the isometry of the near-horizon region as well as that of the low-temperature expansion of the on-shell action should similarly hold for theories in AdS$_5 \times M_5$. Consequently, we expect that the spectrum of near-BPS black holes is also controlled by  the $\cN=2$ super-Schwarzian theory. Nevertheless, a detailed check, regarding the correct value of $r$ and $\theta$-angle, needs to be performed in order to be certain that the spectrum is described by precisely the same super-Schwarzian theory for all Sasaki–Einstein spaces.

\subsection*{Non-perturbative corrections}

In this paper we have focused on the leading quantum corrections around the leading near-BPS black hole geometry in $AdS_5 \times S^5$ with a nearly AdS$_2$ throat. When computing quantities not protected by supersymmetry, such as the partition function with all chemical potentials turned on, there can be a a variety of non-perturbative geometries contributing, including for example spacetime wormholes. 

The situation from the perspective of the gravitational path integral is different when computing protected quantities such as the index \cite{Iliesiu:2021are} or even some quantities that are not protected by supersymmetry, such as the zero temperature partition function which yields the degeneracy of $1/16$-BPS states. In that case, we expect a reduced number of geometries to contribute when they preserve supersymmetry. These have been considered in \cite{Iliesiu:2021are} in the case of throats with emergent $PSU(1,1|2)$ symmetry (in contrast to the $SU(1,1|1)$ near-horizon geometry considered in this paper), where it was shown quantum corrections are described by a deformation of the $\mathcal{N}=4$ Super-Schwarzian theory. This result from the $\mathcal{N}=4$ Super-Schwarzian theory is consistent with examples of black holes in type IIA where the exact index was shown to be reproduced by a sum over these orbifolds \cite{Dabholkar:2014ema}.

Nevertheless, similar results should hold for the black holes discussed in this paper due to similarities between the effective theory of $\cN=2$ JT gravity found in this paper and $\cN=4$ JT gravity analyzed in \cite{Iliesiu:2021are}. Both theories have vacuum BPS states, all with the same spin-statistics, that are separated by a gap for the lightest near-BPS states. In both cases, even though the boundary conditions are supersymmetric for $\alpha=1/2$, the only geometry which preserves supersymmetry in the bulk is the one with $n=0$ (for the case discussed in this paper, see \eqref{N2sumsaddlesAdS55}). A similar calculation as that in \cite{Iliesiu:2021are} shows that, from the perspective of the near-horizon geometry, the only other geometries preserving supersymmetry in $\cN=2$ JT gravity are particular orbifolds of AdS$_2$ (these were considered in bosonic gravity previously in \cite{Mertens:2019tcm, Maxfield:2020ale,Witten:2020wvy}). In particular, the contribution of higher genus geometries or, more broadly, any near-horizon geometry that involves spacetime wormhole can be seen to vanish in $\cN=2$ JT gravity. Consequently, in the full gravitational path integral only orbifolds of $AdS_2$ that can be uplifted to smooth $AdS_2 \times S^3 \times S^5$ geometries contribute.

Have these geometries already been observed from the boundary side? When expanding the $\mathcal{N}=4$ Yang Mills superconformal index in large $N$, while the leading contribution is given by the black hole considered in section \ref{sec:AdS5black hole}, there are subleading contributions which corresponds to supersymmetric orbifolds of the black hole geometry \cite{Aharony:2021zkr}. We leave a detailed comparison of these contributions, including the effect of the $\cN=2$ super-JT one-loop determinant, for future work. We should however stress that as opposed to the case in \cite{Dabholkar:2014ema}, which have a $PSU(1,1|2)$ near-horizon isometry, for the superconformal index in  $\mathcal{N}=4$ Yang Mills there are contributions at large $N$ which cannot be interpreted as orbifolds of the black hole \cite{Aharony:2021zkr}, such as contributions from wrapped D-branes or black hole solutions which are yet unknown. It would be interesting to investigate if these non-perturbative corrections can be analyzed also in the context of the near-BPS limit.

\begin{center}
    \begin{tikzpicture}[baseline={([yshift=-.50ex]current bounding box.center)}, scale=1.0]
 \pgftext{\includegraphics[width=0.8\textwidth]{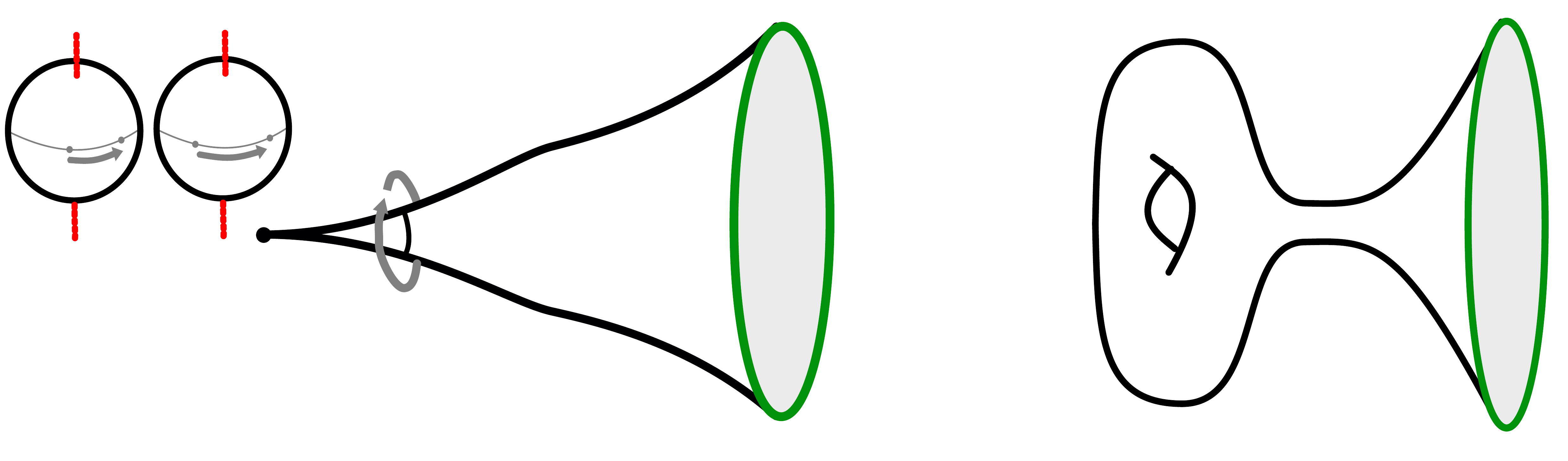}} at (0,0);
  \draw (-6.90,1.6) node  {$S^3$};
    \draw (-4.20,1.6) node  {$S^5$};
     \draw (1.25,0) node  {$\neq\,\, 0$};
        \draw (7.45,0) node  {$=\,\, 0\,.$}; 
  \end{tikzpicture}
  \captionof{figure}{Possible non-perturbative contributions to the index of black holes in $AdS_5 \times S^5$. Using $\cN=2$ super-JT gravity one can show that all geometries involving ``spacetime wormholes'', such as the one on the right side, vanish. The only surviving contributions are orbifolds of the original $AdS_2 \times S^3 \times S^5$ near-horizon geometry which we show on the left side. In such a case the angle of the defect, the rotation on $S^3$ and that on $S^5$ are all related in order for supersymmetry to be preserved and for the spacetime to be smooth. These geometries were also explicitly seen in \cite{Aharony:2021zkr} as subleading corrections to the superconformal index. }
\end{center}

Beyond computing the index, the effect of geometries of higher topology or with a larger number of defects in the near-horizon region (from higher-dimensional generalizations of Seifert geometries \cite{Maxfield:2020ale}) on the $\cN=2$ super-JT path integral, shows that a gap is present in the spectrum associated to each specific geometry.\footnote{The gap was also found to persist when accounting for other topologies in the $\cN=4$ super-JT path integral \cite{Iliesiu:2021are}.} While such geometries are off-shell (for example, the equation of motion for the dilaton cannot be satisfied on such geometries \cite{Saad:2019lba}) they can nevertheless be systematically accounted for using the sum over topologies in the $\cN=2$ super-JT path integral.  Thus, when summing over all geometries we expect that they might yield non-perturbative corrections (in $N^2$) to the value of the gap, but will not affect its existence by inserting additional states between the extremal BPS state and the lightest near-BPS black hole state coming from the original black hole saddle.

One can additionally discuss non-perturbative effects coming from different (possibly supersymmetric) black hole solutions which have not yet been understood analytically. For instance, as previously mentioned, \cite{Markeviciute:2018yal,Markeviciute:2018cqs} found numerical evidence for black hole solutions that support scalar hair and can be supersymmetric even away from $R\neq R^\star$. Such black holes would yield corrections to the spectrum found in figure \ref{fig:regions} that would, once again, be non-perturbatively suppressed in $N^2$. Nevertheless, one might hope to analyze such solutions at low-temperatures in Euclidean signature where we expect there to still be an $AdS_2 \times S^3 \times S^5$ near-horizon region. If at extremality the near-horizon super-isometry is still $SU(1,1|1)$, we expect that the temperature dependence of quantum corrections around these new saddles still be captured by the $\cN=2$ super-Schwarzian. Thus, we expect that even if sectors with $R\neq R^\star$ are populated by new BPS solutions, energy gaps should still be present within each charge sector; moreover, even if the degeneracy of the $1/16$-BPS black states that we have found in the $R=R^\star$ sector is affected by such non-perturbative corrections coming from hairy black holes, the gap should be unaffected within that charged sector.

Thus, this (albeit incomplete) accounting for possible gravitational non-perturbative effects,  all of which preserve the existence of the mass gap, along with the fact that the leading order stringy corrections also did not affect its present, prompts us to conjecture that the gap persists within each large charge sector of $\cN=4$ Yang-Mills.

\subsection*{An effective field theory for near-BPS states from the boundary-side }

We derived the spectrum of nearly $1/16$-BPS black hole states from a bulk $AdS_5\times S^5$ calculation. Due to AdS/CFT, this makes a prediction for the spectrum of $\mathcal{N}=4$ Yang Mills. This raises the question of how to derive the same spectrum from an independent boundary CFT argument. This would be an extremely non-trivial check of holography that would help understand quantum aspects of gravity in these black hole backgrounds better. At weak 't Hooft coupling some of the states with the black hole quantum numbers were constructed in \cite{Berkooz:2006wc, Berkooz:2008gc} in the limit of large spin $J/N^2 \gg 1$, and some interaction effects were considered. We leave for future work to identify the most relevant interactions in this limit, and to derive the emergence of a softly broken $SU(1,1|1)$ symmetry. This would give the first example of a quantum theory describing a local nearly AdS$_2$ background, as opposed to SYK which describes a highly non-local bulk.

\subsection*{Acknowledgements} 
We thank W. Zhao for initial discussion. We also thank O. Aharony, F. Benini, M. Berkooz, P. Caputa, A. Castro, J. Karlsson, M. Kolanowski, Z. Komargodski, S. Komatsu, J. H. Lee, R. Mahajan, J. Maldacena, J. Melo, S. Murthy, B. Post, J. Santos, J.H Schwarz, D. Stanford, L. Tizzano, M. Tomasevic and E. Witten for valuable discussions. MTH is supported by Princeton University and the Institute for Advanced Study under Grant No. DE-SC0009988 and the Corning Glass Foundation. GJT is supported by the Institute for Advanced Study and the National Science Foundation under Grant No. PHY-1911298, and by the Dipal and Rupal Patel funds. LVI was supported by the Simons Collaboration on Ultra-Quantum Matter, a Simons Foundation Grant with No. 651440. JB is supported by the NCN Sonata Bis 9 2019/34/E/ST2/00123 grant. This work was partly done at the  Aspen Center for Physics, which is supported by National Science Foundation grant PHY-1607611. JB wants to thank the organizers and participants of Brussels – Paris – Amsterdam Doctoral School on Quantum Field Theory, Strings and Gravity, for a wonderful workshop, during which a part of this work was done.

\appendix
\section{Details on the low temperature expansion}
\label{app:low-temp-expansion}
In this section we explain in more details how to expand the action 
\be
 I_{\rm ME}(\beta, \mathfrak{j}_{1},\mathfrak{j}_{2}, \alpha) =  I_{\rm GCE}(\mathfrak{q}_1=0,\mathfrak{q}_2=0,  \mathfrak{j}_1 , \mathfrak{j}_2  , \alpha)
+2\pi i \omega_1 \mathfrak{j}_1
+2\pi i \omega_2 \mathfrak{j}_2 ,
\ee
around its BPS limit \cite{Larsen:2019oll}. We start with introducing four expansion parameters $(\epsilon_r , \epsilon_q, \epsilon_a, \epsilon_b)$ such that 
\begin{equation}
r_+^2 = r^{*2} + \epsilon_r  ,~~~q = q^* + \epsilon_q , ~~~a = a^* + \epsilon_a , ~~~b= b^* + \epsilon_b , 
\end{equation}
where parameters $(a^*,b^*)$ are defined through the BPS configuration of fixed charges 
\begin{equation}
\mathfrak{j}_1(a^* , b^*)= \frac{\pi  \left(a^*+b^*\right) \left(a^* +1\right) \left(b^* +1\right)}{4  G_5 \left(a^* -1\right)^2 
\left(1- b^* \right)},~~~\mathfrak{j}_2(a^* , b^*)= \frac{\pi  \left(a^*+b^*\right) \left(a^* +1\right) \left(b^* +1\right)}{4  G_5 \left(1- a^*\right) \left(b^* -1\right)^2} .
\end{equation}
We want to perform the expansion in such a way that the variables $(T,\varphi , \mathfrak{j}_1, \mathfrak{j}_2)$ are fixed. This imposes the following four relations, since the fixed parameters in the PCE ensemble are:
\begin{equation}
\varphi  = \varphi (r_+ , q , a , b) , ~~~T = T (r_+ , q , a , b) , ~~~\mathfrak{j}_1 (a^* , b^*)  
=  
\mathfrak{j}_1 (r_+ , q , a , b),
~~~
\mathfrak{j}_2 (a^* , b^*) 
= 
\mathfrak{j}_2 (r_+ , q , a , b)
.
\end{equation}
Using these relations, we can now express the expansion parameters $(\epsilon_r , \epsilon_q, \epsilon_a, \epsilon_b)$ in terms of $(T,\varphi)$. For our purposes, it will be enough to work to second order in $\epsilon$'s. Inverting the above relations order by order, we find solutions in the form
\begin{align}
\epsilon_X &= \epsilon_{X,\varphi } \varphi +\epsilon_{X, T} T   + 
\epsilon_{X,\varphi^2} \varphi^2 + \epsilon_{X,T^2} T^2 + 
\epsilon_{X,\varphi T}\, (\varphi T)\, 
,
\end{align}
where $X = \, r,\, q, \, a$ or $b$.
The first order coefficients are given through
\begin{align}
\epsilon_{r,\varphi } \varphi +\epsilon_{r, T} T
&= \frac{\left(a^*+b^*\right) \left(a^* +1\right) \left(b^* +1\right) \left(2 \pi  T \sqrt{a^* b^* +a^*+b^*}+ \varphi  \left(a^* b^*
   +a^*+b^*\right)\right)}{2  \left(3 a^*  \left(b^* +1\right)+\left(a^*\right)^2 +\left(b^*\right)^2 +3 b^* +1\right)}
,\\
\epsilon_{q,\varphi } \varphi  +
\epsilon_{q, T}  T  &= \frac{3 \varphi  \left(a^*+b^*\right)^2 \left(a^* +1\right)^2 \left(b^* +1\right)^2}{4 \left(3 a^*  \left(b^* +1\right)+\left(a^*\right)^2 +\left(b^*\right)^2
   +3 b^* +1\right)}
,\\ 
\epsilon_{a,\varphi } \varphi  +
\epsilon_{a, T}  T  &= \frac{\varphi  \left(a^*+b^*\right) \left(a^* -1\right) \left(a^* +1\right) \left(b^* +1\right)}{4 \left(3 a^*  \left(b^* +1\right)+\left(a^*\right)^2
   +\left(b^*\right)^2 +3 b^* +1\right)}
,\\
\epsilon_{b,\varphi } \varphi + 
\epsilon_{b, T} T  &= \frac{\varphi  \left(a^*+b^*\right) \left(a^* +1\right) \left(b^* -1\right) \left(b^* +1\right)}{4 \left(3 a^*  \left(b^* +1\right)+\left(a^*\right)^2
   +\left(b^*\right)^2 +3 b^* +1\right)}
.
\end{align}
Following the same logic we derive second order coefficients. Their form is more complicated for general $(a^*,b^*)$. For reader's convenience we provide simplified formulas for the case $a^* = b^*$ as a potential check
\begin{align}
\epsilon _{r,\varphi ^2} &=\frac{a^{*2} \left(15 a^{*5}+35 a^{*4}+24 a^{*3}-28 a^{*2}+a^*+1\right)}{4 \left(5 a^*+1\right){}^3},
\\
\epsilon _{r,\varphi T}& =\frac{3 \pi  a^{*2} \left(a^*+1\right){}^3 \sqrt{a^* \left(a^*+2\right)}}{\left(5 a^*+1\right){}^3},
\\
\epsilon _{r,T^2}&=\frac{4 \pi ^2 a^{*2} \left(a^{*4}+7
a^{*3}+5 a^{*2}+9 a^*+2\right)}{\left(5 a^*+1\right){}^3}, 
\\
\epsilon _{q,\varphi ^2}&=\frac{a^{*3} \left(30 a^{*5}+84 a^{*4}+105 a^{*3}+7 a^{*2}-11 a^*+1\right)}{2 \left(5
   a^*+1\right){}^3},
\\
\epsilon _{q,\text{$\varphi $T}}&=-\frac{3 \pi  a^{*3} \left(a^*+1\right){}^3 \left(7 a^{*2}-10 a^*-3\right)}{\sqrt{a^* \left(a^*+2\right)} \left(5
   a^*+1\right){}^3},
\\
\epsilon _{q,T^2}&=\frac{2 \pi ^2 a^{*3} \left(6 a^{*4}+51 a^{*3}+13 a^{*2}+a^*+1\right)}{\left(5 a^*+1\right){}^3},
\\
\epsilon _{a,\varphi^2}&=\frac{\left(a^*-1\right){}^2 a^{*3} \left(5 a^{*2}+7 a^*+6\right)}{4 \left(a^*+1\right) \left(5 a^*+1\right){}^3},
\\
\epsilon _{a,\varphi T}&=-\frac{\pi 
   \left(a^*-1\right) a^{*2} \left(a^*+1\right) \left(7 a^{*2}-10 a^*-3\right)}{2 \sqrt{a^* \left(a^*+2\right)} \left(5 a^*+1\right){}^3},
\\
\epsilon _{a,T^2}&=\frac{\pi ^2
   \left(a^*-1\right) a^{*2} \left(2 a^{*3}+15 a^{*2}+6 a^*+1\right)}{\left(a^*+1\right) \left(5 a^*+1\right){}^3},
\end{align}
\begin{align}
\epsilon _{b,\varphi ^2}&=\frac{\left(a^*-1\right){}^2
   a^{*3} \left(5 a^{*2}+7 a^*+6\right)}{4 \left(a^*+1\right) \left(5 a^*+1\right){}^3},
\\
\epsilon _{b,\varphi T}&=-\frac{\pi  \left(a^*-1\right) a^{*2}
\left(a^*+1\right) \left(7 a^{*2}-10 a^*-3\right)}{2 \sqrt{a^* \left(a^*+2\right)} \left(5 a^*+1\right){}^3},
\\
\epsilon _{b,T^2}&=\frac{\pi ^2 \left(a^*-1\right) a^{*2}
\left(2 a^{*3}+15 a^{*2}+6 a^*+1\right)}{\left(a^*+1\right) \left(5 a^*+1\right){}^3}
.
\end{align}
With this we can now expand the quantity of interest to second order in $(\epsilon_r , \epsilon_q , \epsilon_a , \epsilon_b )$ and insert the relations $(\epsilon_r (T,\varphi) , \epsilon_q (T,\varphi), \epsilon_a (T,\varphi), \epsilon_b (T,\varphi) )$ derived above. As a last step, in the resulting expansion we keep only terms up to second order in $(\varphi,T)$ (to get to higher orders we would also need to solve for $\epsilon$'s to higher order). Applying this to procedure to our partial canonical action leads us to the result \eqref{eq:action-low-temp-expansion}.

\section{Details on Killing spinors}
\label{app:details-killing-spinors}
In this section we verify explicitly the Killing spinors \eqref{eq:killing-spinors} and Killing vectors \eqref{eq:killing-vector-J}--\eqref{eq:killing-vector-Em} of the near-horizon geometry lifted to ten-dimensions \eqref{eq:near-horizon-metric-10d-lift}. This has been first derived in \cite{Sinha:2006sh}. Here we review the construction in order to account for differences in conventions and provide reader with necessary formulas. 
\\
\\
Lets recall that we work with a ten dimensional lift \cite{Cvetic:1999xp,Chamblin:1999tk}
\begin{align}
ds^2_{10} 
&=   ds^2_5 +  \sum_{i=1}^3 \left[ d\mu_i^2 + \mu_i^2 \left( d\xi_i^2 - \frac{2}{3} A \right)^2  \right]
, \\
F^{(5)} &=
(1+ \ast_{(10)} )
\left[ 
-4 \, \text{vol}_{(5)} - \frac{1}{3} 
\sum_{i=1}^{3} d(\mu_i^2) \wedge 
d\xi_i \wedge \ast_{(5)} F^{(2)} 
\right]
,
\label{appendix-eq:near-horizon-metric-10d-lift}
\end{align}
where the Hodge star is defined as $\ast_{(n)} \omega_{\mu_1 \dots \mu_{n-p}} = \frac{1}{p!} \epsilon_{\mu_1 \dots \mu_{n-p}}^{\phantom{\mu_1 \dots \mu_{n-p}} \nu_1 \dots \nu_p} \omega_{\nu_1 \dots \nu_p}$, and we work with $\epsilon_{0123456789} = \epsilon_{01234} = 1$. We introduced
\be 
\mu_1 = \sin \tilde{\alpha} , \qquad \mu_2 = \cos \tilde{\alpha} \sin \tilde{\beta} ,
\qquad \mu_3 = \cos \tilde{\alpha} \cos \tilde{\beta} 
,
\ee
and the angles which parametrize the $S^5$ are $\tilde{\alpha} \in [0, \frac{\pi}{2}]$, $\tilde{\beta} \in [0, \frac{\pi}{2}]$, $\xi_i \in [0,2\pi]$. 
The near-horizon 5d metric and the gauge field are given by\footnote{The relation between the near-horizon gauge field used here and the one used in \cite{Sinha:2006sh} is $A_{\text{here}} = - \sqrt{3} A_{\text{there}}$.}
\be 
ds^2_5 &= \left (\frac{\omega}{2\lambda} \right)^2 \left (-r^2 dt^2 
+ \frac{dr^2}{r^2} \right ) + 
3 \left(\frac{\omega ^2}{4}  \sigma _3^L+\frac{\omega}{4 \lambda } r dt\right)^2
+\omega ^2 d\Omega_3^2 , 
\\
A &= -\frac{3}{2} \left(\frac{\omega ^2}{4}  \sigma _3^L+\frac{\omega}{4 \lambda} r dt\right) 
.
\ee
In the above $d\Omega_3^2 =  d\theta ^2 +  \cos ^2 \theta d\tilde{\psi} ^2+\sin ^2 \theta  d\tilde{\phi}^2$ is the metric on $S^3$, $\sigma_3^L = 2 (\cos ^2 \theta  d\tilde{\psi} +\sin ^2 \theta  d\tilde{\phi} )$, $\tilde{\nu} = a \cos ^2 \theta  d\tilde{\psi} +a \sin ^2 \theta  d\tilde{\phi}$, and the parameters are given through $a$ as
\be
\lambda = \sqrt{1+3 \omega^2} , \qquad \omega = \frac{\sqrt{2 a}}{\sqrt{1-a}} \, 
.
\ee
We work with corotating angles $(\tilde{\psi},\tilde{\phi})$, related to original coordinates via $\psi = \tilde{\psi}+t$, $\phi = \tilde{\phi} + t$.
The 5d tetrads of the near-horizon metric are given by 
\begin{align}
e^0 &= \frac{ \omega r  dt}{4  \lambda }-\frac{3}{2} \omega ^2 \left(\cos ^2 \theta  d\tilde{\psi}+\sin ^2 \theta  d\tilde{\phi} \right)
,\\
e^1 &= \frac{\omega  dr}{2 \lambda r}
,\\ 
e^2 &= 
\frac{1}{2} \omega  \sin (2 \theta ) (d\tilde{\psi} -d\tilde{\phi} ) \sin (\tilde{\psi} +\tilde{\phi} )+\omega  d\theta  \cos (\tilde{\psi} +\tilde{\phi} )
,\\ 
e^3 &=
\frac{1}{2} \omega  \sin (2 \theta ) (d\tilde{\psi} -d\tilde{\phi}
   ) \cos (\tilde{\psi} +\tilde{\phi} )-\omega  d\theta  \sin (\tilde{\psi} +\tilde{\phi} )
,\\ 
e^4 &= 
\lambda  \omega  \left(\cos ^2 \theta  d\tilde{\psi} +\sin ^2 \theta  d\tilde{\phi} \right)
.
\end{align}
For tetrads in the lifted 10d metric, in addition to the above tetrads, we also have 
\begin{align}
e^5 &= d\tilde{\alpha}
,\\ 
e^6 &= \cos \tilde{\alpha}  d\tilde{\beta}
,\\ 
e^7 &= \sin \tilde{\alpha}  \cos \tilde{\alpha}  \left( d\xi_1 -\sin ^2 \tilde{\beta}  d\xi_2 -\cos ^2 \tilde{\beta}  d\xi_3  \right)
,\\ 
e^8 &= \cos \tilde{\alpha}  \sin \tilde{\beta}  \cos \tilde{\beta}  (d\xi_2 -d\xi_3 )
,\\ 
e^9 &= - \cos ^2 \tilde{\alpha}  \left(\sin ^2 \tilde{\beta}  d\xi_2 +\cos
^2 \tilde{\beta}  d\xi_3 \right)-\sin ^2 \tilde{\alpha}  d\xi_1 
+ \frac{2 A}{3}
. 
\end{align}
In this basis, the electromagnetic fields are explicitly given by
\begin{align}
F^{(2)} &= 
-\frac{3}{2} \left(
3 e^1 \wedge e^4 - \frac{2\lambda}{\omega} e^0 \wedge e^1 
- e^2 \wedge e^3
\right) ,
\\
F^{(5)} &= -4 (e^0 \wedge e^1 \wedge e^2 \wedge e^3 \wedge e^4 
+ e^5 \wedge e^6 \wedge e^7 \wedge e^8 \wedge e^9 
) \\
& \phantom{=} - 
\frac{2}{3} (e^5 \wedge e^7 + e^6 \wedge e^8) 
\wedge 
(\ast_{(5)} F^{(2)} - e^9 \wedge F^{(2)}) 
.
\end{align} 
Our goal is to verify the solutions to the Killing spinor equation 
\be 
\hat{\nabla}_M \epsilon_K 
\equiv 
(\partial_M + \frac{1}{4} \omega_{M A B} \Gamma^A \Gamma^B) \epsilon_K
+ \frac{i}{1920} F^{(5)}_{A_1 A_2 A_3 A_4 A_5} \Gamma^{A_1 A_2 A_3 A_4 A_5}  \Gamma_M  \epsilon_K = 0 
.
\ee
Here $M,N$ denotes spacetime indices and $A_i, B_i$ denotes frame indices; $\epsilon_K= \epsilon_R + i \epsilon_I$ where $\epsilon_R,\epsilon_I$ are both Majorana-Weyl spinors satisfying the chirality condition
\be 
\Gamma^{11} \epsilon_K \equiv \Gamma^0 \Gamma^1 \dots \Gamma^9 \epsilon_K = \epsilon_K 
.
\ee
A common strategy for solving the Killing spinor equation, is to first use the algebraic integrability condition $[\hat{\nabla}_{M_1} , \hat{\nabla}_{N_1}] \epsilon_K = 0$ to restrict the space of possible solutions, and then solve a simplified Killing spinor equation on the restricted subspace. In our case, the integrability condition takes the form \cite{Gauntlett:2004cm}
\begin{align}
\bigg[ R_{M_1 N_1 S_1 S_2}   &
- \frac{1}{48} F^{(5)}_{\phantom{(5)} M_1 S_1 R_1 R_2 R_3} F_{\phantom{(5)} N_1 S_2}^{(5) \phantom{N_1 S_2} R_1 R_2 R_3} \bigg] \Gamma^{S_1 S_2} \epsilon_K \nonumber
\\
&+\left[
\frac{i}{24} \nabla_{[M_1} F^{(5)}_{\phantom{(5)} N_1] S_1 S_2 S_3 S_4}
+ \frac{1}{96}
F^{(5)}_{\phantom{(5)} M_1 N_1 R_1 R_2 S_1} 
F^{(5) R_1 R_2}_{\phantom{(5) R_1 R_2}S_2 S_3 S_4} 
\right] \Gamma^{S_1 S_2 S_3 S_4} \epsilon_K = 0
.
\label{appendix-eq-integrability condition}
\end{align}
This implies the following projection conditions 
\be 
\Gamma^{23} \epsilon_K = - i \epsilon_K , \qquad \Gamma^{57} \epsilon_K = - i \epsilon_K, \qquad \Gamma^{0149} \epsilon_K = i \epsilon_K .
\ee
Let us now choose a general constant spinor $\epsilon_0=\epsilon_{0,R}+i \epsilon_{0,I}$, subject to all four of the above projection conditions. Such a spinor is now labeled by four independent real parameters. We can split it into two chiralities under projector $P_{\pm} = \frac{1}{2}(1\pm \Gamma^{0 9})$ such that 
\be
\epsilon_0 = \epsilon_0^+ + \epsilon_0^-, \qquad 
\Gamma^{0 9} \epsilon_0^\pm =  \mp \epsilon_0^\pm . 
\ee
With these definitions, one can now verify that two independent spinors 
\begin{align}
\epsilon_1 &= \sqrt{\frac{8 \lambda}{\omega^2 r}} 
e^{-\frac{i}{2}(\xi_1 + \xi_2 + \xi_3)} \left[ 
\epsilon_0^+ - \left(
 \frac{i r t }{2} + \frac{3\omega}{2\lambda}
\right) \Gamma^4 \Gamma^9 \epsilon_0^+ 
\right]
,\\
\epsilon_2 &= \sqrt{\frac{\omega^2 r}{8 \lambda}} e^{-\frac{i}{2}(\xi_1 + \xi_2 + \xi_3)} \epsilon_0^- 
,
\label{appendix-eq:killing-spinors}
\end{align}
satisfy the Killing spinor equation. Because the above spinors are labeled by four independent real parameters, we conclude that the 10d lift of the near-horizon geometry preserves four supersymmetries. Note that this is in contrast with the full black hole geometry, which preserves only two supersymmetries \cite{Gauntlett:2004cm}. This means that there is a supersymmetry enhancement in the near-horizon region.
\\
\\
Knowing the Killing spinors of the near-horizon geometry, we can now find its Killing vectors by computing independent Killing spinor bilinears 
\be 
(\bar{\epsilon}_I \Gamma^A \epsilon_J) \tilde{e}_A , 
\ee
where $\tilde{e}_a$ denotes the dual tetrad basis, explicitly given by 
\begin{align}
\tilde{e}_0 &= \frac{ 4 \lambda \partial _t - r \omega (\partial _{\xi _1}+\partial _{\xi _2}+\partial _{\xi _3})}{r \omega}
,\\
\tilde{e}_1 &= \frac{2 \lambda  r \partial _r}{\omega }
,\\
\tilde{e}_2 &= \frac{ \cos (\tilde{\psi} +\tilde{\phi} )\partial _{\theta } +\tan \theta \, \sin (\tilde{\psi} +\tilde{\phi} ) \left(\partial _{\tilde{\psi} }- \cot ^2 \theta \, \partial _{\tilde{\phi} } \right)}{\omega }
,\\
\tilde{e}_3 &= \frac{- \sin (\tilde{\psi} +\tilde{\phi} )\partial _{\theta }+ \tan \theta \, \cos (\tilde{\psi} +\tilde{\phi} ) \partial _{\tilde{\psi} } - \cot \theta \, \cos (\tilde{\psi} +\tilde{\phi} ) \partial _{\tilde{\phi} }}{\omega }
,\\
\tilde{e}_4 &= \frac{6 \partial_t}{r}
+ 
  \frac{-2  \omega ^2 (\partial _{\xi _1} + \partial _{\xi _2} + \partial _{\xi _3})+\partial _{\tilde{\psi} }+\partial _{\tilde{\phi} }}{\lambda \omega}
,
\end{align}
together with
\begin{align}
\tilde{e}_5 &= \partial _{\tilde{\alpha} }
,\\
\tilde{e}_6 &=  \text{sec} \tilde{\alpha} \,  \partial _{\tilde{\beta} }
,\\
\tilde{e}_7 &= \frac{1}{4} \sin ^2(2 \tilde{\alpha} ) \left( \csc ^3 \tilde{\alpha}  
\, 
\text{sec} \tilde{\alpha} \, \partial _{\xi _1}-\csc \tilde{\alpha} \,  \text{sec} ^3 \tilde{\alpha} \, (\partial _{\xi _2}+\partial _{\xi
  _3})\right)
,\\
\tilde{e}_8 &=  \text{sec} \tilde{\alpha}  \cot \tilde{\beta} \partial _{\xi _2} - \text{sec} \tilde{\alpha}  \tan \tilde{\beta} \partial _{\xi _3}
,\\
\tilde{e}_9 &= -\partial _{\xi _1}-\partial _{\xi _2}-\partial _{\xi _3}
.
\end{align}
The bilinears are found to be 
\begin{align}
(\bar{\epsilon}_2 \Gamma^A \epsilon_1 )\tilde{e}_A &
=\bar{\epsilon}^-_0 \Gamma^4 \epsilon_0^+ \frac{2 i \lambda}{\omega} (\mathcal{Z}+D)  = \mathcal{Z}+D
, 
\label{appendix-eq-killing-spinors-anticommutators1}
\\
(\bar{\epsilon}_1 \Gamma^A \epsilon_2 )\tilde{e}_A &
=\bar{\epsilon}^+_0 \Gamma^4 \epsilon_0^- 
\frac{2 i \lambda}{\omega} (\mathcal{Z}-D) = \mathcal{Z}-D
, 
\\
(\bar{\epsilon}_2 \Gamma^A \epsilon_2 )\tilde{e}_A &
= \bar{\epsilon}^-_0 \Gamma^0 \epsilon_0^- \frac{2 i \lambda}{\omega} E_+ =  E_+
, 
\\
(\bar{\epsilon}_1 \Gamma^A \epsilon_1 )\tilde{e}_A &
= \bar{\epsilon}^+_0 \Gamma^0 \epsilon_0^+ \frac{2 i \lambda}{\omega} E_- =  E_-
,\label{appendix-eq-killing-spinors-anticommutators4} 
\end{align}
where we have chosen a specific normalization for constant spinors $\epsilon_0^\pm$
and introduced four independent Killing vectors $(\mathcal{Z},D,E_+ , E_-)$. They are defined as 
\begin{align}
D &= 
-t \partial_t + r \partial_r 
,\\
\mathcal{Z} &= 
\frac{i \omega^2}{\lambda^2} (
\partial_{\xi_1}+\partial_{\xi_2}+\partial_{\xi_3} 
) - \frac{i}{2\lambda^2} (\partial_{\tilde{\phi}} + \partial_{\tilde{\psi}})
,
\\
E_+ &= 
- \frac{i \omega^2}{4\lambda} \partial_t 
,
\\ 
E_- &= 
\frac{8 i \lambda }{\omega^2} r t \partial_r 
+ \left( 
- \frac{4 i \lambda }{\omega^2}  t^2
+ \frac{ 4i   (9\omega^2 - 4\lambda^2 )}{\omega^2 \lambda}\frac{1}{r^2}
\right) \partial_t 
\\
& 
+ \frac{8i }{\lambda^2 \omega }\frac{1}{r} (\partial_{\xi_1}
+\partial_{\xi_2} + \partial_{\xi_3}
) + \frac{12 i  }{\lambda^2 \omega }\frac{1}{r} 
(\partial_{\tilde{\phi}} + \partial_{\tilde{\psi}} )
.
\end{align}
With these definitions one can verify the algebra 
\begin{align}
[D,E_{\pm}] &= \pm E_{\pm} , \qquad [D,\mathcal{Z}] = 0
,\label{appendix-eq-killing-vectors-commutators1}
\\
[\mathcal{Z}, E_{\pm}] &= 0
,\qquad
[E_+ , E_- ]= 2 D ,
\label{appendix-eq-killing-vectors-commutators2}
\end{align}
which allows us to identify $(D,E_+ ,E_-)$ as $SL(2,\mathbb{R})$ generators and $\mathcal{Z}$ as the $U(1)$ generator of R-symmetry.
\\
We are now in a position to fully determine the isometry superalgebra of the near-horizon geometry \cite{Beck:2017wpm,Gauntlett:1998kc,Gauntlett:1998fz,Figueroa-OFarrill:1999klq}. A standard prescription is to associate bosonic generators $\mathcal{Q}_B (k_i)$ and fermionic generators $\mathcal{Q}_F (\epsilon_I)$ to Killing vectors and Killing spinors respectively. The superalgebra is then determined from the relations
\begin{align}
[\mathcal{Q}_B (k_i) , \mathcal{Q}_B (k_j)] &= \mathcal{Q}_B ([k_i,k_j]) , 
\label{appendix-eq-superalgebra-bosonic}
\\ 
[\mathcal{Q}_B (k) , \mathcal{Q}_F (\epsilon) ] &= \mathcal{Q}_F (\mathbb{L}_k \epsilon) ,
\label{appendix-eq-superalgebra-mixed}
\\
\{ \mathcal{Q}_F (\epsilon_I ) , \mathcal{Q}_F (\epsilon_J ) \} &= \mathcal{Q}_B (\epsilon_I \gamma \epsilon_J) 
\label{appendix-eq-superalgebra-fermionic}
,
\end{align}
where we introduced the spinorial Lie derivative \cite{Figueroa-OFarrill:1999klq} through which Killing vectors act on Killing spinors
\be 
\mathbb{L}_k \epsilon = k^M (\partial_M \epsilon + 
\frac{1}{4} \omega_{M A B} \Gamma^A \Gamma^B \epsilon
) + \frac{1}{4} D_{[A}k_{B]} \Gamma^A \Gamma^A \epsilon .
\label{appendix-eq-spinorial-Lie-derivative}
\ee
In the above $\omega_{M AB}$ denotes the spin connection, and 
\be 
D_A k_B = \tilde{e}^M _A (\partial_M k_B - \omega_{M \phantom{C}B}^{\phantom{M}C} k_C) .
\ee
From \eqref{appendix-eq-superalgebra-bosonic} we immediately read off the bosonic part of isometry algebra as \eqref{appendix-eq-killing-vectors-commutators1},\eqref{appendix-eq-killing-vectors-commutators2}. Similarly, \eqref{appendix-eq-superalgebra-fermionic} together with \eqref{appendix-eq-killing-spinors-anticommutators1}-\eqref{appendix-eq-killing-spinors-anticommutators4} imply non-vanishing anticommutators 
\begin{align}
\{ \bar{\mathcal{Q}}_2,\mathcal{Q}_1 \} &= \mathcal{Z} + D , & \{ \bar{\mathcal{Q}}_1,\mathcal{Q}_2 \} &= \mathcal{Z} - D ,
\\ 
\{ \bar{\mathcal{Q}}_2,\mathcal{Q}_2 \} &= E_+ , & \{ \bar{\mathcal{Q}}_1,\mathcal{Q}_1 \} &= E_- .
\end{align}
Lastly, to determine the commutators of bosonic and fermionic generators we can either compute the spinorial Lie derivatives or try to solve the super Jacobi identity. Here we take the former approach. The Lie derivatives of the Killing spinors in the directions of Killing vectors are found to be 
\begin{align}
\mathbb{L}_Z \epsilon_1 &= \frac{1}{2} \epsilon_1 , & 
\mathbb{L}_Z \bar{\epsilon}_1 &= -\frac{1}{2} \bar{\epsilon}_1 ,&
\mathbb{L}_Z \epsilon_2 &= \frac{1}{2} \epsilon_2 , & 
\mathbb{L}_Z \bar{\epsilon}_2 &= -\frac{1}{2} \bar{\epsilon}_2 ,\\
\mathbb{L}_J \epsilon_1 &= -\frac{1}{2} \epsilon_1 , &
\mathbb{L}_J \bar{\epsilon}_1 &= -\frac{1}{2} \bar{\epsilon}_1 , &
\mathbb{L}_J \epsilon_2 &= \frac{1}{2} \epsilon_2 , &
\mathbb{L}_J \bar{\epsilon}_2 &= \frac{1}{2} \bar{\epsilon}_2 , \\
\mathbb{L}_{E_+} \epsilon_1 &= - \epsilon_2 , &
\mathbb{L}_{E_+} \bar{\epsilon}_1 &=  \bar{\epsilon}_2 , &
\mathbb{L}_{E_+} \epsilon_2 &= 0 , &
\mathbb{L}_{E_+} \bar{\epsilon}_2 &= 0 , \\
\mathbb{L}_{E_-} \epsilon_1 &= 0 , &
\mathbb{L}_{E_-} \bar{\epsilon}_1 &= 0 , &
\mathbb{L}_{E_-} \epsilon_2 &= - \epsilon_1 , &
\mathbb{L}_{E_-} \bar{\epsilon}_2 &=  \bar{\epsilon}_1 ,
\end{align}
from which we identify commutation relations
\begin{align}
[\mathcal{Q}_1,Z] &= -\frac{1}{2}\mathcal{Q}_1 , & [\bar{\mathcal{Q}}_1,Z] &= \frac{1}{2} \bar{\mathcal{Q}}_1 , &
[\mathcal{Q}_2,Z] &= -\frac{1}{2}\mathcal{Q}_2 , & [\bar{\mathcal{Q}}_2,Z] &= \frac{1}{2} \bar{\mathcal{Q}}_2 , 
\\
[\mathcal{Q}_1,J] &= \frac{1}{2}\mathcal{Q}_1 , & [\bar{\mathcal{Q}}_1,J] &= \frac{1}{2} \bar{\mathcal{Q}}_1 , &
[\mathcal{Q}_2,J] &= -\frac{1}{2}\mathcal{Q}_2 , & [\bar{\mathcal{Q}}_2,J] &= -\frac{1}{2} \bar{\mathcal{Q}}_2 ,
\\
[\mathcal{Q}_1,E_+] &= \mathcal{Q}_2 , & [\bar{\mathcal{Q}}_1,E_+] &= -\bar{\mathcal{Q}}_2 , &
[\mathcal{Q}_2,E_+] &= 0 , & [\bar{\mathcal{Q}}_2,E_+] &=0 ,
\\
[\mathcal{Q}_1,E_-] &= 0 , & [\bar{\mathcal{Q}}_1,E_-] &= 0 , &
[\mathcal{Q}_2,E_-] &= \mathcal{Q}_1 , & [\bar{\mathcal{Q}}_2,E_-] &= -\bar{\mathcal{Q}}_1 ,
\end{align}
As a consistency check, we can verify that the above superalgebra indeed satisfies the super Jacobi identity
\be 
(-1)^{|x||z|}[x,[y,z]] + (-1)^{|y||x|}[y,[z,x]] + (-1)^{|z||y|}[z,[x,y]] = 0, 
\ee
where one has to choose $|x|=0$ for bosonic and $|x|=1$ for fermionic generators, and the (anti)commutators are defined through
\be 
[x,y] = -(-1)^{|x||y|} [y,x] .
\ee

\bibliographystyle{utphys2}
{\small \bibliography{Biblio}{}}

\end{document}